\documentclass[12pt,letterpaper]{article}
\usepackage{putex}
\usepackage{graphicx}
\usepackage{latexsym,amsmath,amsfonts,amssymb,amsthm}
\usepackage{empheq}
\usepackage{bbm}
\usepackage{cite}
\usepackage{indentfirst}
\usepackage{booktabs,array}
\usepackage{placeins}
\usepackage{mathtools}
\DeclarePairedDelimiter\ceil{\lceil}{\rceil}

\usepackage{colonequals}
\usepackage{cancel}
\usepackage[mode=image|tex]{standalone}
\usepackage[dvipsnames]{xcolor}
\usepackage{tikz}
\usepackage{tikz-cd}
\usepackage{adjustbox}
\usepackage{enumitem}
\usetikzlibrary{decorations.markings}
\usetikzlibrary{decorations.text}
\usepackage{graphicx}
\usepackage[margin=10pt,font=small,labelfont=bf]{caption}
\usepackage{subcaption}
\usepackage{xcolor}
\usepackage{float}
\theoremstyle{definition}

\usepackage{microtype}
\usepackage{hyperref}
\hypersetup{unicode}
\hypersetup{linktoc = all}
\hypersetup{pdfborderstyle={/S/U/W 0.5}}
\hypersetup{linkbordercolor = gray}
\hypersetup{bookmarksnumbered}
\usepackage[T1]{fontenc}
\usepackage[utf8]{inputenc}
\usepackage{lmodern}

\setlist{itemsep=2pt plus 1pt minus 1pt, topsep=2pt plus 1pt minus 1pt}

\renewenvironment{thebibliography}[1]{%
\begin{oldthebibliography}{#1}%
\setlength\itemsep{-2mm}%
}%
{%
\end{oldthebibliography}%
}

\numberwithin{equation}{section}

\begin{document}


\title{\begin{LARGE}
Surface defects, flavored modular differential equations and modularity
\end{LARGE}}

\authors{Haocong Zheng, Yiwen Pan and Yufan Wang
\medskip\medskip\medskip\medskip
 }

\institution{SYS}{
School of Physics, Sun Yat-Sen University, \cr
$\;\,$ Guangzhou, Guangdong, China}

\abstract{
\begin{onehalfspace}
  Every 4d $\mathcal{N} = 2$ SCFT $\mathcal{T}$ corresponds to an associated VOA $\mathbb{V}(\mathcal{T})$, which is in general non-rational with a more involved representation theory. Null states in $\mathbb{V}(\mathcal{T})$ can give rise to non-trivial flavored modular differential equations, which must be satisfied by the refined/flavored character of all the $\mathbb{V}(\mathcal{T})$-modules. Taking some $A_1$ theories $\mathcal{T}_{g,n}$ of class-$\mathcal{S}$ as examples, we construct the flavored modular differential equations satisfied by the Schur index. We show that three types of surface defect indices give rise to common solutions to these differential equations, and therefore are sources of $\mathbb{V}(\mathcal{T})$-module characters. These equations transform almost covariantly under modular transformations, ensuring the presence of logarithmic solutions which may correspond to characters of logarithmic modules.
\end{onehalfspace}
}

\preprint{}
\setcounter{page}{0}
\maketitle


{
\setcounter{tocdepth}{2}
\setlength\parskip{-0.7mm}
\tableofcontents
}

\newpage

\section{Introduction}

Four dimensional superconformal field theories (SCFTs) with $\mathcal{N} = 2$ supersymmetry are fascinating objects to study, as they are constrained enough to allow various exact computations, and also rich enough to generate numerous physical and mathematical interesting structures.

One remarkable example is the SCFT/VOA correspondence between the 4d $\mathcal{N} = 2$ SCFTs and 2d vertex operator algebras (VOAs) \cite{Beem:2013sza}, which maps the OPE algebra of the Schur operators in any 4d SCFT $\mathcal{T}$ to that of an associated VOA $\mathbb{V}(\mathcal{T})$. According to the correspondence, the Schur limit $\mathcal{I}$ of the 4d $\mathcal{N} = 2$ superconformal index equals the vacuum character of the associated VOA, the $c$ central charge and the flavor central charges of $\mathcal{T}$ are related to the 2d central charge and the levels of affine subalgebras of $\mathbb{V}(\mathcal{T})$ by simple proportionality
\begin{align}
	c_\text{2d} = - 12 c_\text{4d}, \qquad k_\text{2d} = - \frac{1}{2} k_\text{4d} \ .
\end{align}
The minus signs imply that whenever the 4d theory is unitary, the associated VOA will be non-unitary.

Like the Lie algebras, VOAs are interesting objects to study from a representation-theoretic point of view, as they admit many or infinite interesting modules. When a VOA is rational\footnote{A rational VOAs is special case of a lisse/$C_2$-cofinite VOA, which is a VOA with zero dimensional associated variety \cite{zhu1996modular,arakawa2012remark,Arakawa:2015jya,Xie:2019vzr}. In the SCFT/VOA correspondence, the associated variety of an associated VOA equals the Higgs branch of the 4d theory \cite{Beem:2017ooy}. Therefore, rationality of the associated VOA implies the absence of Higgs branch in 4d, and in particular, absence of flavor symmetry.}, namely, when it admits only finitely many irreducible modules whose characters form a vector-valued modular function, it could be considered as the chiral (symmetry) algebra of a rational conformal field theory (RCFT), with its modules corresponding to the primaries of the RCFT. Outside of the realm of RCFT, the representation theory of a VOA could be much more complicated. For instance, logarithmic modules may be present on which $L_0$ does not act diagonally and the corresponding character is logarithmic. In general, the associated VOAs of class-$\mathcal{S}$ theories are not rational\footnote{Theories of class-$\mathcal{S}$ in general have non-trivial Higgs branches. For genus-zero theories, the associated VOAs are shown to be quasi-lisse \cite{Arakawa:2016hkg,Arakawa:2018egx} (i.e., the associated variety has finitely many symplectic leaves).}.

Fortunately, there are tools that may help explore the structure of the modules of the associated VOAs. Crucially, sources of modules can be found in the 4d physics. In a 4d $\mathcal{N} = 2$ SCFT $\mathcal{T}$, one can introduce surface operators that perpendicularly penetrate the VOA plane at the origin while simultaneously preserve a 2d $\mathcal{N} = (2,2)$ superconformal subalgebra of the 4d superconformal algebra \cite{Gukov:2008sn,Gadde:2013dda,Gaiotto:2012xa,Cordova:2017mhb,Bianchi:2019sxz}. It is conjectured that such a defect corresponds to a non-vacuum (twisted) module of the associated VOA $\mathbb{V}(\mathcal{T})$ \cite{Cordova:2016uwk,Cordova:2017mhb,Pan:2017zie,Nishinaka:2018zwq,Bianchi:2019sxz,Pan:2021ulr,Beem:203X}. In particular, the character of such module should coincide with the Schur index in the presence of the defect. In cases where $\mathbb{V}(\mathcal{T})$ have been explicitly known, e.g., when $\mathcal{T}$ is an Argyres-Douglas theory, surface defects that arise from the Higgsing prescription have been identified with the modules of the associated VOAs. However, in general it remains challenging to verify the conjecture.

Another tool that comes in handy is the the modular differential equations. In an RCFT, the chiral symmetry algebra has finitely many irreducible modules, whose characters $\chi_i$ form a vector-valued modular form of weight-zero with respect to $SL(2, \mathbb{Z})$ or a suitable subgroup. As a result, any module character (of the chiral algebra) must satisfy a universal unflavored modular differential equation whose coefficients are ratios of Wronskian matrices made out of characters $\chi_i$ \cite{Eguchi:1987qd,MATHUR1988303}. These coefficients are strongly constrained by their modularity and this fact has been extensively exploited to classify RCFTs with a fixed number of chiral primaries (with or without fermions) \cite{Chandra:2018pjq,Mukhi:2020gnj,Das:2020wsi,Kaidi:2021ent,Das:2021uvd,Bae:2020xzl,Bae:2021mej,Duan:2022ltz}. Modular differential equations also arise in the context of SCFT/VOA correspondence. As shown in \cite{Beem:2017ooy}, the stress tensor of the associated VOA must be nilpotent up to an element $\varphi$ in the subspace $C_2(\mathbb{V}(\mathcal{T}))$ and a null state $\mathcal{N}$, $(L_{-2})^n |0\rangle = \mathcal{N} + \varphi$ for some $n \in \mathbb{N}_{>0}$. Combining with Zhu's recursion formula \cite{zhu1996modular,Gaberdiel:2008pr,Mason:2008zzb} that computes torus one-point functions, the nilpotency may lead to a non-trivial unflavored modular differential equation satisfied by the unflavored Schur index. Such equation has been exploited to classify rank-two 4d $\mathcal{N} = 2$ SCFTs \cite{Kaidi:2022sng}. The same reasoning naturally generalizes to characters of other (twisted) modules of the associated VOA, and one expects the untwisted characters to satisfy the same equation, while the twisted characters to satisfy a twisted version of the equation.

When the 4d theory has flavor symmetry, flavor fugacities can be introduced into the Schur index and defect indices. With the help of the flavored Zhu's recursion formula \cite{Krauel:2013lra,Beem:202X,Pan:2021ulr}, some null states lead to flavored modular differential equations. Note that there are usually additional null states giving rise to a few more equations besides the one corresponding to the nilpotency of the stress tensor. The character of any (twisted) $\mathbb{V}(\mathcal{T})$-module are expected to satisfy all of these (twisted version of) differential equations simultaneously, and therefore are heavily constrained.

This paper aims to further explore the relation between surface defects in a class of 4d $\mathcal{N} = 2$ SCFTs and the module characters of their associated VOAs. For simplicity we will focus on the class-$\mathcal{S}$ theories of type-$A_1$: these theories are the simplest in the sense that their Schur indices and vortex defect indices (from Higgsing) are known in closed-form in terms of some well-known analytic functions \cite{Pan:2021mrw}. We construct the (flavored) modular differential equations that their Schur indices satisfy, and study the common solutions to such equations.

It turns out that there are several physical sources of common solutions: the Schur index $\mathcal{I}_{g,n}$ itself obviously, and vortex defect indices \cite{Gaiotto:2012xa} $\mathcal{I}_{g,n}^\text{defect}(k = \text{even})$ (namely, with even vorticity $k$), the defect indices of Gukov-Witten type surface defects \cite{Gukov:2008sn}, and some surface defects related to modular transformations. Based on these computational results, we conjecture that these surface defects indeed correspond to non-vacuum modules of the associated VOA. Furthermore, vortex defects with odd vorticities are solutions to some twisted version of the differential equations, and therefore it is natural to associate them with the twisted modules.

Although the presence of additional flavored modular differential equations makes the special equation (temporarily called $\text{eq}_\text{Nil}$) from the nilpotency of the stress tensor seem less prominent, in several examples we find that $\text{eq}_\text{Nil}$ actually contains all the information on the allowed flavored characters. The key is modularity. When flavored, the coefficients of the flavor modular differential equations are no longer modular forms, but rather quasi-Jacobi forms. Under suitable modular transformation, $\text{eq}_\text{Nil}$ does transform and it actually generates all the necessary modular differential equations of lower weights. Schematically,
\begin{align}
	S(\text{eq}_\text{Nil}) = \sum_{m,n} \tau^m \mathfrak{b}^n \text{(FMDEs of lower weights)}_{m,n} \ .
\end{align}
They together determine all the allowed non-logarithmic and logarithmic characters.

When unflavored, the presence of logarithmic solutions is expected whenever the indicial roots are integral-spaced, e.g., by the Frobenius method. See also \cite{2002math......6022K,Kaneko:2013up,Arakawa:2016hkg}. In the cases we have studied where the Schur and vortex defect indices have closed-form expressions, these logarithmic solutions are just modular transformations of the non-logarithmic solutions, thanks to the modularity of the coefficients which makes the differential equations covariant (or invariant, up to an overall factor of $\tau^n$) under suitable modular transformations. However, the quasi-Jacobi-ness upon flavoring would naively breaks this logic. Luckily, the covariance can be almost restored by introducing some additional fugacities $y_i$ associated with the flavor central charges, and this leads to the generation of flavored modular differential equations of lower weights we just mentioned.

The organization of this paper is as follows. In section \ref{section:review}, we recall some basics of the SCFT/VOA correspondence and surface defects in 4d $\mathcal{N} = 2$ SCFTs. In particular we review the closed-form expression for the Schur index of all $A_1$ class-$\mathcal{S}$ theories and the defect indices from Higgsing. We also recall how modular differential equation arise in the context of 2d RCFT and 4d SCFT. In section \ref{section:beta-gamma}, we analyze in detail the $\beta \gamma$ system of conformal weighs $\frac{1}{2}$ (also known as the symplectic bosons). In both the untwisted and twisted sector, we construct flavored modular differential equations from trivial null states in the vacuum module, and study their common solutions and modularity. In section \ref{section:a1} we focus on the $A_1$ class-$\mathcal{S}$ theories $\mathcal{T}_{g,n}$, and in simple examples study their associated (flavored) modular differential equaitons and the solutions given in terms of different defect indices.


\section{SCFT/VOA correspondence and MDEs\label{section:review}}

\subsection{The Schur index}

4d $\mathcal{N} = 2$ SCFT has been one of the most interesting subject as it bridges different branches in mathematical physics. These theories are substantially constrained by the symmetry whith allows exact computation of many quantities, yet they retain extremly rich internal mathematical and physical structures. We recall that the 4d $\mathcal{N} = 2$ superconformal algebra $\mathfrak{su}(2,2|2)$ \footnote{In Euclidea signature, the superconformal algebra is $\mathfrak{su}^*(4|2)$ instead.} contains the generators
\begin{align}
	P_{\alpha \dot \alpha}, \quad
	K^{\dot \alpha \alpha}, \quad
	D, \quad
	M_\alpha{^\beta}, \quad
	M^{\dot \alpha}{_{\dot \beta}}, \quad
	R^I{_J}, \quad
	Q^I_\alpha, \quad
	S_I^\alpha, \quad
	\tilde Q_{I \dot \alpha}, \quad
	\tilde S^{I \dot \alpha} \ .
\end{align}
The (anti-)commutation relations can be found in  \cite{Beem:2013sza}.

Two quantities of a 4d $\mathcal{N} = 2$ SCFT attract considerable attention, the $S^4$ partition function \cite{Pestun:2007rz} and the superconformal index $\mathcal{I}(p,q,t)$ (which is also an $S^3 \times S^1$-partition funciton) \cite{Kinney:2005ej,Gadde:2011uv,Rastelli:2014jja},
\begin{align}
  \mathcal{I}(p,q,t)
  \coloneqq \operatorname{tr} (-1)^F
    p^{\frac{1}{2}(\Delta - 2j_1 - 2 \mathcal{R} - r)}
    q^{\frac{1}{2}(\Delta + 2j_1 - 2 \mathcal{R} - r)}
    t^{\mathcal{R} + r}
    e^{- \beta \{\tilde Q_{2 \dot -}, \tilde S^{2 \dot -}\}} a^f\ .
\end{align}
Both quantities participate in certain AGT-type 4d/2d correspondence when the theory under consideration is of class-$\mathcal{S}$ \cite{Alday:2009aq,Gaiotto:2009we,Hosomichi:2010vh,Bonelli:2011wx,Bonelli:2010gk,Gomis:2014eya,Gomis:2016ljm,Gadde:2011ik,Benini:2011nc,Mekareeya:2012tn,Alday:2013kda,Cordova:2015nma,Xie:2016evu,Buican:2017uka,LeFloch:2017lbt}\footnote{The AGT correspondence is also extensively studied in the presence of surface defects. See for example \cite{Alday:2010vg,Alday:2009fs}. In this context, modular property (with respect to the complexified gauge coupling $\tau_\text{gauge} \coloneqq \frac{\theta}{2\pi} + \frac{4\pi i}{g_\text{YM}^2}$) of the effective superpotential $\mathcal{W}$ is also studied \cite{Ashok:2017odt}, where the modular anomaly equation \cite{Minahan:1997if,Billo:2013jba} determines $\mathcal{W}$. See also \cite{Huang:2022bry} for an application of modular anomaly equation to Schur index.}. In comparison, the index is much simpler as it is independent of exactly marginal deformations of the theory. In the context of the 4d/2d correspondence, the superconformal index of a theory $\mathcal{T}[\Sigma]$ equals a topological correlator on the associated Riemann surface $\Sigma$ \cite{Gadde:2011ik}. In different limits, the superconformal index often enjoys supersymmetry enhancement, receiving contributions only from states annihilated by more than two supercharges.

The Schur index of a 4d $\mathcal{N}$ = 2 SCFT is the Schur limit $t \to q$ of the full superconformal index $\mathcal{I}(p,q,t)$ \cite{Gadde:2011uv,Beem:2013sza}, and can be written simply as a (super)trace over the Hilbert space of states in the radial quantization,
\begin{align}\label{def:Schur-index}
	\mathcal{I} = \operatorname{tr}(-1)^F
	e^{- \beta_1\{Q^1_-, S_1^-\}}
	e^{- \beta_2 \{\tilde Q_{2 \dot - },\tilde S^{2 \dot -}\}}
	q^{E - R}	\mathbf{b}^\mathbf{f} \ .
\end{align}
Here $E$ is the conformal dimension, $R$ the $SU(2)_\mathcal{R}$-charge generator defined by $2 R = R^1{_1} - R^2{_2}$, and $F$ the fermion number. Bolded letters $\mathbf{b}$ and $\mathbf{f}$ denote collectively any flavor fugacities and the associated Cartan generators of the flavor group that one may include in the trace. Thanks to the anti-commutivity and the neutrality under $E - R$ (and $\mathbf{f}$) of the two pairs of supercharges $\tilde Q_{2 \dot - },\tilde S^{2 \dot -}$ and $ Q^1_-, S_1^-$, the index is actually independent of $\beta_1$, $\beta_2$, and the $(-1)^F$ insertion leads to vast cancellations between bosonic and fermionic states. The only contributions to the Schur index are from the states satisfying the Schur conditions,
\begin{align}
	\{Q^1_-, S_1^-\} = \{\tilde Q_{2 \dot - },\tilde S^{2 \dot -}\} = 0, \qquad\Leftrightarrow \qquad
	E - 2R - M^\chi = r + M^\text{def} = 0\ .
\end{align}
Here, $M^\chi$ and $M^\text{def}$ denote the spin under the rotations within $\mathbb{R}^2_{x_3, x_4}$ and $\mathbb{R}^2_{x_1, x_2}$, or in other words, the eigenvalues of $M_+{^+} + M^{\dot +}{_{\dot +}}$ and $M_+{^+} - M^{\dot +}{_{\dot +}}$ respectively. The $U(1)_r$ generator $r \equiv \frac{1}{2}(R^1{_1} + R^2{_2})$. These states correspond to the so-called Schur operators in the 4d theory which are typically restricted to the $\mathbb{R}^2_{x_3, x_4}$ plane. 

As a superconformal index, the Schur index is invariant under exactly marginal deformation of the 4d theory. Exploiting such independence, the Schur index of Lagrangian 4d $\mathcal{N} = 2$ SCFTs can be easily computed in the free limit. The result is organized into a contour integral
\begin{align}\label{Schur-index-integral}
	\mathcal{I}
	= & \ \frac{(-i)^{\operatorname{rank}\mathfrak{g} - \dim \mathfrak{g}}}{|W|}\oint \prod_{A = 1}^{\operatorname{rank}\mathfrak{g}} \frac{d a_A}{2\pi i a_A}
	\eta(\tau)^{- \dim \mathfrak{g} + 3 \operatorname{rank}\mathfrak{g}}  \prod_{\alpha \ne 0}\vartheta_1(\alpha(\mathfrak{a}))
	\prod_{w \in \mathcal{R}} \frac{\eta(\tau)}{\vartheta_4(w (\mathfrak{a} + \mathfrak{b}))} \nonumber \\
	\coloneqq & \ \oint \frac{da}{2\pi i a} \mathcal{Z}(\mathfrak{a}) \ .
\end{align}
Here $\mathfrak{g}$ denotes the gauge algebra with the Weyl group $W$, $\mathcal{R}$ denotes the joint representation of the gauge and flavor group in which the hypermultiplets transform. The $\vartheta_i$ are the Jacobi theta functions and $\eta$ the Dedekind $\eta$-function. Their definitions and properties are collected in appendix \ref{app:specialfuctions}. 
Through out this paper, the letters $\mathfrak{a}, \mathfrak{b}$, $\ldots$ in fraktur font are related to the letters $a, b$, $\ldots$ by
\begin{align}
	a_A = e^{2\pi i \mathfrak{a}_A}, \qquad
	b_j = e^{2\pi i \mathfrak{b}_j}, \qquad
	\ldots, \qquad
	q = e^{2\pi i \tau} \ .
\end{align}
The integration contour of each integration over $a_A$ is taken to be the unit circle $|a_A| = 1$, and $|b_j| = 1$. Note that there is no pole along the integration contour, since the zeroes of $\vartheta_4(\mathfrak{z})$ are given by $\mathfrak{z} = \frac{\tau}{2} + m + n \tau$.

The contour integral can be reproduced from a supersymmetric localization computation on $S^3 \times S^1$ \cite{Pan:2019bor,Dedushenko:2019yiw}. In radial quantization, the Euclidean spacetime $\mathbb{R}^4$ is viewed as $S^3 \times \mathbb{R}$ where $\mathbb{R}$ denotes the radial direction. The Schur index as a trace over the Hilbert space, can be equivalently computed by first compactifying the radial $\mathbb{R} \to S^1$ and placing some appropriate background metric (that depends on a complex modulus $\tau$ controlling the relative size and angle between $S^3$ and $S^1$) and $\mathcal{R}$-symmetry gauge fields. Let us parametrize $S^3$ by a coordinate system $\varphi, \chi, \theta$ adapted to the $T^2$-fibration structure of $S^3$, with ranges $\varphi, \chi \in [0, 2\pi]$ and $\theta \in [0, \frac{\pi}{2}]$. The space $S^3_{\varphi, \chi, \theta} \times S^1_t$ has a $T^2_{\varphi, t}$ subspace at $\theta = 0$, and another $T^2_{\chi, t}$ subspace at $\theta = \frac{\pi}{2}$. The path integral of a 4d $\mathcal{N} = 2$ Lagrangian SCFT localizes to an ordinary integral of a 2d path integral on $T^2_{\varphi, t}$ over flat dynamical gauge fields of the form $A \sim \mathfrak{a} dt$,
\begin{align}
	\mathcal{I} = \oint \frac{da}{2 \pi i a} \int D\Phi e^{-S^{T^2_{\varphi, t}}(a)} \ , \qquad
	\mathfrak{a} \in \text{Cartan of the gauge group, } a \sim e^{2\pi i \mathfrak{a}} \ .
\end{align}

The integrand $\mathcal{Z}$ of (\ref{Schur-index-integral}) enjoys a crucial property: ellipticity \cite{Razamat:2012uv}. By that we mean that the integrand, as a meromorphic function of the $\operatorname{rank}\mathfrak{g}$ variables $\mathfrak{a}_A$, is separately doubly periodic under shifts $\mathfrak{a}_A \to \mathfrak{a}_A + \tau$, and $\mathfrak{a}_A \to \mathfrak{a}_A + 1$ of any one variable $\mathfrak{a}_A$,
\begin{align}
	\mathcal{Z}(\mathfrak{a}_A + 1) = \mathcal{Z}(\mathfrak{a}_A + \tau) = \mathcal{Z}(\mathfrak{a}_A) \ .
\end{align}
This fact enables an elementary method to evaluate the integral exactly, and organize the result in terms of finitely many twisted Eisenstein series and Jacobi theta functions \cite{Pan:2021mrw,Beemetal} (see also \cite{Bourdier:2015wda}). 

In this paper, we will mainly focus on the $A_1$ theories $\mathcal{T}_{g, n}$ of class-$\mathcal{S}$ associated to a genus-$g$ Riemann surface with $n$ punctures. Their fully flavored Schur index $\mathcal{I}_{g,n}$ has an elegant compact form given by
\begin{align}\label{Ign}
	\mathcal{I}_{g, n \ge 1} = & \ \frac{i^n}{2} \frac{\eta(\tau)^{n + 2g - 2}}{\prod_{j = 1}^{n} \vartheta_1(2 \mathfrak{b}_j)}
	\sum_{\alpha_j = \pm}\left(\prod_{j = 1}^{n}\alpha_j\right)
	\sum_{k = 1}^{n + 2g - 2} \lambda_k^{(n + 2g - 2)} E_k\left[\begin{matrix}
		(-1)^n \\ \prod_{j = 1}^{n}b_j^{\alpha_j}
	\end{matrix}\right] \ , \\
	\mathcal{I}_{g \ge 1, 0} = & \ \frac{1}{2}\eta(\tau)^{2g - 2}\sum_{k = 1}^{g - 1}
	\lambda_{2k}^{2g - 2}\left(E_{2k} + \frac{B_{2k}}{(2k)!}\right) \ .
\end{align}
Here $\mathcal{I}_{g,n}$ is fully flavored with respect to the class-$\mathcal{S}$ description, and $b_{i = 1, \ldots, n}$ denotes the $SU(2)$ flavor fugacities of the $n$ punctures. coefficients $\lambda$'s are rational numbers determined by the following recursion relations,
\begin{align}
	\lambda_\text{even}^{(\text{odd})} = & \ \lambda_\text{odd}^{(\text{even})} = 0, \qquad
	\lambda_1^{(1)} = \lambda_2^{(2)} = 0 \\
	\lambda_0^{(\text{even})} = & \  0, \qquad \lambda_1^{(2k + 1)} = \sum_{\ell = 1}^{k}\lambda_{2\ell}^{2k}(\mathcal{S}_{2\ell} - \frac{B_{2\ell}}{(2\ell)!})\ ,\\
	\lambda_{2m + 1}^{(2k + 1)} = & \ \sum_{\ell = m}^{k}\lambda_{2\ell}^{(2k)} \mathcal{S}_{2(\ell - m)}, \qquad
	\lambda_{2m + 1}^{(2k + 1)} = \sum_{\ell = m}^{k}\lambda_{2\ell}^{(2k)} \mathcal{S}_{2(\ell - m)}\ .
\end{align}
Here $B_n$ denotes the $n^\text{th}$ Bernoulli number, $\mathcal{S}$ are rational numbers that are given by the $2n^\text{th}$ coefficient of a $y$-series expansion,
\begin{align}
	\mathcal{S}_{2n} \coloneqq \left[\frac{y}{2} \frac{1}{\sinh \frac{y}{2}}\right]_{2n} \ .
\end{align}

\subsection{SCFT/VOA correspondence}

The Schur states contributing to the Schur index are harmonic with respect to the two pairs of supercharges, $(Q^1_-, S_-^1)$ and $(\tilde Q_{2 \dot -}, \tilde S^{2 \dot -})$. By the state/operator correspondence, any Schur state can be created by a Schur operator $\mathcal{O}(0)$ acting on the unique vacuum. This Schur operator at the origin (anti-)commute with all four supercharges. Translating the operator away from the origin typically breaks this BPS condition. However, one may consider moving the operator along the $\mathbb{R}^2_{34} = \mathbb{C}_{z, \bar z}$ plane by the twisted translation \cite{Beem:2013sza}
\begin{align}
	\mathcal{O}(z, \bar z) \coloneqq e^{- z L_{-1} - \bar z \widehat{L}_{-1}} \mathcal{O}(0)e^{+ z L_{-1} + \bar z \widehat{L}_{-1}} \ ,
\end{align}
where
\begin{align}
	L_{-1} = P_{+ \dot +}, \qquad \widehat{L}_{-1} = P_{- \dot -} + R^2{_1}\ .
\end{align}
The translated Schur operator $\mathcal{O}(z, \bar z)$ remains in the kernel of two supercharges $\mathbb{Q}_1 \coloneqq Q^1_- + \tilde S^{2 \dot -}$, $\mathbb{Q}_2 \coloneqq \tilde Q_{2 \dot -} - S^-_1$, and the $\bar z$-dependence is $\mathbb{Q}_{1,2}$-exact. Hence, at the level of cohomology, $\mathcal{O}(z) \coloneqq [\mathcal{O}(z, \bar z)]$ is holomorphic in $z$. Moreover, their OPE coefficients are also holomorphic, forming a 2d vertex operator algebra (VOA)/chiral algebra on the plane $\mathbb{C}_{z, \bar z}$ \cite{Beem:2013sza}. For any local unitary 4d $\mathcal{N} = 2$ SCFT $\mathcal{T}$, the associated VOA $\mathbb{V}(\mathcal{T})$ must be non-trivial and non-unitary, since a component of the $SU(2)_\mathcal{R}$ Noether current must be a non-trivial Schur operator, which gives rise to the stress tensor in the VOA with a negative central charge $c_\text{2d} = -12 c_\text{4d}$. Furthermore, any flavor symmetry $G$ in $\mathcal{T}$ will be associated to an affine subalgebra $\widehat{\mathfrak{g}}_{k_\text{2d}} \subset \mathbb{V}(\mathcal{T})$, whose generators descend from the moment map operator of the symmetry $G$, and they transform in the adjoint representation of $G$. For the $A_1$ theories $\mathcal{T} = \mathcal{T}_{g,n}$, the exact form (\ref{Ign}) of Schur index $\mathcal{I}_{g,n}$ highlights several flavor representations in which the VOA generators transform. In particular, the denominators $\vartheta_1(2 \mathfrak{b}_i)$ are tied to the $SU(2)$-adjoint moment map operators/affine currents of the $n$ puncture, while the $E_k$'s seem to come from the multi-fundamentals.

The associated VOA is an important invariant of 4d $\mathcal{N} = 2$ SCFT, constituting a VOA-valued TQFT for theories of class-$\mathcal{S}$ thanks to the nontrivial associativity properties descending from the class-$\mathcal{S}$ duality \cite{Gadde:2011ik,Beem:2014rza}. Under the correspondence, the Schur index of $\mathcal{T}$ is identified with the character of the vacuum module of $\mathbb{V}(\mathcal{T})$ and it plays a central role in the SCFT/VOA correspondence. (See, for example, \cite{Beem:2014kka,Beem:2014rza,Lemos:2014lua,Song:2015wta,Buican:2015ina,Xie:2016evu,Song:2017oew,Xie:2019zlb,Pan:2020cgc,Dedushenko:2019mzv,Xie:2021omd}.)

2d VOA is an interesing subject in its own right, say, from the representation theoretic persepctive. A VOA typically admits many modules besides the vacuum module. For those constituting an RCFT, there may be finitely many irreducible modules generated from the primaries. Unfortunately, the VOAs that arise in the SCFT/VOA correspondence often don't have nice properties such as rationality, making their modules less straightforward to study. However, one may still hope to access their modules through four dimensional physics.

In a 4d $\mathcal{N} = 2$ SCFT $\mathcal{T}$ one can insert surface defects that perpendicularly intersect at the origin with the $\mathbb{C}$-plane where the VOA resides. In particular, one may consider those preserving $\mathcal{N} = (2,2)$ superconformal symmetry on their support. It is generally believed that such defects correspond to non-vacuum modules of the associated VOA $\mathbb{V}(\mathcal{T})$, since the Schur operators in the 4d theory may act on the defect operators via a bulk-defect OPE. A quantity that captures information of such defect systems is the defect index which we now review.

\subsection{Defect indices \label{section:defect-indices}}

Let us focus on the $\mathcal{N} = (2,2)$ superconformal surface defects supported on $x_1, x_2$-plane which preserve the supercharges $\tilde Q_{2 \dot{-}}, \tilde S^{2 \dot -}$ \cite{Bianchi:2019sxz}. One can compute the full 4d $\mathcal{N} = 2$ superconformal index in the presence of such defect \cite{Gadde:2013dda}, which admits the usual Schur limit $t \to q$ \cite{Gadde:2013dda,Bianchi:2019sxz}. In this section we will briefly review the indices of three types of surface defects. We will see in later section that they all give rise to solutions to the modular differential equations, and therefore potentially correspond to VOA modules.

\subsubsection{Vortex defects}

A vortex defect that we are interested in is labeled by a natural number $k \in \mathbb{N}$. One begines with a 4d $\mathcal{N} = 2$ SCFT which is usually referred to as the $\mathcal{T}_\text{IR}$, and we will take it to be $\mathcal{T}_{g,n}$. First we assume $n \ge 1$, in which case the theory contains at least one $SU(2)_\text{f}$ flavor symmetry. This theory is then coupled to the theory of four hypermultiplets $\mathcal{T}_{0,3}$ by gauging an $SU(2)$ flavor symmetry associated to a puncture, by gauging the diagonal of $SU(2)_\text{f}$ and one $SU(2)$ flavor symmetry of $\mathcal{T}_{0,3}$. The resulting theory is denoted as $\mathcal{T}_\text{UV}$, which has an additional $SU(2)$ flavor symmetry, say, the $(n + 1)^\text{th}$ puncture. One then turns on a position dependent VEV for the corresponding moment map operator with a profile $\sim (x_1 + i x_2)^k$, triggering an RG-flow to the IR. When $k = 0$, the IR fixed point reproduces the original IR theory $\mathcal{I}_\text{IR}$, while for $k \ge 1$, the IR fixed point is $\mathcal{T}_\text{IR}$ coupled to a vortex defect. The Schur index of the resulting IR theory can be computed by \cite{Gaiotto:2012xa,Alday:2013kda,Nishinaka:2018zwq}
\begin{align}\label{def:Higgsing}
	\mathcal{I}_{g, n}^{\text{defect}} (k) = 2 (-1)^k \mathop{\operatorname{Res}}_{b_{n + 1} \to q^{\frac{1}{2} + \frac{k}{2}}} q^{- \frac{(k + 1)}{2}} \frac{\eta(\tau)^2}{b} \mathcal{I}_{g, n + 1} \ .
\end{align}
Inserting the exact formula for $\mathcal{I}_{g, n + 1}$, one arrives at the closed-form of the defect index \cite{Pan:2021mrw}\footnote{Note that $k = 0$ reproduces the original Schur index $\mathcal{I}_{g,n}$,
\begin{align}
	\mathcal{I}^\text{defect}_{g,n}(k = 0) = \mathcal{I}_{g,n} \ .
\end{align}}
\begin{align}\label{def:defect-index}
	\mathcal{I}^{\text{defect}}_{g, n}(k) = (-1)^k & \ \frac{i^n}{2}\frac{\eta(\tau)^{n + 2g - 2}}{\prod_{i = 1}^n\vartheta_1(2 \mathfrak{b}_i)}\\
	& \ \times \sum_{\alpha_i = \pm}\left(\prod_{i = 1}^{n}\alpha_i\right)
	\sum_{\ell = 1}^{n + 1 + 2g - 2}\tilde\lambda^{n + 1 + 2g - 2}_\ell(k + 1) E_\ell \left[\begin{matrix}
    	(-1)^{n + k}\\ \prod_{i = 1}^{n}b_i^{\alpha_i}
	\end{matrix}\right]\ ,
\end{align}
where $\tilde \lambda$ are rational numbers defined by
\begin{align}
	\tilde \lambda_\ell^{(n + 1 + 2g -2)}(k) \colonequals \sum_{\ell' = \ell}^{n + 1 + 2g -2} \left(\frac{k}{2}\right)^{\ell' - \ell} \frac{1}{(\ell' - \ell)!} \lambda_{\ell'}^{(n + 1 + 2g - 2)}\ .
\end{align}
Here we list a few values of $\tilde \lambda$ for readers' convenience.

The exact formula of $\mathcal{I}^\text{defect}_{g,n}(k)$ suggests that the defect indices are (combinations of) spectral-flowed vacuum character $\mathcal{I}_{g,n}$ \cite{Pan:2021mrw}. In particular, when $k = $ odd, the corresponding flowed modules are twisted modules where the multi-fundamental generators in the VOA have their conformal weights shifted by half-integers, while the affine currents associated to the $n$ punctures keep their weights (mod $1$). We will observe such pattern again when we discuss the flavored modular differential equations.

Alghough the above construction of a vortex defect generalizes to $g \ge 1, n = 0$ by cutting a handle and inserting a $\mathcal{T}_{0,3}$, the formula (\ref{def:defect-index}) is only valid for $n \ge 1$.  Indeed, with $n = 1$ the poles $b \to q^{\frac{1}{2} + \frac{k = \text{even}}{2}}$ are actually double poles due to the twisted Eisenstein series\footnote{See also \cite{Gaiotto:2012uq} for a discussion on higher order poles in the context of the Hall-Littlewood index.}. In more details,
\begin{align}
	\mathcal{I}_{g, 1} = \frac{i}{2} \frac{\eta(\tau)^{2g - 1}}{ \vartheta_1(2 \mathfrak{b}_1)}
	\sum_{\alpha_1 = \pm}\alpha_1
	\sum_{\ell = 1}^{2g - 1} \lambda_\ell^{(2g - 1)} E_\ell\left[\begin{matrix}
		-1 \\ b_1^{\alpha_1}
	\end{matrix}\right] \ .
\end{align}
The twisted Eisenstein series involved have simple poles at $b_1 = q^{\frac{1}{2} + \frac{k = \text{even}}{2}}$ (see \ref{Eisenstein-residues}), which collide with the simple poles of the $\vartheta_1(2 \mathfrak{b}_1)$. Still, the residue in (\ref{def:Higgsing}) can be computed using
\begin{align}
	\mathop{\operatorname{Res}}_{b \to q^{\frac{1}{2}}}\frac{1}{b}\frac{1}{\vartheta_1(2 \mathfrak{b})} E_{2n + 1} \begin{bmatrix}
		-1 \\ b
	\end{bmatrix}
	= e_n + \sum^{n - 1}_{\ell = 0} \frac{1}{4^\ell (2\ell+1)!} E_{2n - 2\ell}\ ,
\end{align}
where $e_n$ are some rational number. Combining with (\ref{Eisenstein-shift-property}), we have the following defect indices,
\begin{align}\label{def:defect-index-g0}
	\mathcal{I}_{g, 0}^\text{defect}(k = \text{even}) = & \ \eta(\tau)^{2g - 2} \sum_{\ell = 0}^{g - 1} c_\ell(k) E_{2\ell} \ ,\\
	\mathcal{I}_{g, 0}^\text{defect}(k = \text{odd}) = & \ \eta(\tau)^{2g - 2} \sum_{\ell = 0}^{g - 1} c'_\ell(k) E_{2\ell}\begin{bmatrix}
		-1 \\ 1
	\end{bmatrix} \ .
\end{align}
Here $c_\ell(k), c'_\ell(k)$ are some rational numbers that can be worked out explicitly.

Given $g $ and $ n \ge 1$, there are infinitely many vortex defects corresponding to $k \in \mathbb{N}$. However, we see from the exact form of $\mathcal{I}^\text{defect}_{g,n}(k)$ that they are all linear combinations of the fixed structures with different $\ell$,
\begin{align}
	\sum_{\alpha} \prod_{i = 1}^{n}\alpha_i E_\ell \begin{bmatrix}
		\pm 1 \\ \prod_{i = 1}^{n}b_i^{\alpha_i}
	\end{bmatrix} \ .
\end{align}
From the symmetry property (\ref{Eisenstein-symmetry}), such structure is non-zero only when $\ell = n$ mod 2 and $\ell > 0$ \footnote{Here we have assumed $n \ge 1$, and we also define $E_0\big[  \substack{\theta \\ \phi}  \big] = -1$; see appendix \ref{app:specialfuctions}.}. In particular, when $n = $ even, only the following $\ceil{\frac{n + 2g - 2}{2}}$ Eisenstein series contribute,
\begin{align}
	E_2 \begin{bmatrix}
		\pm 1 \\ \prod_{i} b_i^{\alpha_i}
	\end{bmatrix}, \qquad
	E_4 \begin{bmatrix}
		\pm 1 \\ \prod_{i} b_i^{\alpha_i}
	\end{bmatrix}, \qquad
	\ldots
	\qquad
	E_{n + 2g - 2} \begin{bmatrix}
		\pm 1 \\ \prod_{i} b_i^{\alpha_i}
	\end{bmatrix}\ ,
\end{align}
while when $n = $ odd, only
\begin{align}
	E_1 \begin{bmatrix}
		\pm 1 \\ \prod_{i} b_i^{\alpha_i}
	\end{bmatrix}, \qquad
	E_3 \begin{bmatrix}
		\pm 1 \\ \prod_{i} b_i^{\alpha_i}
	\end{bmatrix}, \qquad
	\ldots
	\qquad
	E_{n + 2g - 2} \begin{bmatrix}
		\pm 1 \\ \prod_{i} b_i^{\alpha_i}
	\end{bmatrix}\ 
\end{align}
contribute, and again there are $\ceil{\frac{n + 2g - 2}{2}}$ of them. The $\pm 1$ in the Eisenstein series is given by $(-1)^{n + k}$. These Eisenstein structures are linear independent functions of $b_1, \ldots, b_n$, hence for even $k$ or odd $k$ (so that $(-1)^{n + k}$ is fixed), there are $\ceil{\frac{n + 2g - 2}{2}}$ linear independent vortex defect indices (including the original Schur index corresponding to $k = 0$).

A similar analysis can be applied to defect indices with $n = 0$. There, the defect indices $\mathcal{I}^\text{defect}_{g, 0}(k)$ are all linear combinations of $g$ Eisenstein series (multiplied by $\eta(\tau)^{2g - 2}$ which is omitted here),
\begin{align}
	& E_0 = -1, \quad E_2, \quad E_4, \quad \ldots \quad E_{2g - 2} \, &  \text{when }k \text{ is even} \ , \\
	& E_0 \begin{bmatrix}
		-1 \\ 1
	\end{bmatrix} = -1, \quad
	E_2 \begin{bmatrix}
		-1 \\ 1
	\end{bmatrix} , \quad E_4 \begin{bmatrix}
		-1 \\ 1
	\end{bmatrix}, \quad E_{2g - 2}\begin{bmatrix}
		-1 \\ 1
	\end{bmatrix} \ , &  \text{when }k \text{ is odd} \ .
\end{align}
Hence, for $n = 0$, there will be $g$ independent defect indices for either parities of $k$.

\subsubsection{Gukov-Witten defects}
Another type of superconformal surface defects the will be relevant are the Gukov-Witten surface defects \cite{Gukov:2008sn}, where the dynamical gauge fields are prescribed with some singular background profile at a defect plane orthogonal to the VOA plane.
\begin{figure}
	\centering
	\includegraphics{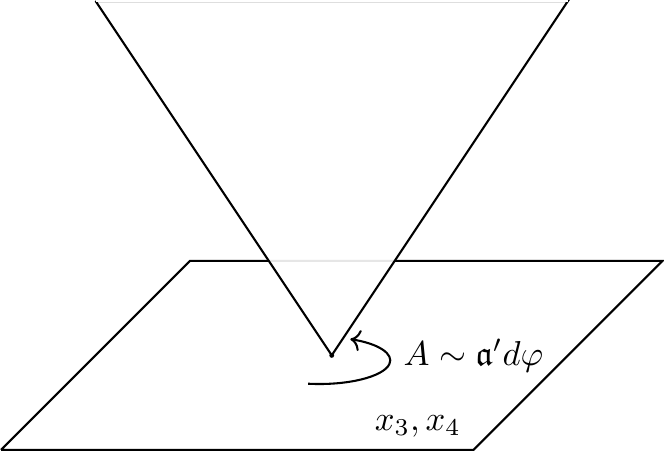}
	\caption{The dynamic gauge field with a prescribed singular behavior near the defect plane drawn vertically. $\varphi$ is the angular coordinate in the $x_3, x_4$ plane where the associated VOA lives. The wedge denotes the $x_1, x_2$ plane on which the surface defect is supported.}
\end{figure}
Upon mapping to $S^3_{\varphi, \chi, \theta} \times S^1_t$, the defect plane is mapped to the torus $T^2_{\chi, t} \subset S^3 \times S^1$ at $\theta = \frac{\pi}{2}$, linking (not intersecting) the VOA torus $T^2_{\varphi, t}$ at $\theta = 0$. The singular profile in flat space then translates to a background gauge field $A \sim \mathfrak{a}' d\varphi$ which is singular at the locus $T^2_{\chi, t} $, since the $\varphi$-circle is contractible in $S^3$. Once the supersymmetric localization is performed, the final integral over flat gauge fields $\mathfrak{a}$ will be shifted to an integral over $\mathfrak{a} + \mathfrak{a}' \tau$, and the Schur index reads
\begin{align}
	\mathcal{I} = \oint \frac{da}{2\pi i a} \mathcal{Z}(\mathfrak{a} + \mathfrak{a}' \tau) \ .
\end{align}
This effectively shift (expand or shrink) the integration contour away from the unit circles. For small $\mathfrak{a}'$, the integral does not change, until the integration contour hits the poles of the integrand $\mathcal{Z}$ and residues $\operatorname{Res}_i$ are picked up. Schematically, in the presence of different Gukov-Witten type surface defects \cite{Pan:2021ulr},
\begin{align}
	\mathcal{I}^\text{defect} \sim \mathcal{I} + \sum_{i} c_i \operatorname{Res}_i \ ,
\end{align}
where $c_i$ are numbers that depend on the precise configuration of the singular background value. Therefore, the residues of the integrand $\mathcal{Z}$ can be identified with the Schur index in the presence of Gukov-Witten type surface defects. These residues can be also interpreted as free field characters of some $bc \beta \gamma$ systems since the residues are just ratios of $\eta(\tau)$ and $\vartheta_i$ functions.

\subsubsection{Defect indices from $S$-transformation}
There is yet another type of surface defects that we will encounter. We have explained that the Schur index $\mathcal{I}_{g,n}(b)$ can be computed as a $S^3_{\varphi, \chi, \theta} \times S^1_t$ partition function \cite{Pan:2019bor}. In such an interpretation, the flavor fugacities $b$ correspond to flat background gauge fields $B$ of the flavor symmetry, roughly of the form $B \sim \mathfrak{b} dt$ (where $b \sim e^{2\pi i \mathfrak{b}}$), leading to a smooth and vanishing background field strength.

Suppose one performs an $S$-transformation on the index $\mathcal{I}_{g,n}$, mapping
\begin{align}
	\mathfrak{b} \to \mathfrak{b}' = \frac{\mathfrak{b}}{\tau}, \qquad
	\tau \to \tau' = - \frac{1}{\tau}\ .
\end{align}
The new index $S\mathcal{I}(\mathfrak{b}, \tau) \coloneqq \mathcal{I}_{g,n}(\frac{\mathfrak{b}}{\tau}, - \frac{1}{\tau})$ can also be reinterpreted as an $S^3 \times S^1$ partition function with a new background flavor gauge field. Now that the background gauge field $\mathfrak{b}' = - \mathfrak{b}( - \frac{1}{\tau})$ is proportional to the new complex modulus $\tau' = - \frac{1}{\tau}$, the background flavor gauge field will be of the form $B' \sim \mathfrak{b} d\varphi$. Although the gauge field $B'$ is flat almost everywhere, it is singular along the torus $T^2_{\chi, t}$ at $\theta = \frac{\pi}{2}$, where the flavor background field strength has a $\delta$-function profile. Therefore, an $S$-transformed Schur index $S\mathcal{I}(\mathfrak{b}, \tau)$ can be interpreted as a flavor defect partition function on the geometry with complex modulus $\tau'$.

\subsection{Modular differential equations \label{section:MDE}}

Modular differential equations will play an important role in our subsequent analysis, and it have already been a useful tool to study both 2d CFT and 4d $\mathcal{N} = 2$ SCFTs. There are two major ways where such an object comes into play.

Rational CFTs are CFTs with finitely many primaries. Each primary generates an irreducible module of the chiral symmetry algebra (namely, VOA) with character $\operatorname{ch}_{i = 1, \ldots, N}$. The full partition function $Z \coloneqq \operatorname{tr} q^{L_0 - \frac{c}{24}} \bar q^{\bar L_0 - \frac{c}{24}}$ can be expanded in these module characters,
\begin{align}
	Z = \sum_{i, j} M_{ij} \operatorname{ch}_i(\tau) \operatorname{ch}_j(\bar \tau) \ , \qquad q \coloneqq e^{2\pi i \tau} \ ,
\end{align}
where $M_{ij}$ is the paring matrix independent of $q, \bar q$. The full partition function $Z$ is also a $T^2$-partition function where the torus has complex structure labeled by $\tau$, and therefore $Z$ is expected to be invariant under the modular group $SL(2, \mathbb{Z})$. Consequently, the characters $\operatorname{ch}_i$ are required to form vector-valued modular form of weight-zero under $SL(2, \mathbb{Z})$ (or its subgroups, if fermions are present \cite{Bae:2020xzl,Bae:2021mej}). For example, for bosonic theory,
\begin{align}
	\operatorname{ch}_i\left(- \frac{1}{\tau}\right) = \sum_{j} S_{ij}\operatorname{ch}_j(\tau)\ .
\end{align}
The $S_{ij}$ form the well-known modular $S$-matrix, from which one can compute fusion coefficients between the said primaries by the Verlind formula \cite{Verlinde:1988sn}.

Using the $N$ characters $\operatorname{ch}_i$, one can write down a ``trivial'' ordinary linear differential equation\cite{Mathur:1988na}
\begin{align}
	D_q^{(N)} \operatorname{ch}_i + \sum_{r = 0}^{N - 1} \phi_r D^{(r)}_q \operatorname{ch}_i = 0\ ,
\end{align}
using the Wronskian matrices $W_r$
\begin{align}
	W_r \coloneqq \begin{pmatrix}
		\operatorname{ch}_1 & \operatorname{ch}_2 & \cdots & \operatorname{ch}_N \\
		D_q^{(1)}\operatorname{ch}_1 & D_q^{(1)}\operatorname{ch}_2 & \cdots & D_q^{(1)}\operatorname{ch}_N \\
		\vdots & \vdots& \vdots & \vdots\\
		D_q^{(r - 1)} \operatorname{ch}_1 & D_q^{(r - 1)} \operatorname{ch}_2 & \cdots & D_q^{(r - 1)} \operatorname{ch}_N\\
		D_q^{(r + 1)} \operatorname{ch}_1 & D_q^{(r + 1)} \operatorname{ch}_2 & \cdots & D_q^{(r + 1)} \operatorname{ch}_N\\
		\vdots & \vdots& \vdots & \vdots\\
		D_q^{(N)} \operatorname{ch}_1 & D_q^{(N)} \operatorname{ch}_2 & \cdots & D_q^{(N)} \operatorname{ch}_N
	\end{pmatrix} \ , \qquad
	\phi_r \coloneqq (-1)^{N - r} \frac{W_r}{W_N} \ .
\end{align}
What is non-trivial about this equation, however, is the fact that the coefficients $\phi_r$ must be weight-$(2N - 2r)$ modular forms, and therefore are severly constrained by modularity. Note also that the differential equation is homogeneous in modular weight, and therefore transforms covariantly under suitable modular transformations. Any reducible module of the given rational VOA is a direct sum of the above finitely many irreducible modules, and the corresponding module character must also be a solution to the above modular differential equation.

This fact has been exploited to classify (bosonic/fermionic) RCFTs \cite{Chandra:2018pjq,Mukhi:2020gnj,Das:2020wsi,Kaidi:2021ent,Das:2021uvd,Bae:2020xzl,Bae:2021mej,Duan:2022ltz}. A modular differential equation can be labeled by its order $N$, and the ``total order $\ell$ of zeros'' of the Wronskian $W_N$. The parameter $\tau$ takes value in the fundamental region of the $SL(2, \mathbb{Z})$ (or suitable subgroups when fermions are involved), and the zeros can sit at the orbifold points, internal points or the cusp $i\infty$. The total order of the zeros equals $\ell/6$, $\ell = \mathbb{N} - \{1\}$. Fow low values of $(N, \ell)$, there are only a small number of free coefficients in the modular differential equation due to modularity. In these situations, it is possible to scan a large range of values for these coefficients and look for ``admissible character solutions'' with non-negative integral coefficients when expanded in $q$-series. These solutions are then tested against more stringent conditions, e.g., by demending non-negative and integral fusion coefficients.

Another way in which the modular differential equaitons arise is through some null states in the vacuum module of a VOA $\mathbb{V}$. One can insert the zero mode $\mathcal{N}_{[0]}$ (if non-zero; see appendix \ref{app:voa} for a brief review on notations) of a null state $\mathcal{N}$ into the (super-)trace that computes the (super-)character of a module $M$ of $\mathbb{V}$,
\begin{align}
	0 = \operatorname{str}_{M} \mathcal{N}_{[0]} q^{L_0 - \frac{c}{24}} \mathbf{b}^\mathbf{f} \ .
\end{align}
Here we have included flavor fugacities $\mathbf{b}$ associated to the Cartan of the flavor symmetries $\mathbf{f}$. When $\mathcal{N}$ takes certain form, the zero mode $\mathcal{N}_{[0]}$ can be ``pulled out'' of the trace using Zhu's recursion formula, and the equation turns into a (flavored) modular differential equation \cite{zhu1996modular,Gaberdiel:2008pr,Krauel:2013lra,Beem:2017ooy}.

For any 4d $\mathcal{N} = 2$ SCFT $\mathcal{T}$, the associated VOA $\mathbb{V}(\mathcal{T})$ descend from the Schur operators in 4d. These operators may originate from different superconformal multiplets, and some are outside of the Higgs branch chiral ring $\mathcal{R}_\text{H}$. In particular, the 2d stress tensor $T$ of $\mathbb{V}(\mathcal{T})$ descends from a component of the $SU(2)_\mathcal{R}$ Noether current in 4d, which does not belong to $\mathcal{R}_\text{H}$. The chiral ring $\mathcal{R}_\text{H}$ is identical to the associated variety of $\mathbb{V}(\mathcal{T})$, and as a result, $T$ must be nilpotent up to $C_2(\mathbb{V}(\mathcal{T}))$ and a null state $\mathcal{N}$ of the VOA \cite{Beem:2017ooy},
\begin{align}
	(L_{-2})^n |0\rangle = \mathcal{N} + \varphi, \qquad \varphi \in C_2(\mathbb{V}(\mathcal{T})) \ , \qquad n \in \mathbb{N}_{\ge 1} \ .
\end{align}
Inserting this equation to the supertrace, it is believed to turn into an unflavored modular differential equation for the unflavored Schur index of $\mathcal{T}$/vacuum character of $\mathbb{V}(\mathcal{T})$. Such an equation plays an important role in a recent classification of rank-two 4d $\mathcal{N} = 2$ SCFT \cite{Kaidi:2022sng}.

\section{\texorpdfstring{$\beta \gamma$ system}{}}\label{section:beta-gamma}

Although our focus will be on the $\mathfrak{a}_1$-type class-$\mathcal{S}$ theories, it proves helpful to begin with a detailed analysis of the simple theory of $\beta \gamma$ system with conformal weights $\frac{1}{2}$. The theory is also the associated VOA of a free hypermultiplet in four dimensions. The $\beta, \gamma$ OPE reads
\begin{align}
	\beta(z)\gamma(w) \sim \frac{1}{z - w} \ \Rightarrow \ [\beta_m, \gamma_n] = \delta_{m + n, 0} \ ,
\end{align}
where the two fields are expanded in the traditional manner,
\begin{align}
	\beta(z) = \sum_{n \in \mathbb{Z} - h[\beta]} \beta_n z^{-n - h[\beta]}, \qquad
	\gamma(z) = \sum_{n \in \mathbb{Z} - h[\gamma]} \beta_n z^{-n - h[\gamma]} \ .
\end{align}
The theory possesses a stress tensor $T$ and a $U(1)$ current $J$, given by
\begin{align}
	T \colonequals \frac{1}{2} (\beta \partial \gamma) - \frac{1}{2} (\gamma \partial \beta), \qquad
	J \colonequals (\gamma \beta)\ .
\end{align}
Note that the stress tensor $T$ is defined as a composite, and also as an element in the subspace $C_2(\beta \gamma)$, since
\begin{align}
	L_{-2} |0\rangle \propto (\beta_{-\frac{3}{2}}\gamma_{- \frac{1}{2}} - \gamma_{-\frac{3}{2}}\beta_{- \frac{1}{2}}) |0\rangle \ ,
\end{align}
where $|0\rangle$ is the vacuum state of the VOA.

The $\beta$ and $\gamma$ fields carry charges under $L_0$ and $J_0$,
\begin{center}
	\begin{tabular}{c|c|c}
		& $\beta$ & $\gamma$\\
		\hline
		$L_0$ & $1/2$ & $1/2$\\
		$J_0$ & $-1$ & $+1$
	\end{tabular}
\end{center}
The vacuum module of the $\beta \gamma$ system is simply the Fock module from acting $\beta_{- n -\frac{1}{2}}$, $\gamma_{-n - \frac{1}{2}}$ on the vacuum $|0\rangle$, $n \in \mathbb{N}$. We recall that $\beta_{n - \frac{1}{2}}$, $\gamma_{n - \frac{1}{2}}$ annihilate $|0\rangle$, $\forall n > 0$. The vacuum character, or the Schur index of a free hypermultiplet in 4d, is thus
\begin{align}
	\operatorname{ch} = \operatorname{tr} q^{L_0 - \frac{c}{24}} b^{J_0}
	= q^{\frac{1}{24}} \operatorname{PE}\left[
	\frac{q^{\frac{1}{2}} b^{-1} + q^{\frac{1}{2}} b}{1-q}
	\right]
	= \frac{\eta(\tau)}{\vartheta_4(\mathfrak{b})} \ .
\end{align}

\subsection{Untwisted sector}

Following \cite{Gaberdiel:2008pr}, we consider inserting the zero modes of the null states $J - (\gamma \beta)$ and $T - \frac{1}{2} ((\beta \partial \gamma) - (\gamma \partial \beta))$ into the trace. For example,\footnote{All modes in the trace are taken to be the ``square-modes''; see \cite{zhu1996modular,Beem:2017ooy} for more detail.}
\begin{align}
	0 = q^{-\frac{c}{24}} \operatorname{tr} o(J - (\gamma \beta)) q^{L_0} b^{J_0}
	= q^{-\frac{c}{24}}\operatorname{tr}\left[
	    J_0 - o(\gamma_{-\frac{1}{2}}\beta_{- \frac{1}{2}}|0\rangle)
	  \right]q^{L_0} b^{J_0} \ .
\end{align}
The first trace involving $J_0$ is nothing but $D_b \operatorname{ch}$, where $D_b \colonequals b \partial_b$. The second term can be easily computed by Zhu's recursion relations \cite{zhu1996modular,Mason:2008zzb},
\begin{align}
	q^{-\frac{c}{24}}\operatorname{tr}o(\gamma_{-\frac{1}{2}}\beta_{- \frac{1}{2}}|0\rangle) & \ q^{L_0} b^{J_0} \nonumber\\
 	= & \ \sum_{n = 1}^{+\infty}E_n\left[\begin{matrix}
    	-1 \\ b
	\end{matrix}\right]\operatorname{tr}o(\gamma_{n - \frac{1}{2}}\beta_{-\frac{1}{2}} |0\rangle) q^{L_0 - \frac{c}{24}} b^{J_0}
	= - E_1\left[\begin{matrix}
      -1 \\ b
	\end{matrix}\right] \operatorname{ch} \ .
\end{align}
Altogether, the vacuum character satisfy a weight-one flavored modular differential equation,
\begin{align}\label{MDE-betagamma-1}
	D_b \operatorname{ch} + E_1\left[\begin{matrix}
    	-1 \\ b
	\end{matrix}\right] \operatorname{ch} = 0 \ .
\end{align}

A similar computation can be performed for the $T - \frac{1}{2} ((\beta \partial \gamma) - (\gamma \partial \beta))$ insertion,
\begin{align}
	0 = \operatorname{tr} o\left(T - \frac{1}{2} ((\beta \partial \gamma) - (\gamma \partial \beta))\right)q^{L_0 - \frac{c}{24}} b^{J_0} \ .
\end{align}
Zhu's recursion relations leads to a weight-two equation,
\begin{align}\label{MDE-betagamma-2}
	D_q^{(1)} \operatorname{ch} - E_2\left[\begin{matrix}
    	-1 \\ b
	\end{matrix}\right] \operatorname{ch} = 0 \ .
\end{align}
Here we used the symmetry property $E_\text{even}\big[\substack{\pm 1 \\ b^{-1}}\big] = E_\text{even}\big[\substack{\pm 1 \\ b}\big]$, and $D_q^{(n)}$ denotes the $n$-th modular differential operator given in (\ref{def:MDO}). In particular, $D_q^{(1)} = q \partial_q$. Note that this null state expresses the stress tensor in terms of an element $(\beta \partial\gamma) - (\gamma \partial \beta) \in C_2 (\beta \gamma)$, which precisely corresponds to the nilpotency of $T$ \cite{Arakawa:2016hkg,Beem:2017ooy}.

The above computation generalized to all kinds of normal ordered products of $T$ and $J$. One simply subtracts from it its explicit expression in terms of the free fields $\beta \gamma$. For example, $(JJ) - (\beta (\beta(\gamma \gamma))) - (\beta \partial \gamma) + (\partial\beta \gamma) = 0$ gives rise to
\begin{align}\label{MDE-betagamma-2-extra}
	\left(D_b^2 - E_2 - 2 E_1\bigg[\begin{matrix}
	 	     	-1 \\ b
	 	 	\end{matrix}\bigg]^2 - 2 E_2\bigg[\begin{matrix}
	 	     	-1 \\ b
	 	 	\end{matrix}\bigg]\right)\operatorname{ch} = 0 \ .
\end{align}
Note that the square of $E_1$ can be eliminated by combining the weight-two null descending from $(\beta(J - (\gamma \beta)))$, leaving
\begin{align}
	\left(
		D_b^2
	    + E_1\bigg[\begin{matrix}
	    	-1 \\ b
		\end{matrix}\bigg] D_b
		- E_1 \begin{bmatrix}
			-1 \\ b
		\end{bmatrix}^2
		- 2 E_2 \bigg[  \begin{matrix}
	    	-1 \\ b
		\end{matrix}\bigg]
		- E_2 
	\right)\operatorname{ch} = 0 \ .
\end{align}
However, these higher-weight equations are not independent equations, since they can be derived from the (\ref{MDE-betagamma-1}) and (\ref{MDE-betagamma-2}). For example, (\ref{MDE-betagamma-2-extra}) can be obtained from (\ref{MDE-betagamma-1}) by taking a $D_b$ derivative
\begin{align}
	D_b^2 \operatorname{ch} + D_b E_1 \begin{bmatrix}
		-1 \\ b
	\end{bmatrix} \operatorname{ch} + E_1 \begin{bmatrix}
		-1 \\ b
	\end{bmatrix}D_b\operatorname{ch} = 0 \ .
\end{align}
Applying the identity
\begin{align}
	D_b E_1 \begin{bmatrix}
		-1 \\ b
	\end{bmatrix}
	= -2 E_2 \begin{bmatrix}
		-1 \\ b
	\end{bmatrix}
	- E_1 \begin{bmatrix}
		-1 \\ b
	\end{bmatrix}^2 - E_2 \ ,
\end{align}
one recovers (\ref{MDE-betagamma-2-extra}).

The vacuum character admits a smooth unflavoring limit,
\begin{align}
	\operatorname{ch} \xrightarrow{b \to 1} \frac{\eta(\tau)}{\vartheta_4(0)} \ .
\end{align}
Consequently, some of the flavored modular differential equations reduce to the more familiar unflavored modular differential equations when unflavoring. For example, the weight-two equation (\ref{MDE-betagamma-2}) reduces directly to
\begin{align}
	\left(D_q^{(1)} - E_2 \begin{bmatrix}
 		-1 \\ + 1
 	\end{bmatrix}\right) \operatorname{ch}(b = 1) = 0 \ .
\end{align}
However, the weight-one equation (\ref{MDE-betagamma-1}) does not reduce to a non-trivial unflavored equation. Instead,
\begin{align}
	D_b \operatorname{ch} \xrightarrow{b \to 1} 0 \ , \qquad
	E_1 \begin{bmatrix}
		- 1 \\ b
	\end{bmatrix} \operatorname{ch} \xrightarrow{b \to 1} 0 \ ,
\end{align}
hence its unflavoring limit is trivial.

\subsection{Twisted sector}

Besides the vacuum module in the untwisted sector, one may also consider twisted modules of the $\beta \gamma$ VOA. For our purpose, we consider the $\frac{1}{2}$-twisted sector based on a twisted vacuum $|0\rangle_{\frac{1}{2}}$. The two fields expand in the following form,
\begin{align}
	\beta(z) = \sum_{n \in \mathbb{Z}} \beta_n z^{- n - \frac{1}{2}}, \qquad
	\gamma(z) = \sum_{n \in \mathbb{Z}} \gamma_n z^{-n - \frac{1}{2}} \ .
\end{align}
such that the twisted vacuum $|0\rangle_{\frac{1}{2}}$ is annihilated by $\gamma_{n \in \mathbb{Z}_{\ge 0}}$, $\beta_{n \in \mathbb{Z}_{>0}}$. In this sector, $T$ and $J$ still have integer moding, however, their precise relations with modes of $\beta$, $\gamma$ are shifted. In particular \cite{Eholzer:1997se},
\begin{align}
	J_0 = \sum_{k \in \mathbb{Z}_{< 0}} \gamma_k \beta_{-k} + \sum_{k \in \mathbb{Z}_{\ge 0}} \beta_{-k}\gamma_k - \frac{1}{2} \ .
\end{align}
and
\begin{align}
	L_0 = & \ - \frac{1}{2} \left(\sum_{k < 0}(k - 1) \beta_k \gamma_{- k} + \sum_{k \ge 0} (k - 1) \gamma_{- k}\beta_k\right)\\
	& \ + 1\left(
	\sum_{k < 0}(k - 1) \gamma_k \beta_{- k} + \sum_{k \ge 0} (k - 1) \beta_{- k}\gamma_k
	\right) + \frac{3}{8} \ . \nonumber
\end{align}
The relevant charges are
\begin{center}
	\begin{tabular}{c|c|c|c}
		& $\beta$ & $\gamma$ & $|0\rangle_{\frac{1}{2}}$ \\
		\hline
		$L_0$ & $0$ & $1$ & $- \frac{1}{8}$ \\
		$J_0$ & $-1$ & $+1$ & $- \frac{1}{2}$
	\end{tabular}
\end{center}
With these charges, the character of the twisted Fock module built from $|0\rangle_{\frac{1}{2}}$ reads
\begin{align}
	\operatorname{ch}_{\frac{1}{2}} = q^{- \frac{1}{8}} b^{- \frac{1}{2}} \operatorname{tr} q^{L_0 - \frac{c}{24}} b^{J_0}
	= q^{- \frac{1}{8}} b^{- \frac{1}{2}} q^{\frac{1}{24}} PE\left[
	\frac{q^0 b^{-1} + q^1 b}{1 - q}
	\right]
	= -i\frac{\eta(\tau)}{\vartheta_1(\mathfrak{b})} \ .
\end{align}

As before, one can insert the same null states discussed above into the trace to produce flavored modular differential equations satisfied by the twisted character $\operatorname{ch}_{\frac{1}{2}}$. The only difference from the untwisted case is that now the conformal weights of $\beta, \gamma$ are integers, and therefore all $E_n\left[\substack{-1\\b}\right]$ should be replaced by $E_n\left[\substack{+1 \\ b}\right]$. For example,
\begin{align}
	0 = & \ \left(D_b + E_1\begin{bmatrix}
		    	+ 1 \\ b
			\end{bmatrix}\right)\operatorname{ch}_{\frac{1}{2}} \ , \\
	0 = & \ \left(D_q^{(1)} - E_2\begin{bmatrix}
		    	+ 1 \\ b
			\end{bmatrix}\right)\operatorname{ch}_{\frac{1}{2}} \ , \\
	0 = & \ \left(D_b^2 - E_2 + E_1\left[\begin{matrix}
		      + 1 \\ b
			\end{matrix}\right] D_b - 2 E_2\begin{bmatrix}
		      + 1 \\ b
			\end{bmatrix}\right)\operatorname{ch}_{\frac{1}{2}} \ .
\end{align}

\subsection{Unique character(s)}
Now that the (twisted) characters are constrained by an infinitely many (with only two independent) partial differential equations, it is natural to ask if there are additional solutions. It turns out that the equations in the untwisted and twisted sector uniquely determine (up to a numerical coefficient) the corresponding characters. For instance, the weight-two equation (\ref{MDE-betagamma-2}) in the untwisted sector is an ordinary differential equation in $q$. Recall
\begin{align}
	E_2 \begin{bmatrix}
		-1 \\ b
	\end{bmatrix}
	= \frac{1}{8\pi^2} \frac{\vartheta_4''(\mathfrak{b})}{\vartheta_4(\mathfrak{b})} - \frac{1}{2}E_2
	=  D_q^{(1)}\left[\frac{1}{3} \ln \frac{\vartheta_1'(0)}{\vartheta_4(\mathfrak{b})^3}\right] \ ,
\end{align}
and therefore the equation (\ref{MDE-betagamma-2}) can be solved by (as an analytic function)
\begin{align}
	\operatorname{ch} = C(b) \frac{\eta(\tau)}{\vartheta_4(\mathfrak{b})} \ .
\end{align}
Finally, the weight-one equation (\ref{MDE-betagamma-1}) further fixes $C(b)$ to be independent of $b$. Similar arguments show that the twisted character is also uniquely fixed by the weight-one and two equations. At the end of this section we will see that the weight-one equation (\ref{MDE-betagamma-1}) is actually redundant, in the sense that it can be generated from the weight-two equation (\ref{MDE-betagamma-2}) through a modular transformation, and (\ref{MDE-betagamma-2}) (or its twisted version) alone actually encodes all the character information.

\subsection{Modular properties of the equations}

The coefficients of the unflavored modular differential equations in \cite{Beem:2017ooy,Chandra:2018pjq,Mukhi:2020gnj,Das:2020wsi,Kaidi:2021ent,Das:2021uvd,Bae:2020xzl,Bae:2021mej,Duan:2022ltz} are modular forms with respect to suitable modular groups. Consequently, the equations transform covariantly under $SL(2, \mathbb{Z})$ (or a subgroup). In contrast, the coefficients of the flavored modular differential equations (\ref{MDE-betagamma-1}), (\ref{MDE-betagamma-2}) are quasi-Jacobi forms, and their modular properties are less straightforward. For simplicity, we first look at the twisted sector. The simpler equation is the weight-one equation (\ref{MDE-betagamma-1}),
\begin{align}
	\left(D_b + E_1 \begin{bmatrix}
		+ 1 \\ b
	\end{bmatrix}\right) \operatorname{ch} = 0 \ .
\end{align}
Consider the naive $S$-transformation that acts on the $\tau$ and $\mathfrak{b}$  parameter,
\begin{align}
	\tau \to - \frac{1}{\tau}, \qquad
	\mathfrak{b} \to \frac{\mathfrak{b}}{\tau} \ .
\end{align}
The modular differential equation transforms non-trivially and non-covariantly under $S$,
\begin{align}
	D_b \to \tau D_b, \
	E_1 \begin{bmatrix}
		1 \\ b
	\end{bmatrix} \to \mathfrak{b} + 
	\tau E_1 \begin{bmatrix}
		1 \\ b
	\end{bmatrix} \quad \Rightarrow \quad
	D_b + E_1 \begin{bmatrix}
		1 \\ b
	\end{bmatrix}
	\to D_b + E_1 \begin{bmatrix}
		1 \\ b
	\end{bmatrix} + \mathfrak{b} \ .
\end{align}
Even so, the $S$-transformed solution remains a solution: in the case at hand, the $S$-transformed twisted character differs from the original by a simple exponential factor,
\begin{align}
	\frac{\eta(\tau)}{\vartheta_1(\mathfrak{b})} \to i e^{- \frac{i \pi \mathfrak{b}^2}{\tau}} \frac{\eta(\tau)}{\vartheta_1(\mathfrak{b})} \ .
\end{align}
This transformed twisted character is annihilated by the transformed equation,
\begin{align}
	& \ \left(
	\tau D_b + \tau E_1 \begin{bmatrix}
		1 \\ b
	\end{bmatrix} + \mathfrak{b} 
	\right) \left(e^{- \frac{i \pi \mathfrak{b}^2}{\tau}} \frac{\eta(\tau)}{\vartheta_1(\mathfrak{b})}\right)\\
	= & \ \tau \left(
	D_b + \tau E_1 \begin{bmatrix}
		1 \\ b
	\end{bmatrix}
	\right)\frac{\eta(\tau)}{\vartheta_1(\mathfrak{b})}
	+ \frac{1}{2\pi i}\tau \frac{-2 i \pi \mathfrak{b}}{\tau}\frac{\eta(\tau)}{\vartheta_1(\mathfrak{b})}
	+ \mathfrak{b}\frac{\eta(\tau)}{\vartheta_1(\mathfrak{b})}
	= \tau \left(
	D_b + \tau E_1 \begin{bmatrix}
		1 \\ b
	\end{bmatrix}
	\right)\frac{\eta(\tau)}{\vartheta_1(\mathfrak{b})} = 0 \ . \nonumber
\end{align}
Phrased differently, after stripping off the exponential factor in the $S$-transformed character, the remaining object is the solution to the original modular differential equation. At the moment, this statement holds true trivially in the current case (since the $S$-transformation merely introduces a simple factor).

A more systematic approach to deal with the modular properties of flavored modular differential equations is to introduce an additional fugacity that couples to the affine level of the current $J$, in this case, $k = - 1/2$. Define the $y$-extended character (using the same symbol)
\begin{align}
	\operatorname{ch} \colonequals \operatorname{tr} q^{L_0 - \frac{c}{24}} y^{k} b^{J_0} \ .
\end{align}
This is essentially the original index, since $y^k$ is merely a constant that can be pulled out of the trace.

We further define the $SL(2, \mathbb{Z})$-transformation of the fugacities $(\mathfrak{y}, \mathfrak{b}, \tau)$ in the following way \cite{Ridout:2008nh},
\begin{align}
	(\mathfrak{y}, \mathfrak{b}, \tau) \xrightarrow{S} (\mathfrak{y} - \frac{\mathfrak{b}^2}{\tau}, \frac{\mathfrak{b}}{\tau}, - \frac{1}{\tau})\ , \qquad
	(\mathfrak{y}, \mathfrak{b}, \tau) \xrightarrow{T} (\mathfrak{y}, \mathfrak{b}, \tau + 1)\ .
\end{align}
In particular, under the $S$-transformation, the derivatives transform as
\begin{align}
	D_q^{(1)} \to & \ D_{q'}^{(1)} = \tau^2 D_q^{(1)} + \mathfrak{b}\tau D_b + \mathfrak{b}^2 D_y, \qquad D_b \to D_{b'} = \tau D_b + 2 \mathfrak{b} D_y \ ,\\
	\text{where} \qquad D_y \colonequals & \ y \partial_y \ , \qquad D_y \operatorname{ch} = k \operatorname{ch} \ .
\end{align}
Therefore, the weight-one modular differential equation transforms under $S$ as
\begin{align}
	D_b + E_1 \begin{bmatrix}
		+ 1 \\ b
	\end{bmatrix} \xrightarrow{S} \tau D_b + 2 \mathfrak{b} D_y + \tau E_1 \begin{bmatrix}
		+ 1 \\ b
	\end{bmatrix} + \mathfrak{b} = \tau \left(D_b + E_1 \begin{bmatrix}
		+ 1 \\ b
	\end{bmatrix}\right) \ ,
\end{align}
showing that the weight-one modular differential equation transforms covariantly once we incorporate the additional $y$-fugacity. Here we have used the fact that $D_y \operatorname{ch} = k \operatorname{ch} = -\frac{1}{2} \operatorname{ch}$.

The transformation of the weight-two equation can be similarly analyzed,
\begin{align}
	D_{q}^{(1)} - E_2 \begin{bmatrix}
		+ 1 \\b
	\end{bmatrix}
	\xrightarrow{S} & \ \tau^2 D_q^{(1)} + \mathfrak{b}\tau D_b + \mathfrak{b}^2 D_y - \tau^2 E_2 \begin{bmatrix}
		+ 1 \\ b
	\end{bmatrix}
	+ \mathfrak{b}\tau E_1 \begin{bmatrix}
		+ 1 \\ b
	\end{bmatrix} + \frac{\mathfrak{b}^2}{2} \nonumber\\
	= & \ \tau^2\left( D_q^{(1)} - E_2 \begin{bmatrix}
			+ 1 \\ b
		\end{bmatrix}\right)
		+ \mathfrak{b}\tau \left(D_b 
		+ E_1 \begin{bmatrix}
			+ 1 \\ b
		\end{bmatrix}\right) \ .
\end{align}
In going to the second line we have applied $D_y = k = - \frac{1}{2}$. Clearly, the weight-two equation transforms almost covariantly under $S$, up to a term proportional to the weight-one equation. 

The analysis of the equations in the untwisted sector is similar but slightly more involved, where the relevant modular group is $\Gamma^0(2)$. Note that under $STS \in \Gamma^0(2)$,
\begin{align}
	E_1 \begin{bmatrix}
		- 1 \\ b
	\end{bmatrix} \xrightarrow{STS} \tau E_1 \begin{bmatrix}
		- 1 \\ b
	\end{bmatrix}
	+ \mathfrak{b}
	- E_1 \begin{bmatrix}
		- 1 \\ b
	\end{bmatrix} \ .
\end{align}
Collecting everything, we find covariance for the weight-one equation,
\begin{align}
	D_b + E_1 \begin{bmatrix}
		- 1 \\ b
	\end{bmatrix} \to (\tau - 1)\left(
	D_b + E_1 \begin{bmatrix}
		- 1 \\ b
	\end{bmatrix}
	\right) \ .
\end{align}
and almost covariance for the weight-two equation,
\begin{align}
	& \ D_q^{(1)} - E_2 \begin{bmatrix}
		-1 \\ b
	\end{bmatrix}\\
	\xrightarrow{STS} & \ 
	(\tau - 1)\mathfrak{b} \left(D_b + E_1 \begin{bmatrix}
		-1 \\ b
	\end{bmatrix}\right)
	+ \frac{(\tau - 1)^2}{\tau} \left(
	\tau \left(D_q^{(1)} - E_2 \begin{bmatrix}
		-1 \\ b
	\end{bmatrix}\right)
	+ \mathfrak{b}\left(D_b + E_1 \begin{bmatrix}
		-1 \\ b
	\end{bmatrix}\right)
	\right) \ . \nonumber
\end{align}
Again, the weight-one equation appears on the right hand side of the transformed weight-two equation.

From the earlier discussions, the weight-one and weight-two equations are enough to determine the unique character of the $\beta \gamma$ system in either the twisted or untwisted sector. Now we also learn that the weight-two equation alone generates the weight-one equation through the  modular transformation $S$ or $STS$. Therefore it appears that the weight-two flavored equation, which reflects the nilpotency of $T$ up to $C_2(\beta \gamma)$, holds all the information of the characters of the $\beta \gamma$ system.

\section{\texorpdfstring{Class $\mathcal{S}$ theories of type $A_1$}{}\label{section:a1}}

Each $A_1$ theory $\mathcal{T} = \mathcal{T}_{g,n}$ is associated to a vertex operator algebra $\mathbb{V}(\mathcal{T})$ that consists of the Schur operators on the VOA-plane $\mathbb{R}^2_{x^3 x^4}$. The fact that a vortex defect in section (\ref{section:defect-indices}) preserves the supercharges $\tilde Q_{2 \dot -}, \tilde S^{2 \dot -}, Q^1_-, S_1^-$ implies that the Schur index in their presence equals the character of a non-trivial $\mathbb{V}(\mathcal{T})$-module \cite{Bianchi:2019sxz}. As discussed in section \ref{section:MDE} and the example in section \ref{section:beta-gamma}, null states in $\mathbb{V}(\mathcal{T})$ may lead to (flavored) modular differential equations that the Schur index must satisfy. By the same logic, the character of any module $M$ of $\mathbb{V}(\mathcal{T})$ must also satisfy the same differential equations (or twisted version of them in case of a twisted module).

It is therefore natural to expect that the defect indices mentioned in section \ref{section:defect-indices} also share such feature, since they are supposed to be module characters of $\mathbb{V}(\mathcal{T})$. We will show that this is indeed the case by examining several simple examples by studying the (flavored) modular differential equations the Schur indices satisfy and look for their common solutions.

\subsection{\texorpdfstring{$\mathcal{T}_{0,3}$}{}}\label{section:T03}

The trinion theory $\mathcal{T}_{0,3}$ is the simplest $A_1$ theory of class-$\mathcal{S}$, which consists of four free hypermultiplets. The associated vertex operator algebra is just the product of four $\beta \gamma$ systems \cite{Beem:2013sza}. Since we have discussed in detail the $\beta \gamma$ system in section \ref{section:beta-gamma}, here we will be brief. The Schur index of the $A_1$ trinion theory is given by
\begin{align}
	\mathcal{I}_{0,3} = \prod_{\pm \pm} \frac{\eta(\tau)}{\vartheta_4(\mathfrak{b}_1 \pm \mathfrak{b}_2 \pm \mathfrak{b}_3)}
	= \frac{1}{2i} \frac{\eta(\tau)}{\prod_{i = 1}^3 \vartheta_1(2 \mathfrak{b}_i)}\sum_{\alpha_i = \pm} \left(\prod_{i = 1}^{3}\alpha_i\right)E_1\left[
	\begin{matrix}
    	-1 \\ \prod_{i = 1}^{3} b_i^{\alpha_i}
	\end{matrix}
	\right]\ .
\end{align}
It is straightforward to find the following modular differential equations for the Schur index $\mathcal{I}_{0,3}$,
\begin{align}
	\left(D_{b_i} + \sum_{\alpha, \beta = \pm} E_1 \begin{bmatrix}
				-1 \\ b_i b_j^\alpha b_k^\beta
	\end{bmatrix}\right) \mathcal{I}_{0,3} = & \ 0 \ , \qquad
	i \ne j \ne k \ , \label{MDE-I03-weight-1}\\
	\left(D_q^{(1)} - \frac{1}{2} \sum_{\alpha_i = \pm} E_2 \begin{bmatrix}
				-1 \\ \prod_{i=1}^{3}b_i^{\alpha_i}
			\end{bmatrix} \right) \mathcal{I}_{0,3} = & \ 0 \ .\label{MDE-I03-weight-2}
\end{align}
Equations of higher weights can be similarly deduced. Note that the second equation (reflecting the nilpotency of the stress tensor) has an unflavoring limit $b_i \to 1$ which reproduces the first order equation in \cite{Beem:2017ooy}, while the weight-one equation does not have a non-trivial limit.

The vortex defects labeled by $k$ have indices given by a simple formula (\ref{def:defect-index}),
\begin{align}
	\mathcal{I}_{0,3}^{\text{defect}}(k) = & \ \frac{(k + 1)\eta(\tau)}{2\prod_{i = 1}^{3} \vartheta_i(2 \mathfrak{b}_i)} \sum_{\alpha_i = \pm} \left(\prod_{i = 1}^{3}\alpha_i\right)E_1 \begin{bmatrix}
		- 1 \\ \prod_{i = 1}^{3}b_i^{\alpha_i}
	\end{bmatrix}  = (k + 1)\mathcal{I}_{0,3} \ , \quad k = \text{even}\ ,\\
	\mathcal{I}_{0,3}^{\text{defect}}(k) = & \ \frac{-i (k + 1) \eta(\tau)}{2\prod_{i = 1}^{3} \vartheta_i(2 \mathfrak{b}_i)} \sum_{\alpha_i = \pm} \left(\prod_{i = 1}^{3}\alpha_i\right)E_1 \begin{bmatrix}
		+ 1 \\ \prod_{i = 1}^{3}b_i^{\alpha_i}
	\end{bmatrix} \ , \quad k = \text{odd}, \ldots \ .
\end{align}
The vortex defects labeled by even $k \in \mathbb{N}$ have indices identical to $\mathcal{I}_{0,3}$ up to some numerical factors, and therefore they all satisfy exactly the same modular equations. For odd $k$, the above defect indices can be equivalently rewritten as
\begin{align}
	\mathcal{I}^\text{defect}(k = \text{odd}) \sim \prod_{\pm \pm}\frac{\eta(\tau)}{\vartheta_1(\mathfrak{b}_1 \pm \mathfrak{b}_2 \pm \mathfrak{b}_3)} \ .
\end{align}
Obviously, this corresponds to nothing but the (product of) $\frac{1}{2}$-twisted module of the four $\beta \gamma$ systems. Immediately, one derives the modular differential equations that they satisfy, for instance,
\begin{align}
	\left(D_{b_i} + \sum_{\alpha, \beta = \pm} E_1 \begin{bmatrix}
				+1 \\ b_i b_j^\alpha b_k^\beta 
	\end{bmatrix}\right) \mathcal{I}_{0,3} = & \ 0 \ , \qquad
	i \ne j \ne k \ , \\
	\left(D_q^{(1)} - \frac{1}{2} \sum_{\alpha_i = \pm} E_2 \begin{bmatrix}
				+1 \\ \prod_{i=1}^{3}b_i^{\alpha_i}
			\end{bmatrix} \right) \mathcal{I}_{0,3} = & \ 0 \ .
\end{align}

Similar to the discussion in section \ref{section:beta-gamma}, the weight-two modular differential equation (\ref{MDE-I03-weight-2}) uniquely determines the relevant characters up to numerical factors. Recall again that
\begin{align}
	E_2 \begin{bmatrix}
		-1 \\ b
	\end{bmatrix}
	= \frac{1}{8\pi^2} \frac{\vartheta_4''(\mathfrak{b})}{\vartheta_4(\mathfrak{b})} - \frac{1}{2}E_2
	=  q \partial_q\left[\frac{1}{3} \ln \frac{\vartheta_1'(0)}{\vartheta_4(\mathfrak{b})^3}\right] \ ,
\end{align}
and therefore by the weight-two equation (\ref{MDE-I03-weight-2})
\begin{align}
	\mathcal{I}_{0,3} = C(b_1, b_2, b_3) \prod_{\pm \pm} \frac{\eta(\tau)}{\vartheta_4(\mathfrak{b}_1 \pm \mathfrak{b}_2 \pm \mathfrak{b}_3)} \ ,
\end{align}
The weight-two equation (\ref{MDE-I03-weight-1}) further fixes $C$ to be constant in $b_i$. Similarly, the solution in the twisted sector is also uniquely fixed to be
\begin{align}
	\mathcal{I}_{0,3}^{\text{defect}(k = \text{odd})} =  \prod_{\pm \pm} \frac{\eta(\tau)}{\vartheta_1(\mathfrak{b}_1 \pm \mathfrak{b}_2 \pm \mathfrak{b}_3)} \ .
\end{align}

\subsection{\texorpdfstring{$\mathcal{T}_{0,4}$}{}}\label{section:T04}

The theory $\mathcal{T}_{0,4}$ describes an $SU(2)$ theory coupled to $N_\text{F} = 4$ fundamental hypermultiplets \cite{Gaiotto:2009we}. The flavor symmetry of the theory is $SO(8) = SO(2N_\text{F})$ that rotates the 8 half-hypermultiplets $Q^i_a$ transforming in the (pseudoreal) fundamental representation of the $SU(2)_\text{g}$ gauge group, where $i = 1, \ldots, 8$ is the $SO(8)$ vector indices, $a = 1,2$ is the $SU(2)_\text{g}$-fundamental indices. The moment map $M$ associated to the $SO(8)$ flavor symmetry transforms under the adjoint $\mathbf{adj}$ of $SO(8)$, and is a gauge-invariant composite of the scalars in the 8 half-hypermultiplets,
\begin{align}
	M^{ij} = Q_a^iQ^{aj} \ .
\end{align}
This simple structure of $M^{ij}$ implies that within the Higgs branch chiral ring \cite{Argyres:1996eh},
\begin{align}\label{Joseph-ideal-1}
	(M \otimes M)_{\mathbf{35}_s} = (M\otimes M)_{\mathbf{35}_c} = 0 \ .
\end{align}
Moreover the $\mathcal{N} = 1$ superpotential of the $\mathcal{N} = 2$ theory imposes
\begin{align}\label{Joseph-ideal-2}
	(M \otimes M)_{\mathbf{35}_v} = (M\otimes M)_{\mathbf{1}} = 0\ .
\end{align}
Note that $\operatorname{symm}^2\mathbf{adj} = \mathbf{35}_{v}\oplus\mathbf{35}_s\oplus\mathbf{35}_c \oplus \mathbf{1} \oplus 2 \mathbf{adj}$, and the above four relations give the Joseph ideal.

The associated VOA of $\mathcal{T}_{0,4}$ is given by the affine algebra $\widehat{\mathfrak{so}}(8)_{-2}$ with central charge $c = -14$, whose generators descend from the moment map operator in the 4d theory. The algebra $\widehat{\mathfrak{so}}(8)_{-2}$ is a member of the affine Lie algebras associated to the Deligne-Cvitanovic exceptional series 
\begin{align}
	\mathfrak{a}_1 \subset \mathfrak{a}_2 \subset \mathfrak{g}_2 \subset \mathfrak{d}_4 \subset \mathfrak{f}_4 \subset \mathfrak{e}_6 \subset \mathfrak{e}_7 \subset \mathfrak{e}_8 , \qquad k = - \frac{h^\vee}{6} - 1\ .
\end{align}
This set of current algebras are quasi-lisse \cite{Arakawa:2016hkg}, and their (unflavored) characters satisfy a (unflavored) modular differential equation of the form
\begin{align}
	(D_q^{(2)} - 5 (h^\vee + 1)(h^\vee - 1)E_4) \mathcal{I} = 0 \ ,
\end{align}
where $h^\vee$ denotes the dual Coxeter number. For $\widehat{\mathfrak{so}}(8)_{-2}$, this equation reads
\begin{align}\label{MDE-I04-unflavored-weight-4}
	(D_q^{(2)} - 175 E_4) \mathcal{I}_{0,4} = 0 \ .
\end{align}
It is known that $\widehat{\mathfrak{so}}(8)_{-2}$ with the central charge $c = -14 = -4 - 2 \times 5$ is not rational \cite{Kaneko:2013up}. Therefore one would expect its representation theory to be more involved than rational VOAs.

The stress-tensor $T$ of $\widehat{\mathfrak{so}}(8)_{-2}$ is a composite given by the Sugawara construction \cite{Beem:2013sza},
\begin{align}
	T = \frac{1}{2(k_\text{2d} + h^\vee)} \sum_{A,B} K_{AB} (J^A J^B) \ ,
\end{align}
where $K_{AB}$ is the inverse of the Killing form $K^{AB} \coloneqq K(J^A, J^B)$. This equation corresponds to the Joseph ideal relation in the trivial representation in (\ref{Joseph-ideal-2}), since $T$ is not in the Higgs branch chiral ring.

\subsubsection{The equations in the untwisted sector}

The Joseph ideal relations (\ref{Joseph-ideal-1}), (\ref{Joseph-ideal-2}) descend to nontrivial null states $\mathcal{N}_a$ in the associated VOA $\widehat{\mathfrak{so}}(8)_{-2}$, and can be inserted into the supertrace $\operatorname{str} o(\mathcal{N}) q^{L_0} \mathbf{b}^\mathbf{f}$. Null states charged under the Cartan of $SO(8)$ do not have interesting outcomes, however, those uncharged can lead to non-trivial modular differential equations.

As a warm-up, let us first consider a simpler partially unflavored limit where all $b_i \to b$. In this limit, the index corresponds to the supertrace over the vacuum module
\begin{align}
	\mathcal{I}_{0,4} = \operatorname{str} b^{h_1 + h_2 + h_3 + h_4}q^{L_0 - \frac{c}{24}} = \frac{\eta(\tau)^2}{\vartheta_1(2 \mathfrak{b})^4}\left(
	3 E_2 - 4E_2 \begin{bmatrix}
		1 \\ b^2
	\end{bmatrix}
	+ E_2 \begin{bmatrix}
		1 \\ b^4
	\end{bmatrix} 
	\right)\ ,
\end{align}
where $h_I$ are the Cartan generators of the four $SU(2)$ flavor groups associated to the four punctures.

The simplest null state associated with the Joseph ideal is the Sugawara construction $T - \sum_{a,b} K_{ab} (J^a J^b) = 0$. Upon inserting into the supertrace, the equation translates to a weight-two modular differential equation
\begin{align}\label{MDE-weight-two-I04}
	0 = \left[D_q^{(1)} - \frac{1}{16}D_b^2 - \frac{1}{2}\left(
	E_1\left[\begin{matrix}
    	1 \\ b^2
	\end{matrix}\right]
	+ E_1\left[\begin{matrix}
		1 \\ b^4    
	\end{matrix}\right]
	\right)D_b
	+ \left(E_2 + 4 E_2\left[\begin{matrix}
    	1 \\ b^2
	\end{matrix}\right]
	+ 2 E_2\left[\begin{matrix}
		1 \\ b^4    
	\end{matrix}\right]
	\right)\right] \mathcal{I}_{0,4} \ . \nonumber 
\end{align}
The remaining three relations corresponding to $\mathbf{35}_{v,s,c}$ each lead to three uncharged null states, however, in the partial unflavoring limit they do not give rise to any non-trivial modular differential equation.

At weight-three, there are new null states besides the descendants of the above Joseph relations. In the partial unflavoring limit, they give rise to three modular differential equations, for example,
\begin{align}
	\left(D_q^{(1)}D_b + \left(E_2 - 4 E_2 \begin{bmatrix}
		1 \\ b^2
	\end{bmatrix}
	- 2 E_2 \begin{bmatrix}
		1 \\ b^4
	\end{bmatrix}
	\right)D_b
	+ 16\left(4 E_3 \begin{bmatrix}
		1 \\ b^2
	\end{bmatrix}
	+ E_3 \begin{bmatrix}
		1 \\ b^4
	\end{bmatrix}
	\right)\right)\mathcal{I}_{0,4} = 0 \ .
\end{align}
Finally, at weight-four there is
\begin{align}\label{MDE-weight-four-I04}
	\bigg[D^{(2)}_q + \left(4 E_3 \begin{bmatrix}
		1 \\ b^2
	\end{bmatrix}
	+ E_3 \begin{bmatrix}
		1 \\ b^4
	\end{bmatrix}
	\right) D_b
	- \left(
	96E_4 \begin{bmatrix}
		1 \\ b^2
	\end{bmatrix}
	+ 12 E_4 \begin{bmatrix}
		1 \\ b^4
	\end{bmatrix}
	+ 67 E_4
	\right) \bigg] \mathcal{I}_{0,4} = 0 \ .
\end{align}

All the above partially-unflavored modular differential equations can be fully refined to depend on four generic flavor $SU(2)$ fugacities $b_i$ \cite{Peelaers}. Now there are additional equations which were unavailable in the partially unflavoring limit. For example, the Sugawara condition gives rise to a weight-two fully flavored modular differential equation,
\begin{align}\label{MDE-I04-Sugawara}
	0 = \Bigg[D^{(1)}_q & \ - \frac{1}{4}\left(D_{b_3}D_{b_2} + D_{b_4}D_{b_2} + D_{b_4}D_{b_3} + D_{b_4}^2 
	\right) \nonumber \\
	& \ - \frac{1}{2}E_1 \begin{bmatrix}
		1 \\ \frac{b_1}{b_2 b_3 b_4} 
	\end{bmatrix}\left(D_{b_1} - D_{b_2} - D_{b_3} - D_{b_4}\right) \nonumber\\
	& \ - \frac{1}{2}E_1 \begin{bmatrix}
		1 \\ b_1 b_2 b_3 b_4 
	\end{bmatrix}\left(D_{b_1} + D_{b_2} + D_{b_3} + D_{b_4}\right)
	- E_1\begin{bmatrix}
		1 \\ b_4^2
	\end{bmatrix} D_{b_4} \nonumber \\
	& \ + \left(E_2 + 2 E_2 \begin{bmatrix}
		1 \\ \frac{b_1}{b_2 b_3 b_4}
	\end{bmatrix}
	+ 2E_2
	\begin{bmatrix}
		1 \\ b_1 b_2 b_3 b_4
	\end{bmatrix}
	+ 2 E_2
	\begin{bmatrix}
		1 \\ b_4^2
	\end{bmatrix}
	\right)  \Bigg] \mathcal{I}_{0,4} \ .
\end{align}
For later convenience, we denote the differential operator acting on the index as $\mathcal{D}^\text{Sug}$.

Additional Joseph ideal relations corresponding to $\mathbf{35}_{v,s,c}$ lead to in total nine equations \cite{Peelaers}. Three null states at weight-two are associated to the $\mathbf{35}_v$ relations. They are
\begin{align}
&J^{[1j]}J^{[1j]}+J^{[5j]}J^{[5j]}-\frac{1}{4}J^{[mn]}J^{[mn]} \ ,\\
&J^{[2j]}J^{[2j]}+J^{[6j]}J^{[6j]}-\frac{1}{4}J^{[mn]}J^{[mn]} \ ,\\
&J^{[3j]}J^{[3j]}+J^{[7j]}J^{[7j]}-\frac{1}{4}J^{[mn]}J^{[mn]}\ .
\end{align}
The three states lead to the following flavored modular differential equations,
\begin{align}
	\sum_{i = 1}^{2} \left(\frac{1}{4}D_{b_i}^2 + E_1 \begin{bmatrix}
		1 \\ b_i^2
	\end{bmatrix}D_{b_i} - 2 E_2 \begin{bmatrix}
		1 \\ b_i^2
	\end{bmatrix}\right) \mathcal{I}_{0,4}
	= \sum_{i = 3}^{4} \left(\frac{1}{4}D_{b_i}^2 + E_1 \begin{bmatrix}
		1 \\ b_i^2
	\end{bmatrix}D_{b_i} - 2 E_2 \begin{bmatrix}
		1 \\ b_i^2
	\end{bmatrix}\right) \mathcal{I}_{0,4} \ ,\nonumber
\end{align}
and
\begin{align}
	& \ \frac{1}{2}D_{b_1}D_{b_2}
	+ \sum_{j = 1}^{2} \left(
	+ E_2 \begin{bmatrix}
		1 \\ b_j^{-2} b_1 b_2 b_3 b_4
	\end{bmatrix}
	- E_1 \begin{bmatrix}
		1 \\ b_j^{-2} b_1 b_2 b_3 b_4
	\end{bmatrix}\sum_{i = 1}^4 (-1)^{\delta_{ij}}D_{b_i}
	\right) \\
	= & \ \frac{1}{2}D_{b_3}D_{b_4}
	+ \sum_{j = 3}^{4} \left(
	+ E_2 \begin{bmatrix}
		1 \\ b_j^{-2} b_1 b_2 b_3 b_4
	\end{bmatrix}
	- E_1 \begin{bmatrix}
		1 \\ b_j^{-2} b_1 b_2 b_3 b_4
	\end{bmatrix}\sum_{i = 1}^4 (-1)^{\delta_{ij}}D_{b_i}
	\right) \ ,
\end{align}
as well as
\begin{align}
	& \ \frac{1}{2}D_{b_1}D_{b_2}
	+ \sum_{j = 1}^{2} \left(
	+ E_2 \begin{bmatrix}
		1 \\ b_j^{-2} b_1 b_2 b_3^{-1} b_4
	\end{bmatrix}
	- E_1 \begin{bmatrix}
		1 \\ b_j^{-2} b_1 b_2 b_3^{-1} b_4
	\end{bmatrix}\sum_{i = 1}^4 (-1)^{\delta_{ij}+\delta_{i3}}D_{b_i}
	\right) \\
	= & \ - \frac{1}{2}D_{b_3}D_{b_4}
	+ \sum_{j = 3}^{4}
	\left(
	+ E_2 \begin{bmatrix}
		1 \\ b_j^{-2} b_1 b_2 b_3^{-1} b_4
	\end{bmatrix}
	- E_1 \begin{bmatrix}
		1 \\ b_j^{-2} b_1 b_2 b_3^{-1} b_4
	\end{bmatrix}\sum_{i = 1}^4 (-1)^{\delta_{ij}+\delta_{i3}}D_{b_i}
	\right) \ .
\end{align}
From the $\mathbf{35}_s$ and $\mathbf{35}_c$ relations there are also six other similar equations, For example, from the null state $J^{[12]}J^{[56]}-J^{[15]}J^{[26]}+J^{[25]}J^{[16]}$ it follows that
\begin{align}
	\left(D_{b_1}^2 + 4 E_1 \begin{bmatrix}
		1 \\ b_1^2
	\end{bmatrix} D_{b_1}
	- 8 E_2 \begin{bmatrix}
		1 \\ b_1^2
	\end{bmatrix}\right) \mathcal{I}_{0,4}
	= \left(D_{b_2}^2 + 4 E_1 \begin{bmatrix}
		1 \\ b_2^2
	\end{bmatrix} D_{b_2}
	- 8 E_2 \begin{bmatrix}
		1 \\ b_2^2
	\end{bmatrix}\right) \mathcal{I}_{0,4} \ .
\end{align}
It will be convenient to reorganize the nine equations from the three $\mathbf{35}$'s in the following more compact form,
\begin{align}\label{MDE-I04-35}
	0 = & \ \sum_{i < j}a_{ij}D_{b_i}D_{b_j} \mathcal{I}_{0,4} + \sum_{i = 1}^{4} a_i \left(
	D_{b_i}^2 + 4 E_1 \begin{bmatrix}
		1 \\ b_i^2
	\end{bmatrix}D_{b_i}
	- 8 E_2 \begin{bmatrix}
		1 \\ b_i^2
	\end{bmatrix}
	\right)\mathcal{I}_{0,4} \nonumber \\
	& \ \qquad + \sum_{\alpha_i = \pm} \Big(\sum_{i<j}\alpha_i \alpha_j a_{ij}\Big)
	\Big(- E_2 \begin{bmatrix}
		1 \\ \prod_{k = 1}^{4}b_k^{\alpha_k}
	\end{bmatrix} + \frac{1}{4}E_1 \begin{bmatrix}
		1 \\ \prod_{k = 1}^{4}b_k^{\alpha_k}
	\end{bmatrix}
	\sum_{i = 1}^{4}\alpha_i D_{b_i}
	\Big) \mathcal{I}_{0,4} \ . 
\end{align}
Here $a_i$ and $a_{i, j}$ are 9 arbitrary constants with constraint $a_1 + a_2 + a_3 + a_4 = 0$. Let us denote the differential operator acting on $\mathcal{I}_{0,4}$ as $\mathcal{D}^\mathbf{35}$.

At weight-three there are four independent modular differential equations, where the index is annihilated by the differential operator
\begin{align}\label{MDE-I04-weight-3}
	& \ \sum_{i = 1}^4 c_i 
	  \left(D_q^{(1)}D_{b_i} + E_2D_{b_i} - 2 E_2 \begin{bmatrix}
	  	  	  	1 \\ b_i^2
	  	  	  \end{bmatrix}D_{b_i} + 8 E_3 \begin{bmatrix}
	  	  	  	1 \\ b_i^2
	  	  	  \end{bmatrix}\right)\\
	& \ \qquad  - \frac{1}{4}\sum_{\alpha_i = \pm}E_2 \begin{bmatrix}
	  	1 \\ \prod_{k = 1}^{4}b_k^{\alpha_k}
	  \end{bmatrix}
	  \sum_{i, j = 1}^4\alpha_i \alpha_j c_iD_{b_j}
	  + 2 \sum_{\alpha_i = \pm} \sum_{i = 1}^{4}\alpha_i c_i E_3 \begin{bmatrix}
	  	1 \\ \prod_{k = 1}^{4}b_k^{\alpha_k}
	  \end{bmatrix} \ .
\end{align}
here $c_i$ are four arbitrary constants. Finally, at weight-four there is one equation
\begin{align}\label{MDE-I04-weight-4}
	0 = \bigg( D_q^{(2)} - 31 E_4
	+ \frac{1}{2}\sum_{\alpha_i = \pm} E_3 \begin{bmatrix}
		+ 1 \\ \prod_{i = 1}^{4} b_i^{\alpha_i}
	\end{bmatrix}& \ \Big(  \sum_{i = 1}^{4}\alpha_iD_{b_i}  \Big)
	+ 2 \sum_{i = 1}^{4} E_3 \begin{bmatrix}
		+ 1 \\ b_i^2
	\end{bmatrix} D_{b_i} \\
	& \ - 12 \sum_{i = 1}^{4}E_4 \begin{bmatrix}
		+ 1 \\ b_i^2
	\end{bmatrix}
	-6 \sum_{\alpha_i = \pm} E_4 \begin{bmatrix}
		+ 1 \\ \prod_i b_i^{\alpha_i}
	\end{bmatrix} \bigg) \mathcal{I}_{0,4} \ . \nonumber
\end{align}
In the $b_i \to 1$ limit, equation (\ref{MDE-I04-weight-4}) reduces to the unflavored equation (\ref{MDE-I04-unflavored-weight-4}) where one sends $E_\text{odd}\big[  \substack{\pm 1 \\ 1}  \big] \to 0$, corresponding to the nilpotency of the stress tensor $T$ in $\widehat{\mathfrak{so}}(8)_{-2}$ \cite{Arakawa:2016hkg,Beem:2017ooy}.

\subsubsection{The solutions and modular property}

It is natural to ask if there are additional solutions to all these modular differential equations besides the flavored Schur index. We begin by noting that the second order unflavored equation (\ref{MDE-I04-unflavored-weight-4}) has an additional solution which is logarithmic \cite{Kaneko:2013up,2002math......6022K,Arakawa:2016hkg,Beem:2017ooy}. This can be seen from the integral spacing of the indicial roots for the anzatz $\mathcal{I} = q^\alpha(1 + \ldots)$,
\begin{align}
	\alpha^2 - \frac{\alpha}{6} - \frac{35}{144} = 0 \qquad \Rightarrow \qquad \alpha = - \frac{5}{12}, \frac{7}{12} \ .
\end{align}
Indeed, the unflavored index can be written as \cite{Arakawa:2016hkg,Pan:2021mrw}
\begin{align}
	\mathcal{I}_{0,4}(b = 1) = 3 \frac{q \partial_q E_4}{\eta(\tau)^{10}} \ .
\end{align}
Under $S$-transformation,
\begin{align}
	S\mathcal{I}_{0,4} = \frac{1}{960\pi^7 \eta(\tau)^{13}}\bigg(
	& \ 
	\frac{5 \vartheta_1^{(3)}(0)^2}{\eta(\tau)^3} - 6\pi \vartheta_1^{(5)(0)} \nonumber\\
	& \ + \frac{5i \tau \vartheta_1^{(3)}(0)^3}{16\pi^2 \eta(\tau)^6}
	- \frac{13i \tau \vartheta_1^{(3)}(0)\vartheta_1^{(5)(0)}}{16\pi\eta(\tau)^3}
	+ \frac{3}{8}i \tau \vartheta_1^{(7)}(0)
	\bigg)\ ,
\end{align}
which is precisely the additional logarithmic solution of the form if expanded in $q$ series,
\begin{align}
	q^{-\frac{5}{12}} (a + \ldots) + q^{\frac{7}{12}} (\log q)(a' + \ldots)\ .
\end{align}
The fact that the modular transformation of a solution leads to another solution is guaranteed by the covariance of the unflavored modular differential equation under $SL(2, \mathbb{Z})$ (or a suitable subgroup).

Next we turn to the fully flavored case and consider additional solutions to all the equations of weight-two, three and four discussed above. Similar to the unflavored case, we will see that there is also a logarithmic solutions that arises from the $S$ transformations of the $y$-extended Schur index given by $S \mathcal{I}_{0,4}$. 
Here, following the discussions in section \ref{section:beta-gamma}, we assume an $\mathbf{y}$-extension for $\mathcal{I}_{0,4}$ by a factor $\prod_{i = 1}^{4}y_i^{k_i}$, where $k_i = -2$ being the critical affine level of each $\widehat{\mathfrak{su}}(2)$ affine subalgebra. The presence of a logarithmic solution can be seen by studying the modular properties of the modular differential equations. First of all, all ten of the weight-two equations are covariant separately under the $S$-transformation,
\begin{align}
	S(\text{weight-}2) = \tau^2 (\text{weight-}2) \ .
\end{align}
Here we apply the transformation of the derivatives
\begin{align}\label{Dqn-S-transformation}
	D_q^{(n)} \to \Big(\tau^2 \partial_{(2n - 2)} & \ + \tau \sum_{i}\mathfrak{b}_i D_{b_i} + \mathfrak{b}_i^2 k_i - (2n - 2) \frac{\tau}{2\pi i}\Big) \circ \ldots\\
	& \ \circ \Big(\tau^2 \partial_{(2)} + \tau \sum_{i}\mathfrak{b}_i D_{b_i} + \mathfrak{b}_i^2 k_i - 2 \frac{\tau}{2\pi i}\Big)
	\circ \Big(\tau^2 \partial_{(0)} + \tau \sum_{i}\mathfrak{b}_i D_{b_i} + \mathfrak{b}_i^2 k_i\Big) \ , \nonumber
\end{align}
and $D_{b_i} \to \tau D_{b_i} + 2 \mathfrak{b}_i k_i$. The Eisenstein series transform under $S$ following (\ref{Eisenstein-S-transformation}).

The weight-three equations (\ref{MDE-I04-weight-3}) are almost covariant under $S$-transformation up to combinations of the weight-two equations. For example, the weight-three equation with $c_1 = 1$, $c_2 = c_3 = c_4 = 0$ transforms under $S$ as
\begin{align}
	S(\text{weight-}3) = & \ \tau^3 (\text{weight-}3) \nonumber\\
	& \ - \tau^2 \mathfrak{b}_1(4 \mathcal{D}^\text{Sug} + \mathcal{D}^\mathbf{35}(a_{1i} = 0, a_{23} = a_{24} = a_{34} = 1, a_1 = -1, a_2 = a_3 = 0)) \nonumber\\
	& \ - \tau^2 \mathfrak{b}_2(\mathcal{D}^\mathbf{35}(a_{12} = -1, \text{all other } a's = 0)) \nonumber \\
	& \ - \tau^2 \mathfrak{b}_3(\mathcal{D}^\mathbf{35}(a_{13} = -1, \text{all other } a's = 0)) \nonumber\\
	& \ - \tau^2 \mathfrak{b}_4(\mathcal{D}^\mathbf{35}(a_{14} = -1, \text{all other } a's = 0))\ .
\end{align}
Finally, the weight-four equation is almost covariant under $S$-transformation, up to combinations of weight-two and weight-three equations,
\begin{align}
	S(\text{weight-}4) = & \ \tau^4 (\text{weight-4}) \\
	& + \tau^3 \sum_{i = 1}^{4} (\text{weight-3})(c_i = 2, c_{j \ne i} = 0) \\
	& + \tau^2 \sum_{i = 1}^{3} \mathfrak{b}_i^2 (-4 \mathcal{D}^\text{Sug} + \mathcal{D}^\mathbf{35}(a_{23} = a_{24} = a_{34} = -1, a_i = 1, \text{other }a = 0)) \nonumber\\
	& + \tau^2 \mathfrak{b}_4^2 (-4 \mathcal{D}^\text{Sug} + \mathcal{D}^\mathbf{35}(a_{23} = a_{24} = a_{34} = -1, \text{other }a = 0)) \nonumber\\
	& + \tau^2 \sum_{i < j}^4 \mathfrak{b}_i\mathfrak{b}_j (\mathcal{D}^\mathbf{35}(a_{ij} = 2, a_i = 1, \text{other }a = 0)) \nonumber \ .
\end{align}
The (almost) covariance implies that the $S$-transformation of a solution must also be a solution which is logarithmic in this case, and therefore it is potentially a logarithmic module character of $\widehat{\mathfrak{so}}(8)_{-2}$. As was discussed in section \ref{section:defect-indices}, $S\mathcal{I}_{0,4}$ can be interpreted as a type of surface defect index.

There are additional four non-logarithmic solutions to all of the above equations \cite{Peelaers}. Recall that the fully flavored Schur index can be computed by the contour integral
\begin{equation}
	\mathcal{I}_{0, 4} = - \frac{1}{2}\oint_{|a| = 1} \frac{da}{2\pi i a} \vartheta_1(2 \mathfrak{a})\vartheta_1(- 2 \mathfrak{a}) \prod_{j = 1}^4\prod_{\pm}\frac{\eta(\tau)}{\vartheta_4(\pm\mathfrak{a} + \mathfrak{m}_j)} \colonequals
	\oint \frac{da}{2\pi i a} \mathcal{Z}(a) \ ,
\end{equation}
where $m_j = e^{2\pi i \mathfrak{m}_j}$ are related to the standard flavor fugacities in the class-$\mathcal{S}$ description,
\begin{equation}
  m_1 =  b_1 b_2 , \qquad m_2 = \frac{b_1}{b_2}, \quad
  m_3 =  b_3 b_4, \qquad m_4 = \frac{b_3}{b_4}\ .
\end{equation}
The integrand $\mathcal{Z}(a)$ has four residues,
\begin{align}\label{I04-residues}
	R_{j = 1,\ldots, 4} \colonequals \mathop{\operatorname{Res}}_{a \to m_jq^{\frac{1}{2}}} \mathcal{Z}(a) = \frac{i}{2}\frac{\vartheta_1(2 \mathfrak{m}_j)}{\eta(\tau)}
	\prod_{\ell \ne j} \frac{\eta(\tau)}{\vartheta_1(\mathfrak{m}_j + \mathfrak{m}_\ell)} \frac{\eta(\tau)}{\vartheta_1(\mathfrak{m}_j - \mathfrak{m}_\ell)} \ .
\end{align}
These residues also appear in the modular transformation of $\mathcal{I}_{0,4}$,
\begin{align}
	S \mathcal{I}_{0,4} = i \tau \mathcal{I}_{0,4} + 2i \sum_{j = 1}^{4}  \mathfrak{m}_j R_j \ , \qquad
	(\mathfrak{y}_i, \mathfrak{b}_i, \tau) \xrightarrow{S} (\mathfrak{y}_i - \frac{\mathfrak{b}_i^2}{\tau}, \frac{\mathfrak{b}_i}{\tau}, - \frac{1}{\tau})\ .
\end{align}
From the discussions in section \ref{section:defect-indices}, $R_j$ can be interpreted as indices of Gukov-Witten defects with specific singular boundary behavior at the defect plane.

These four residues $R_j$ are actually additional linear independent non-logarithmic solutions to the modular differential equations. They are conjectured to be the characters of the four highest weight modules of $\widehat{\mathfrak{so}}(8)_{-2}$, whose (finite) highest weights are given by $\lambda = w(\omega_1 + \omega_3 + \omega_4) - \rho$, with Weyl reflections $w = 1, s_{1,3,4}$ \cite{Arakawa:2015jya,Pan:2021mrw}. Given that $R_j$ are just simple ratios of $\vartheta_i$ and $\eta(\tau)$, they can also be viewed as free field characters of four $bc \beta \gamma$ systems, and provide new free field realization of $\widehat{\mathfrak{so}}(8)_{-2}$ \cite{Peelaers2}. Unlike the index $\mathcal{I}_{0,4}$ and $S \mathcal{I}_{0,4}$, the residues $R_j$ do not have smooth unflavoring limit, and therefore they are invisible if one only studies unflavored modular differential equations.

It can be shown that the $R_j$ and $\mathcal{I}_{0,4}$ are the only non-logarithmic solutions to all the weight-two, three and four equations that we have discussed, by solving them through an anzatz $\mathcal{I} = q^h \sum_{n} a_n(b_1, \ldots, b_n)q^{n}$ \cite{Peelaers}. It is therefore natural to conjecture, with the logarithmic solution $S\mathcal{I}_{0,4}$, that we have found all the (untwisted) module characters of $\widehat{\mathfrak{so}}(8)_{-2}$, and the tools required were simply the flavored modular differential equations (\ref{MDE-I04-Sugawara}), (\ref{MDE-I04-35}), (\ref{MDE-I04-weight-3}) and (\ref{MDE-I04-weight-4}).

In fact, the modular differential equation (\ref{MDE-I04-weight-4}) alone, which corresponds to the nilpotency of the stress tensor $T$, actually generates all the equations of lower weights by modular transformations. Subsequently they together determine all the allowed characters of $\mathfrak{so}(8)_{-2}$. So it appears that all the information about the untwisted characters of $\widehat{\mathfrak{so}}(8)_{-2}$ is encoded in one single equation (\ref{MDE-I04-weight-4}).

The above solutions transform nicely under modular transformations. First of all, the residues $R_j$ transform in the one-dimensional representation of $SL(2, \mathbb{Z})$,
\begin{align}
	S R_j = i R_j, \qquad
	T R_j = e^{\frac{7\pi ii}{6}} R_j \ .	
\end{align}
They satisfy $S^2 = (ST)^3 = - \operatorname{id}$.

On the other hand, $\mathcal{I}_{0,4}$ and $S\mathcal{I}_{0,4}$ transform in a two-dimensional representation of $SL(2, \mathbb{Z})$. Denote $\text{ch}_0 \colonequals \mathcal{I}_{0,4}$, $\text{ch}_1 \colonequals S\mathcal{I}_{0,4}$. Then we have in this basis
\begin{align}
	S = \begin{pmatrix}
		0 & 1 \\
		1 & 0
	\end{pmatrix}, \qquad
	T = \begin{pmatrix}
		e^{\frac{7 \pi i}{6}} & 0\\
		e^{- \frac{\pi i}{3}} & e^{- \frac{5\pi i}{6}}
	\end{pmatrix} \ .
\end{align}
Here we use the convention that $g \operatorname{ch}_i = \sum_{j = 0,1} g_{ij}\operatorname{ch}_j$ for $g \in SL(2, \mathbb{Z})$. The two matrices satisfy $(S T)^3 = S^2 = 1$, as they should. The two characters $\operatorname{ch}_{0,1}$ form an $SL(2, \mathbb{Z})$ invariant partition function
\begin{align}
	Z = n (\operatorname{ch}_0 \overline{\operatorname{ch}_1} + \operatorname{ch}_1 \overline{\operatorname{ch}_0}) \coloneqq \sum_{i, j = 0,1} M_{ij} \operatorname{ch}_i \overline{\operatorname{ch}_j}\ ,
\end{align}
where $n$ denotes possible multiplicity. The $S$-matrix is symmetric and unitary, however, it does not give rise to sensible fusion coefficients since $S_{01} = 0$. We atempt to fix this by considering a different basis
\begin{align}
	\operatorname{ch}'_0 \coloneqq \operatorname{ch}_0, \qquad
	\operatorname{ch}'_1 \coloneqq a\operatorname{ch}_0 + b\operatorname{ch}_1 \ , \qquad \operatorname{ch}'_i = \sum_{j = 0, 1} U_{ij} \operatorname{ch}_j\ .
\end{align}
Doing so, the $SL(2, \mathbb{Z})$ invairnat partition function reads
\begin{align}
	Z = \sum_{i,j}M'_{ij} \operatorname{ch}'_i \overline{\operatorname{ch}'_j}
	= \sum_{k,\ell} \Big(\sum_{i,j} U^{-1}_{ik}\overline{U^{-1}_{j\ell}} M_{jk}\Big) \operatorname{ch}'_{k} \overline{\operatorname{ch}'_\ell} \ ,
\end{align}
and
\begin{align}
	g\operatorname{ch}'_j = \sum_{\ell} \Big(\sum_{j, k} U_{ij}g_{jk}U^{-1}_{k\ell}\Big) \operatorname{ch}'_\ell \ , \qquad \forall g \in SL(2, \mathbb{Z})\ .
\end{align}
With the $S'$-matrix in the new basis, we tentatively define
\begin{align}
	N_{ij}^k = \sum_{\ell} \frac{S'_{i\ell} S'_{j\ell} \overline{S'_{\ell k}}}{S_{0\ell}} \ ,
\end{align}
and require $N_{ij}^k$ to be non-negative integers and $M'$ to be integral. The minimal solution to such problem is given by
\begin{align}
	U_{11} = 1, \qquad U_{12} = 0, \qquad U_{21} = 1, \qquad U_{22} = \pm 1 \ ,
\end{align}
such that
\begin{align}
	M' = \mp  \begin{bmatrix}
		2 & -1\\
		-1 & 0
	\end{bmatrix}\ ,
	\qquad
	[\operatorname{ch}'_0] \times [\operatorname{ch}'_i] = [\operatorname{ch}'_i], \qquad
	[\operatorname{ch}'_2] \times [\operatorname{ch}'_2] = [\operatorname{ch}'_2] \ .
\end{align}
However it is unclear if such fusion algebra bears any sensible mathematical or physical meaning.

\subsubsection{The twisted sector}

In the case at hand, $\lceil \frac{n + 2g - 2}{2} \rceil = 1$, and therefore among all defect indices, there is only one independent index with even $k$, and one with odd $k$. For those with odd vorticity, we focus on $\mathcal{I}^\text{defect}_{0,4}(k = 1)$, 
\begin{align}
	\mathcal{I}^\text{defect}_{0,4}(k = 1) = \frac{\eta(\tau)^2}{\prod_{i = 1}^{4}\vartheta_1(2 \mathfrak{b}_i)} \sum_{\vec\alpha = \pm} \left(\prod_{i = 1}^4 \alpha_i\right)E_2\left[\begin{matrix}
			-1 \\ \prod_{i = 1}^{4} b_i^{\alpha_i}
		\end{matrix}\right] \ .
\end{align}
The defect index $\mathcal{I}^\text{defect}_{0,4}(k = 1)$ does not satisfy the equations discussed in the previous seciton.However, they do satisfy equations that belong to the twisted sector. To begin, the defect index has a smooth unflavoring limit,
\begin{align}
	\mathcal{I}^\text{defect}_{0,4}(k = 1, b_i = 1) = \frac{\eta(\tau)^2}{8\pi^2 \vartheta_4(0)^3 \vartheta'_1(0)^4}
	\left[6 \vartheta''_4(0)^3 - 7 \vartheta_4(0)\vartheta_4''(0)\vartheta_4^{(4)}(0)
	+ \vartheta_4(0)^2 \vartheta_4^{(6)(0)}
	\right] \ . \nonumber
\end{align}
It is easy to check that it satisfies a weight-four equation
\begin{align}
	0 = \left[D_q^{(2)} - 79 E_4 - 96 E_4 \begin{bmatrix}
				-1 \\ 1
			\end{bmatrix}\right] \mathcal{I}^\text{defect}_{0,4}(k = 1, b = 1) \ .
\end{align}
Apparently, this is the twisted version of the unflavored equation (\ref{MDE-I04-unflavored-weight-4}).

Next we consider the fully flavored defect index. As discussed in section \ref{section:defect-indices}, the exact form (\ref{def:defect-index}) suggests that the multi-fundamentals in the VOA have half-integer conformal weight. One can insert all the weight-two, three and four null states that we mentioned above into the supertrace that compute the twisted module character $\mathcal{I}_{0,4}^\text{defect}(k = 1)$ and turn them into flavored modular differential equations. The equations will be almost identical to the ones in the untwisted sector, except that all the Eisenstein series $E_n \big[  \substack{+1\\\ldots}  \big]$ associated to the multi-fundamentals should be modified to $E_n \big[  \substack{-1\\\ldots}  \big]$. Therefore, the Sugawara condition leads to an equation
\begin{align}\label{MDE-I04-Sugawara-twisted}
	0 = \Bigg[D^{(1)}_q & \ - \frac{1}{4}\left(D_{b_3}D_{b_2} + D_{b_4}D_{b_2} + D_{b_4}D_{b_3} + D_{b_4}^2 
	\right) \nonumber \\
	& \ - \frac{1}{2}E_1 \begin{bmatrix}
		- 1 \\ \frac{b_1}{b_2 b_3 b_4} 
	\end{bmatrix}\left(D_{b_1} - D_{b_2} - D_{b_3} - D_{b_4}\right)\nonumber\\
	& \ - \frac{1}{2}E_1 \begin{bmatrix}
		- 1 \\ b_1 b_2 b_3 b_4 
	\end{bmatrix}\left(D_{b_1} + D_{b_2} + D_{b_3} + D_{b_4}\right)
	- E_1\begin{bmatrix}
		1 \\ b_4^2
	\end{bmatrix} D_{b_4}\nonumber \\
	& \ + \left(E_2 + 2 E_2 \begin{bmatrix}
		- 1 \\ \frac{b_1}{b_2 b_3 b_4}
	\end{bmatrix}
	+ 2E_2
	\begin{bmatrix}
		- 1 \\ b_1 b_2 b_3 b_4
	\end{bmatrix}
	+ 2 E_2
	\begin{bmatrix}
		1 \\ b_4^2
	\end{bmatrix}
	\right)  \Bigg] \mathcal{I}_{0,4}^\text{defect}(1) \ ,
\end{align}
while all the nulls from the three $\mathbf{35}$-relations give rise to
\begin{align}\label{MDE-I04-35-twisted}
	0 = & \ \sum_{i < j}a_{ij}D_{b_i}D_{b_j} \mathcal{I}_{0,4} + \sum_{i = 1}^{4} a_i \left(
	D_{b_i}^2 + 4 E_1 \begin{bmatrix}
		1 \\ b_i^2
	\end{bmatrix}D_{b_i}
	- 8 E_2 \begin{bmatrix}
		1 \\ b_i^2
	\end{bmatrix}
	\right)\mathcal{I}_{0,4}  \\
	& \ \qquad + \sum_{\alpha_i = \pm} \Big(\sum_{i<j}\alpha_i \alpha_j a_{ij}\Big)
	\Big(- E_2 \begin{bmatrix}
		- 1 \\ \prod_{k = 1}^{4}b_k^{\alpha_k}
	\end{bmatrix} + \frac{1}{4}E_1 \begin{bmatrix}
		- 1 \\ \prod_{k = 1}^{4}b_k^{\alpha_k}
	\end{bmatrix}
	\sum_{i = 1}^{4}\alpha_i D_{b_i}
	\Big) \mathcal{I}_{0,4}^\text{defect}(1) \ .  \nonumber
\end{align}
Higher weight equations can be similarly obtained. These equations are almost covariant under $STS$ transformations, and therefore $STS$-transformation of the defect index $\mathcal{I}^\text{defect}_{0,4}(k = 1)$ provides a logarithmic solution to this set of equations. There may be additional non-logarithmic solutions that resemble the residues/free field characters $R_j$, however, we leave their existence to future study.

\subsection{\texorpdfstring{$\mathcal{N} = 4$ $SU(2)$ theory}{}}\label{section:TN4}

The $\mathcal{N} = 4$ theory with an $SU(2)$ gauge group has an $SU(2)_\text{f}$ flavor symmetry. The corresponding moment map operator $M$ transforming in the adjoint of $SU(2)_\text{f}$ satisfies the Joseph relation in the Higgs branch chiral ring,
\begin{align}
	(M \otimes M)_\mathbf{1} = 0 \ .
\end{align}
Additional relations with the Hall-Littlewood chiral ring operators $\omega, \tilde\omega$ also exist,
\begin{align}
	(M \otimes \omega)_\mathbf{2} = (M \otimes \tilde \omega)_\mathbf{2} = \omega \otimes \omega = \tilde \omega \otimes \tilde\omega = 0 \ .
\end{align}

The $\mathcal{N} = 4$ theory with $SU(2)$ gauge group has the small $\mathcal{N} = 4$ superconformal algebra as its associated VOA, with the central charge $c = - 9$. The $SU(2)$ flavor symmetry leads to an $\widehat{\mathfrak{su}}(2)_{k_\text{2d} = - 3/2}$ subalgebra with generators $J^A$. The Sugawara stress tensor $T_\text{Sug} = \frac{1}{2(k_\text{2d} + h^\vee)} \sum_{A,B} K_{AB}(J^A J^B)$ equals the full stress tensor $\mathcal{T}$ of $\mathbb{V}_{\mathcal{N} = 4}$. Since the stress tensor descends from 4d outside the Higgs branch chiral ring, the Sugawara construction reflects the aforementioned Joseph relation. The above Hall-Littlewood chiral ring relations are also reflected by four null states in the VOA at conformal weight $5/3$ and $3$ \cite{Beem:2017ooy}, which are charged under the Cartan of $SU(2)_\text{f}$. There is an additional neutral null state (with $A = 3$) at conformal weight-$3$,
\begin{align}
	\mathcal{N}_3^A = (\sigma^A_{\alpha \dot \beta}G^\alpha_{- 3/2}\tilde G^{\dot\beta}_{-3/2} + 2 f^A{_{BC}}J^B_{-2} J^C_{-1} + 2J^A_{-3} - 2L_{-2}J^A_{-1})|0\rangle \ .
\end{align}

The stress tensor $T$ is outside of the chiral ring. As analyzed in \cite{Beem:2017ooy}, it must be nilpotent up to terms in the subalgebra $C_2(\mathbb{V}_{\mathcal{N} = 4})$. This is concretely realized by a null at conformal weight-4,
\begin{align}\label{null-N4-weight-4}
	\mathcal{N}_4 = \left((L_{-2})^2 + \epsilon_{\alpha\beta}(\tilde G^\alpha_{-5/2}G^\beta_{-3/2} - G^\alpha_{- 5/2}\tilde G^\beta_{-3/2}) - K_{AB}J^A_{-2}J^B_{-2} - \frac{1}{2}L_{-4}\right)|0\rangle \ ,
\end{align}
corresponding to a relation in the Higgs branch,
\begin{align}
	(TT) \sim (K_{AB}J^A J^B)^2 = 0 \ .
\end{align}

Now we turn to the modular differential equations that follow from the null states. The Schur index of the $\mathcal{N} = 4$ theory is given by the simple expression
\begin{align}
	\mathcal{I}_{\mathcal{N} = 4} = \operatorname{tr}(-1)^F q^{L_0 - \frac{c}{24}}b^{J_0}
	= \frac{1}{2\pi} \frac{\vartheta'_4(\mathfrak{b})}{\vartheta_1(2 \mathfrak{b})}
	= \frac{i\vartheta_4(\mathfrak{b})}{\vartheta_1(2\mathfrak{b})}E_1 \begin{bmatrix}
		-1 \\ b
	\end{bmatrix} \ .
\end{align}
The factor in front of the Eisenstein series can be interpreted as a character of a $bc\beta\gamma$ system \cite{Bonetti:2018fqz,Pan:2021ulr},
\begin{align}
	\mathcal{I}_{bc\beta \gamma} \coloneqq \frac{i \vartheta_4(\mathfrak{b})}{\vartheta_1(2 \mathfrak{b})} \ .
\end{align}
On the other hand, the Schur index of the $\mathcal{N} = 4$ theory is a simple contour integral
\begin{equation}\label{N=4resc}
	\mathcal{I}_{\mathcal{N} = 4}
	= \frac{1}{2} \oint \frac{da}{2\pi i a} \frac{\eta(\tau)^3}{\vartheta_4(\mathfrak{b})} \prod_{\pm} \frac{\vartheta_1(\pm \mathfrak{a})}{\vartheta_4(\pm \mathfrak{a} + \mathfrak{b})}
	\coloneqq \oint \frac{da}{2\pi i a}\mathcal{Z}(a)\ ,
\end{equation}
Character $\mathcal{I}_{bc \beta\gamma}$ coincides with the residue of the integrand,
\begin{align}
	\operatorname{Res}_{\mathfrak{a} \to \mathfrak{b} + \frac{\tau}{2}} \mathcal{Z}(a) \sim \mathcal{I}_{bc \beta \gamma} \ ,
\end{align}
and can be viewed physically as related to the Schur index in the presence of a Gukov-Witten surface defect in the $\mathcal{N} = 4$ theory.

Here $J$ denotes the Cartan generator of the flavor group. Various null states above can be inserted into the trace, leading to non-trivial modular differential equations. The Sugawara construction $0 = T - T_\text{Sug}$ is the simplest example, giving a modular weight-two equation \cite{Peelaers,Pan:2021ulr},
\begin{align}\label{MDE-N4-weight-two}
	\left[D_q^{(1)} - \frac{1}{2(k + h^\vee)}\left(
		\frac{1}{2}D_b^2 + kE_2 + 2k E_2 \begin{bmatrix}
			1 \\ b^2
		\end{bmatrix}
		+ 2 E_1 \begin{bmatrix}
			1 \\ b^2
		\end{bmatrix}D_b
	\right)\right] \mathcal{I}_{\mathcal{N} = 4} = 0 \ .
\end{align}
Also, the weight-three and weight-four null states $\mathcal{N}^{A = 3}_3$, $\mathcal{N}_4$ lead to
\begin{align}\label{MDE-N4-weight-three}
	0 = \bigg[   D_q^{(1)} D_b + E_1 \begin{bmatrix}
		-1 \\ b
	\end{bmatrix}D_q^{(1)} 
	& \ - 3 E_3 \begin{bmatrix}
		-1 \\ b
	\end{bmatrix}
	+ 6 E_3 \begin{bmatrix}
		1 \\ b^2
	\end{bmatrix}\\
	& \ + \left(
	E_2 + E_2 \begin{bmatrix}
		-1 \\ b
	\end{bmatrix}
	- 2 E_2 \begin{bmatrix}
		1 \\ b^2
	\end{bmatrix}
	\right)D_b \bigg]\mathcal{I}_{\mathcal{N} = 4}  = 0 \ ,
\end{align}
and
\begin{align}\label{MDE-N4-weight-four}
  0 = & \ (D_q^{(2)} + \frac{c_\text{2d}}{2} E_4 )\mathcal{I}_{\mathcal{N} = 4} + \left(-2 E_2\left[
    \begin{matrix}
      -1\\b
    \end{matrix}
    \right]D_q^{(1)}
    - 4E_3\left[
    \begin{matrix}
      -1 \\b
    \end{matrix}
    \right]D_b
    + 18 E_4\left[
    \begin{matrix}
      -1 \\ b
    \end{matrix}
    \right]
  \right) \mathcal{I}_{\mathcal{N} = 4} \nonumber\\
  & + \left(
  3k_\text{2d}E_4 + 2 E_3\left[
  \begin{matrix}
    1 \\ b^2
  \end{matrix}
  \right] D_b
  - 9 E_4\left[
  \begin{matrix}
    1 \\ b^2
  \end{matrix}
  \right]
  \right) \mathcal{I}_{\mathcal{N} = 4} \  . 
\end{align}

There are additional solutions to the above three modular differential equations. First we recall that in the unflavoring limit, the Schur index is given by
\begin{align}
	\mathcal{I}_{\mathcal{N} = 4}(b = 1) = \frac{y^k}{4\pi} \frac{\vartheta_4''(0)}{\vartheta_1'(0)} \ ,
\end{align}
and it satisfies a $\Gamma^0(2)$-modular equation following from the null state (\ref{null-N4-weight-4}),
\begin{align}\label{MDE-N4-weight-two-unflavored}
	\left(D^{(2)}_q - 18E_4
	- 2 E_2 \begin{bmatrix}
		-1 \\ 1
	\end{bmatrix}D_q^{(1)}
	+ 18E_4 \begin{bmatrix}
		-1 \\ 1
	\end{bmatrix}\right)\mathcal{I}_{\mathcal{N} = 4} = 0 \ ,
\end{align}
This equation reflects the nilpotency of the stress tensor, and is also the unflavoring limit of (\ref{MDE-N4-weight-four}), where
\begin{align}
	E_3 \begin{bmatrix}
		\pm 1 \\ b^2
	\end{bmatrix} \xrightarrow{b \to 1} 0, \qquad
	D_b \operatorname{\mathcal{I}} \xrightarrow{b \to 1} 0 \ .
\end{align}
The corresponding indicial equation predicts $\alpha = - 1/8$ and $\alpha = 3/8$ for the standard anzatz $\mathcal{I} = q^\alpha \sum_n a_n q^{n/2}$. The half-integer spacing between $\alpha$'s implies that among the two linear independent solutions, one is logarithmic of the form
\begin{align}
	q^{-1/8}\sum_{n} a_n q^{n/2} + q^{3/8} \log q \sum_{n}a'_n q^{n/2} \ .
\end{align}
Such logarithmic solution is given by the $STS$-transformation of the unflavored Schur index,
\begin{align}
	\mathcal{I}_{\log} = - \frac{i}{2}y^k \frac{\vartheta_4(0)}{\vartheta_1'(0)} + (1 - \tau) \mathcal{I}_{\mathcal{N} = 4}(b = 1) \ .
\end{align}

When flavored, there are three equations (\ref{MDE-N4-weight-two}), (\ref{MDE-N4-weight-three}) and (\ref{MDE-N4-weight-four}) to be concerned with. One can assume an anzatz for non-logarithmic solutions of the form
\begin{align}
	\mathcal{I} = q^\alpha \sum_{n \ge 0} a_n(b) q^{\frac{n}{2}} \ .
\end{align}
The coefficients $a_n(b)$ can be solved order by order, and one finds that the Schur index $\mathcal{I}_{\mathcal{N} = 4}$ and the Gukov-Witten defect index $\mathcal{I}_{bc\beta \gamma}$ are the only two solutions (corresponding to $\alpha = 3/8$ and $\alpha = - 1/8$ respectively) to the three flavored modular equations \cite{Peelaers}. The flavored Schur index $\mathcal{I}_{\mathcal{N} = 4}$ in the $b \to 1$ limit reproduces the non-logarithmic unflavored solution of unflavored equation (\ref{MDE-N4-weight-two-unflavored}), while the $\alpha = - \frac{1}{8}$ solution $\mathcal{I}_{\beta \gamma bc}$ does not have a smooth $b \to 1$ limit and is invisible from (\ref{MDE-N4-weight-two-unflavored}). As discussed in \cite{Adamovic:2014lra}, irreducible modules from the category-$\mathcal{O}$ of the small $\mathcal{N} = 4$ superconformal algebra $\mathbb{V}_{\mathcal{N} = 4}$ were classified: there are only two, one being the vacuum module, and the other will be called $M$. From \cite{Adamovic:2014lra,Bonetti:2018fqz}, the associated VOA $\mathbb{V}_{\mathcal{N} = 4}$ is a sub-VOA of the $bc\beta \gamma$ system, and the quotient gives precisely the irreducible module $M = \mathbb{V}_{bc\beta\gamma}/\mathbb{V}_{\mathcal{N} = 4}$. The two non-logarithmic solutions that we have found precisely correspond to the vacuum module and the reducible but indecomposable module $\mathbb{V}_{bc\beta \gamma}$ of $\mathbb{V}_{\mathcal{N} = 4}$. 

There is also a logarithmic solution to the flavored modular differential equations, given by the $STS$-transformation of $\mathcal{I}_{\mathcal{N} = 4}$. To see the presence of such a solution, let us first analyze the modular property of the equation (\ref{MDE-N4-weight-two}). As discussed in section \ref{section:beta-gamma}, we consider the $y$-extended character
\begin{align}
	\mathcal{I}_{\mathcal{N} = 4}(\mathfrak{y}, \mathfrak{b}, \tau) = y^k \mathcal{I}_{\mathcal{N} = 4}(\mathfrak{b}, \tau)\ , \qquad
	k = - \frac{3}{2}\ .
\end{align}
Recall that the auxiliary variable $y$ is associated to the flavor $SU(2)$ fugacity $b$, whose affine level is $k = \frac{-3}{2}$. We consider again the $S$-transformation
\begin{align}
	S: (\mathfrak{y}, \mathfrak{b}, \tau) \to (\mathfrak{y} - \frac{\mathfrak{b}^2}{\tau}, \frac{\mathfrak{b}}{\tau}, - \frac{1}{\tau}) \ .
\end{align}
The weight-two equation (\ref{MDE-N4-weight-two}) is actually covariant under $S$,
\begin{align}
	& \ D_q^{(1)} - \frac{1}{2(k + h^\vee)}\left(
		\frac{1}{2}D_b^2 + kE_2 + 2k E_2 \begin{bmatrix}
			1 \\ b^2
		\end{bmatrix}
		+ 2 E_1 \begin{bmatrix}
			1 \\ b^2
		\end{bmatrix}D_b
	\right)\\
	\to & \ \tau^2\left(
	D_q^{(1)} - \frac{1}{2(k + h^\vee)}\left(
		\frac{1}{2}D_b^2 + kE_2 + 2k E_2 \begin{bmatrix}
			1 \\ b^2
		\end{bmatrix}
		+ 2 E_1 \begin{bmatrix}
			1 \\ b^2
		\end{bmatrix}D_b
	\right)
	\right) \ .
\end{align}
This implies that $S\mathcal{I}_{\mathcal{N} = 4}$ must be a solution. It is also invariant under $T$. A similar analysis extends to (\ref{MDE-N4-weight-three}), (\ref{MDE-N4-weight-four}) which can be shown to be almost covariant under $STS \in \Gamma^0(2)$, up to equations in lower modular weights. Explicitly,
\begin{align}
	STS (\text{weight-3}) = & \ (\tau - 1)^3 (\text{weight-3}) - 2 \mathfrak{b} (\tau - 1)^2 (\text{weight-2}) \ , \\
	STS (\text{weight-4}) = & \ (\tau - 1)^4 (\text{weight-4}) + 2 \mathfrak{b} (\tau - 1)^3 (\text{weight-3}) - 2 \mathfrak{b}^2 (\tau - 1)^2 (\text{weight-2}) \ . \nonumber
\end{align}

To conclude, $STS \mathcal{I}_{\mathcal{N} = 4}$ furnishes a logarithmic solution to all three modular equations. We conjecture that $STS \mathcal{I}_{\mathcal{N} = 4}$ together with the two non-logarithmic solutions $\mathcal{I}_{\mathcal{N} = 4}$ and $\mathcal{I}_{bc\beta \gamma}$ form the complete set of solutions. The logarithmic solution $STS\mathcal{I}_{\mathcal{N} = 4}$ may correspond to a logarithmic module of $\mathbb{V}_{\mathcal{N} = 4}$ where the Virasoro zero mode $L_0$ does not have a diagonalizble action \cite{Adamovic:2014lra}, though the precise relation will be left for future study.

Like in the case of $\widehat{\mathfrak{so}}(8)_{-2}$, by a modular transformation $STS$, the weight-four equation (\ref{MDE-N4-weight-four}) generates all the lower weight equations that are necessary for determining the allowed characters. Hence, all the character information are encoded in one equation (\ref{MDE-N4-weight-four}), which reflects the nilpotency of the stress tensor.

Finally, it is straightforward to show that $\Gamma^0(2)$ acts linearly on the two-dimensional space spanned by $\text{ch}_0 \colonequals \mathcal{I}_{\mathcal{N} = 4}$ and $\text{ch}_1 = STS \mathcal{I}_{\mathcal{N} = 4}$:
\begin{align}
	S^2 = 1, \qquad
	T^2 = \begin{pmatrix}
		-i & 0 \\
		2i & -i 
	\end{pmatrix}, \qquad
	STS = \begin{pmatrix}
		0 & 1 \\
		-1 & 2
	\end{pmatrix} \ .
\end{align}
Here the matrix $g_{ij}$ of an element $g \in \Gamma^0(2)$ are defined through action $g \text{ch}_i = \sum_{j = 0,1} g_{ij} \text{ch}_j$.

\subsection{\texorpdfstring{$\mathcal{T}_{1,1}$}{}}\label{section:T11}

\subsubsection{Untwisted sector}

The $\mathcal{T}_{1,1}$ theory is the product of the $\mathcal{N} = 4$ $SU(2)$ theory and a free hypermultiplet. In general the two sectors each have their own $SU(2)$ flavor symmetry with seperate flavor fugacities $b_{1,2}$, on which the Schur index depends. Let us first look at the naive class-$\mathcal{S}$ limit $b_{1} = b_2 = b$. The Schur index is given by the formula (\ref{Ign}),
\begin{align}
	\mathcal{I}_{1,1} = i\frac{\eta(\tau)}{\vartheta_1(2 \mathfrak{b})} E_1 \begin{bmatrix}
		-1 \\ b
	\end{bmatrix} \ .
\end{align}
It satisfies one weight-two, two weight-three and three weight-four equations which are collected in the appendix \ref{app:fmldes}. For example, there are equations mirroring those in the $\mathcal{N} = 4$ theory at weight-two
{\small\begin{align}\label{MDE-I11-weight-2}
	0 = \left(D_q^{(1)} - \frac{1}{2}D_b^2 - E_1 \begin{bmatrix}
				-1 \\ b
			\end{bmatrix} D_b
			-2 E_1 \begin{bmatrix}
				1 \\ b^2
			\end{bmatrix} D_b
			+ 3 E_2 \begin{bmatrix}
				1 \\ b^2
			\end{bmatrix} + 2 E_2
			- 2 E_1 \begin{bmatrix}
				-1 \\ b
			\end{bmatrix}
			E_1 \begin{bmatrix}
					1 \\ b^2
				\end{bmatrix}
	\right) \mathcal{I}_{1,1} \ ,
\end{align}}%
weight-three
\begin{align}\label{MDE-I11-weight-3}
	0 = \left(D_q^{(1)}D_b + E_2 D_b + E_1 \begin{bmatrix}
				-1 \\ b
			\end{bmatrix} D_q^{(1)}
			- 2 E_2 \begin{bmatrix}
				1 \\ b^2
			\end{bmatrix} D_b
			- 2 E_1 \begin{bmatrix}
				-1 \\ b
			\end{bmatrix}E_2 \begin{bmatrix}
				1 \\ b^2
			\end{bmatrix}
			+ 6 E_3 \begin{bmatrix}
				1 \\ b^2
			\end{bmatrix}\right) \mathcal{I}_{1,1} \ ,
\end{align}
and at weight-four,

\begin{align}\label{MDE-I11-weight-4}
	0 = \bigg(D_q^{(2)} - & 4 E_2 \begin{bmatrix}
		-1 \\ b
	\end{bmatrix}D_q^{(1)} 
	- 4 E_3 \begin{bmatrix}
		-1 \\ b
	\end{bmatrix}D_b
	+ 2 E_3 \begin{bmatrix}
		1 \\ b^2
	\end{bmatrix}D_b 
	\\
	& \ + \frac{8}{3} E_1 \begin{bmatrix}
		-1 \\ b
	\end{bmatrix}E_3 \begin{bmatrix}
		-1 \\ b
	\end{bmatrix} 
	+ \frac{2}{3}E_1 \begin{bmatrix}
		-1 \\ b
	\end{bmatrix}E_3 \begin{bmatrix}
		1 \\ b^2
	\end{bmatrix} + 16 E_4 \begin{bmatrix}
		-1 \\ b
	\end{bmatrix}
	- 11 E_4 \begin{bmatrix}
		1 \\ b^2
	\end{bmatrix}\bigg) \mathcal{I}_{1,1} \ . \nonumber
\end{align}
Note that this weight-four equation reduces in the $b \to 1$ limit the 2$^\text{nd}$ order unflavored modular differential equation the reflects the nilpotency of the stress tensor \cite{Beem:2017ooy},
\begin{align}
	0 = \left(D_q^{(2)} - 4 E_2 \begin{bmatrix}
				-1 \\ 1
			\end{bmatrix}D_q^{(1)} - 11 E_4 \begin{bmatrix}
				+ 1 \\ 1
			\end{bmatrix}
			+ 16 E_4 \begin{bmatrix}
				-1 \\ 1
			\end{bmatrix}\right) \mathcal{I}_{1,1}(b = 1) \ .
\end{align}

There are additional solutions to this set of equations. First of all, the factor $\mathcal{I}_{\beta \gamma} \coloneqq \frac{\eta(\tau)}{\vartheta_1(2 \mathfrak{b})}$ in front of $\mathcal{I}_{1,1}$ is a non-logarithmic solution, and also coincides with the residue of the integrand that computes $\mathcal{I}_{1,1}$ in a contour integral. Like in the $\mathcal{N} = 4$ case, the three equaitons (\ref{MDE-I11-weight-2}), (\ref{MDE-I11-weight-3}) and (\ref{MDE-I11-weight-3}) can be solved order by order through an anzatz $\mathcal{I} = q^h\sum_{n = 0}a_n(b)q^{\frac{n}{2}}$. It turns out that $\mathcal{I}_{1,1}$ and $\mathcal{I}_{\beta \gamma}$ are the only two non-logarithmic solutions.

There are also logarithmic solutions. This can be seen by working out the modular transformation of the (\ref{MDE-I11-weight-2}), (\ref{MDE-I11-weight-3}), (\ref{MDE-I11-weight-3}). For example, (\ref{MDE-I11-weight-2}) is covariant under $STS$, while (\ref{MDE-I11-weight-4}) is almost covariant
\begin{align}
	STS (\text{weight-4}) = (\tau-1)^4 (\text{weight-4}) - 2(\tau - 1)^3 \mathfrak{b} (\text{weight-3}) + 2 (\tau - 1)^2 \mathfrak{b}^2 (\text{weight-2}) \ . \nonumber
\end{align}
Therefore, $STS \mathcal{I}_{1,1}$ is also a solution, and we believe we have exhausted all the solutions to the flavored modular differential equations. Once again, the flavored modular differential equation corresponding to the nilpotency of $T$ encodes all the information of all the flavored characters, as it generates all the necessary equations of lower weight.

Next we turn on separate flavor fugacities for the two $SU(2)$ flavor symmetries. The fully flavored Schur index is just the product of the Schur indices of the two theories,
\begin{align}\label{I11-fully-flavored}
	\mathcal{I}_{1,1}(\mathfrak{b}_1, \mathfrak{b}_2) = \frac{1}{2\pi} \frac{\vartheta_4'(\mathfrak{b}_1)}{\vartheta_1(2 \mathfrak{b}_1)} \frac{\eta(\tau)}{\vartheta_4(\mathfrak{b}_2)} = \frac{i \vartheta_4(\mathfrak{b}_1)}{\vartheta_1(2 \mathfrak{b}_1)}\frac{\eta(\tau)}{\vartheta_4(\mathfrak{b}_2)} E_1 \begin{bmatrix}
		- 1 \\ b_1
	\end{bmatrix}\ .
\end{align}
It is straightforward to derive the fully-flavored modular differential equations for this theory by combining the results in section \ref{section:T03}, \ref{section:TN4}. For instance, at weight-two one has the equation by combining the Sugawara condition (\ref{MDE-N4-weight-two}) and \eqref{MDE-betagamma-2},
\begin{align}
	0 = \left[D_q^{(1)} - \frac{1}{2(k + h^\vee)}\left(
		\frac{1}{2}D_{b_1}^2 + kE_2 + 2k E_2 \begin{bmatrix}
			1 \\ b_1^2
		\end{bmatrix}
		+ 2 E_1 \begin{bmatrix}
			1 \\ b_1^2
		\end{bmatrix}D_{b_1}
	\right)
	- E_2 \begin{bmatrix}
		-1 \\ b_2
	\end{bmatrix}
	\right] \mathcal{I}_{1,1} \ .
\end{align}
Equations of weight-three and four in section \eqref{section:TN4} can be easily carried over from the previous section. Equations in section \eqref{section:T03} without any $D_q^{(n)}$ also appear here naturally.

Again there are additional solutions to these equations. The coefficient of the fully flavored index is given by
\begin{align}
	\mathcal{I}_{bc(\beta \gamma)^2} = \frac{i \vartheta_4(\mathfrak{b}_1)}{\vartheta_1(2 \mathfrak{b}_1)}
	\frac{\eta(\tau)}{\vartheta_4(\mathfrak{b}_2)}\ ,
\end{align}
which coincides with the residue of the integrand that computes \eqref{I11-fully-flavored}. It satisfies all the above mentioned modular differential equations. The equations are also (almost) covariant under $STS$ transformation, and hence the transformed index $STS\mathcal{I}_{1,1}(\mathfrak{b}_1, \mathfrak{b}_2)$ gives a logarithmic solution. We believe that $\mathcal{I}_{1,1}$, $STS \mathcal{I}_{1,1}$ and $\mathcal{I}_{bc(\beta \gamma)^2}$ form the complete set of solutions.

\subsubsection{Untwisted sector}

Moving onto the twisted sector where we consider again the $b_i = b$ limit. Since $\ceil{\frac{n + 2g - 2}{2}} = 1$ in this case, there is only one linear independent vortex defect index with odd vorticity, which we choose to be $\mathcal{I}^\text{defect}_{1,1}(k = 1)$. In this sector, we should change all the flavor fundamentals to have integer conformal weights. Indeed, one can check that the defect index $\mathcal{I}^\text{defect}_{1,1}(k = 1)$ satisfies all the above equations with $E_n \Big[\substack{- 1 \\ b}\Big] \to E_n \Big[\substack{+ 1 \\ b}\Big]$. For example, at weight-two, there is
{\small{\begin{align}
	0 = \left(D_q^{(1)} - \frac{1}{2}D_b^2 - E_1 \begin{bmatrix}
				1 \\ b
			\end{bmatrix} D_b
			-2 E_1 \begin{bmatrix}
				1 \\ b^2
			\end{bmatrix} D_b
			+ 3 E_2 \begin{bmatrix}
				1 \\ b^2
			\end{bmatrix} + 2 E_2
			- 2 E_1 \begin{bmatrix}
				1 \\ b
			\end{bmatrix}
			E_1 \begin{bmatrix}
					1 \\ b^2
				\end{bmatrix}\right) \mathcal{I}^\text{defect}_{1,1}(1) \ .
\end{align}}}%
At weight-four, there is also
\begin{align}\label{MDE-I11-weight-4-twisted}
	0 = \bigg(D_q^{(2)} - & 4 E_2 \begin{bmatrix}
		1 \\ b
	\end{bmatrix}D_q^{(1)} 
	- 4 E_3 \begin{bmatrix}
		1 \\ b
	\end{bmatrix}D_b
	+ 2 E_3 \begin{bmatrix}
		1 \\ b^2
	\end{bmatrix}D_b 
	\\
	& \ + \frac{8}{3} E_1 \begin{bmatrix}
		1 \\ b
	\end{bmatrix}E_3 \begin{bmatrix}
		1 \\ b
	\end{bmatrix} 
	+ \frac{2}{3}E_1 \begin{bmatrix}
		1 \\ b
	\end{bmatrix}E_3 \begin{bmatrix}
		1 \\ b^2
	\end{bmatrix} + 16 E_4 \begin{bmatrix}
		1 \\ b
	\end{bmatrix}
	- 11 E_4 \begin{bmatrix}
		1 \\ b^2
	\end{bmatrix}\bigg) \mathcal{I}^\text{defect}_{1,1}(1) \ . \nonumber
\end{align}

What's surprising to observe is that the free field character $\mathcal{I}_{\beta \gamma}$ is actually an additional non-logarithmic solution to all the equations in both the untwisted and the twisted sector\footnote{Among all the examples we have examined, this is the only instance where the free field character walks between both worlds. We leave its physical or mathematical implication to future study.}. In fact, there are precisely two linear independent non-logarithmic solutions to all the equations in the twisted sector, the free field character $\mathcal{I}_{\beta \gamma}$ and the defect index $\mathcal{I}^\text{defect}_{1,1}(k = 1)$ \footnote{However, unlike that in the untwisted sector, neither of these solutions has smooth unflavoring limit. In particular,
\begin{align}
	\mathcal{I}_{1,1}^\text{defect}(k = 1) = - \frac{\eta(\tau)\vartheta_1'(\mathfrak{b})}{\pi \vartheta_1(\mathfrak{b},q) \vartheta_1(2\mathfrak{b})}
\end{align}
has a double pole at $\mathfrak{b} = 0$. Therefore there is no unflavored modular differential equation in the twisted sector.}. This can be similarly shown by solving them order by order. Finally, the equations in the twisted sector are all $SL(2, \mathbb{Z})$ (almost) covariant, and therefore logarithmic solutions are present given by the modular transformations of the vortex defect index.

\subsection{\texorpdfstring{$\mathcal{T}_{2,0}$}{}}

The genus-two theory $\mathcal{T}_{2,0}$ admits two weak-coupling limits. One limit is given by gauging two $\mathcal{T}_{1,1}$ by an $SU(2)$ vector multiplet, where one reads off a $U(1)$ flavor symmetry invisible in the class-$\mathcal{S}$ description \cite{Beem:2013sza}. The chiral algebra of $\mathcal{I}_{2,0}$ has been constructed in \cite{Kiyoshige:2020uqz} and later a free field realization was proposed in \cite{Beem:2021jnm}. The associated flavored Schur index is given by the exact formula \cite{Pan:2021mrw},
\begin{align}\label{I20}
	\mathcal{I}_{2,0} = \frac{i \vartheta_1(\mathfrak{b})^2}{\eta(\tau) \vartheta_1(2 \mathfrak{b})} \left(
	E_3 \begin{bmatrix}
		+1\\b
	\end{bmatrix}
	+ E_1 \begin{bmatrix}
		+1\\b
	\end{bmatrix}
	E_2 \begin{bmatrix}
		+1\\b
	\end{bmatrix}
	+ \frac{1}{12}E_1\begin{bmatrix}
		+1\\b
	\end{bmatrix}
	\right) \ .
\end{align}
Here $b$ denotes the $U(1)$ flavor fugacity. Note that the factor in front can be viewed as a free field character of a $(bc)^2\beta \gamma$ system,
\begin{align}
	\mathcal{I}_{(bc)^2 \beta \gamma} \colonequals \frac{i\vartheta_1(\mathfrak{b})^2}{\eta(\tau)\vartheta_1(2 \mathfrak{b})} \ .
\end{align}

The Schur index satisfies several modular differential equations. At weight-four, we have
\begin{align}
	\Bigg[ D_q^{(2)}& \ +\frac{1}{4}D_{b}^{4}-2D_q^{(1)}D_b^2+2E_1\begin{bmatrix}
		1\\
		b^2\\
	\end{bmatrix} D_{b}^{3}
	-2\left( E_1\begin{bmatrix}
		1\\
		b\\
	\end{bmatrix} +2E_1\begin{bmatrix}
		1\\
		b^2\\
	\end{bmatrix} \right) D_q^{(1)}D_b \nonumber\\
	& \ +4\left( E_2+E_2\begin{bmatrix}
		1\\
		b\\
	\end{bmatrix} +E_2\begin{bmatrix}
		1\\
		b^2\\
	\end{bmatrix} \right) D_q^{(1)}
	+\left( -7E_2-2E_2\begin{bmatrix}
		1\\
		b\\
	\end{bmatrix} -8E_2\begin{bmatrix}
		1\\
		b^2\\
	\end{bmatrix} \right) D_b^2\\
	& \ \left( -2E_2\left( E_1 \begin{bmatrix}
		1\\
		b\\
	\end{bmatrix} +8E_1 \begin{bmatrix}
		1\\
		b^2\\
	\end{bmatrix} \right) +12E_3 \begin{bmatrix}
		1\\
		b\\
	\end{bmatrix} +18E_3 \begin{bmatrix}
		1\\
		b^2\\
	\end{bmatrix} \right) D_b\nonumber\\
	& \ +\left( 7E_{2}^{2}+4E_2\left( E_2 \begin{bmatrix}
		1\\
		b\\
	\end{bmatrix} +4E_2 \begin{bmatrix}
		1\\
		b^2\\
	\end{bmatrix} \right) -2\left( 14E_4+12E_4 \begin{bmatrix}
		1\\
		b\\
	\end{bmatrix} +9E_4 \begin{bmatrix}
		1\\
		b^2\\
	\end{bmatrix} \right) \right)  \Bigg] \mathcal{I}_{2,0} = 0 \ .\nonumber
\end{align}
This modular differential equation comes from the null \cite{Kiyoshige:2020uqz}
\begin{align}
	(B^+ B^- - J^4) + 2 (D^{+I}\bar D_I^-) = 4T^2 -6 \partial^2 T - 8J^2 T & \ + 12\partial(JT) - 4\partial J^3 \nonumber \\
	& \ + 9 (\partial J)^2 + 14 J \partial^2 J - 5 \partial^3 J \ .
\end{align}

There is no weight-five equation, even though there are several null states at conformal weight-five. At weight-six, the Schur index satisfies two modular differential equaitons. The first one is
\begin{align}
	\Bigg[ & \   D_{q}^{(3)}+\frac{3}{2}D_q^{(2)}D_b^2-3E_1\begin{bmatrix}
	1\\b
	\end{bmatrix} D_q^{(2)}D_b+6\left( E_2-2E_2\begin{bmatrix}
	1\\b
	\end{bmatrix} \right) D_q^{(1)}D_b^2 \\
		& \ 
		+3\left( E_2+2E_2\begin{bmatrix}
		1\\b
	\end{bmatrix} \right) D_q^{(2)}-12\left( E_2E_1\begin{bmatrix}
	1\\b
	\end{bmatrix} -5E_3\begin{bmatrix}
	1\\b
	\end{bmatrix} +E_3 \begin{bmatrix}
	1\\b^2
	\end{bmatrix} \right) D_q^{(1)}D_b\\
	& \ +6\left( -3E_3\begin{bmatrix}
		1\\b
	\end{bmatrix} +E_3 \begin{bmatrix}
	1\\b^2
	\end{bmatrix} \right) D_b^3\\
		& \ +\left( 9{E_2}^2-\frac{111}{2}E_4-24E_2E_2\begin{bmatrix}
		1\\b
	\end{bmatrix} +180E_4\begin{bmatrix}
	1\\b
	\end{bmatrix} -72E_4 \begin{bmatrix}
	1\\b^2
	\end{bmatrix} \right) D_b^2\\
	& \ +D_q^{(1)}\left( \begin{array}{c}
			-6{E_2}^2-49E_4+24E_2E_2\begin{bmatrix}
			1\\b
	\end{bmatrix} -72E_4\begin{bmatrix}
	1\\b
	\end{bmatrix} +36E_4 \begin{bmatrix}
	1\\b^2
	\end{bmatrix}\\
		\end{array} \right) \nonumber \\
	& \ -3\left( 4{E_2}^2E_1\begin{bmatrix}
		1\\b
	\end{bmatrix} +5E_4E_1\begin{bmatrix}
	1\\b
	\end{bmatrix} +E_2\left( -44E_3\begin{bmatrix}
	1\\b
	\end{bmatrix} +16E_3 \begin{bmatrix}
	1\\b^2
	\end{bmatrix} \right) +216E_5\begin{bmatrix}
		1\\b
	\end{bmatrix} -108E_5 \begin{bmatrix}
	1\\b^2
	\end{bmatrix} \right) \nonumber \\
	& \ -\left( 6{E_2}^3-24{E_2}^2E_2\begin{bmatrix}
		1\\b
	\end{bmatrix} +3E_2\left( E_4+72E_4\begin{bmatrix}
	1\\b
	\end{bmatrix} -48E_4 \begin{bmatrix}
	1\\b^2
	\end{bmatrix} \right) \right) \\
	& \ -\left( 10\left( 11E_6-3E_4E_2\begin{bmatrix}
		1\\b
	\end{bmatrix} -72E_6\begin{bmatrix}
	1\\b
	\end{bmatrix} +54E_6 \begin{bmatrix}
	1\\b^2
	\end{bmatrix} \right) \right)   \Bigg]\mathcal{I}_{2,0} = 0
\end{align}
This equation arises from the weight-six null in \cite{Kiyoshige:2020uqz}. Another modular differential equation takes the schematic form
\begin{align}
	(D^6_b - 14 D^{(1)}_qD_b^4 + \ldots) \mathcal{I}_{2,0} = 0\ .
\end{align}
Note however that this equation does not follow from any null states in \cite{Kiyoshige:2020uqz} and we shall discard.

Let us turn to additional solutions to the modular differential equations. It is easy to verify that the the free field character $\mathcal{I}_{(bc)^2 \beta \gamma}$ in front of the Schur index is an additional solution to the above equations with associated null states. Like in the previous examples, it also coincides with the residue of the integrand that computes the original Schur index \eqref{I20}. Recall that the theory has a duality frame of gluing two copies of $\mathcal{T}_{0,3}$, where the Schur index is written as
\begin{align}
	\mathcal{I}_{2, 0} = & \ - \frac{1}{8}\oint \prod_{i = 1}^3 \left[\frac{da_i}{2\pi i a_i}\vartheta_1(2\mathfrak{a}_i)\vartheta_1(- 2\mathfrak{a}_i)\right]
	\prod_{\pm, \pm, \pm} \frac{\eta(\tau)}{\vartheta_4(\pm \mathfrak{a}_1 \pm \mathfrak{a}_2 \pm \mathfrak{a}_3 + \frac{\mathfrak{b}}{2})} \\
	\colonequals & \ \oint \prod_{i = 1}^{3}\frac{da_i}{2\pi i a_i} \mathcal{Z}(a) \ .
\end{align}
The free field character appears as the following residue,
\begin{align}
	\mathcal{I}_{(bc)^2\beta \gamma} = \mathop{\operatorname{Res}}_{\mathfrak{a}_1 \to \mathfrak{a}_2 + \mathfrak{a}_3 + \frac{\mathfrak{b}}{2} + \frac{\tau}{2}}
	\mathop{\operatorname{Res}}_{\mathfrak{a}_2 \to - \mathfrak{a}_3}
	\mathop{\operatorname{Res}}_{\mathfrak{a}_3 \to - \frac{1}{2}\mathfrak{b}} \mathcal{Z}(a, b) = \frac{i}{32} \frac{\vartheta_1(\mathfrak{b})^2}{\eta(\tau)\vartheta_1(2 \mathfrak{b})} \ .
\end{align}
Note that $\mathcal{I}_{(bc)^2 \beta \gamma}$ is not annihilated by the second weight-six modular differential equation which has no associated null state, which is somewhat expected. Unfortunately, although one would expect that the $SL(2, \mathbb{Z})$ transformation of the index $\mathcal{I}_{2,0}$ gives an additional logarithmic solution, the weight-six equation does not transform properly under $SL(2, \mathbb{Z})$. As a result, the $S$-transformed $\mathcal{I}_{2,0}$ only satisfies the weight-four equation. The physical meaning of the lack of a logarithmic solution remains unclear.

\subsection{\texorpdfstring{$\mathcal{T}_{1,2}$}{}}

The Schur index of $\mathcal{T}_{1,2}$ in exact form is given by
\begin{align}
	\mathcal{I}_{1,2} = - \frac{\eta(\tau)^2}{\prod_{i = 1}^{2}\vartheta_1(2\mathfrak{b}_i)}
	\left(
	E_2 \begin{bmatrix}
		1 \\ b_1 b_2
	\end{bmatrix}
	- E_2 \begin{bmatrix}
		1 \\ b_1 / b_2
	\end{bmatrix}
	\right) \ .
\end{align}
Note that the factor in front can be interpreted as a free field character of a pair of $\beta \gamma$ system,
\begin{align}
	\mathcal{I}_{\beta \gamma\beta \gamma} = \frac{\eta(\tau)}{\vartheta_1(2\mathfrak{b}_1)}\frac{\eta(\tau)}{\vartheta_1(2\mathfrak{b}_2)} \ .
\end{align}
Noting that $\lceil \frac{n + 2g - 2}{2} \rceil = 1$, there is only one defect index $\mathcal{I}_{1,2}^\text{defect}(k = 0) = \mathcal{I}_{1,2}$ with even $k$ and one $\mathcal{I}^\text{defect}_{1,2}(1)$ with odd $k$.

\subsubsection{The untwisted sector}

At weight-two there are two modular differential equations satisfied by the index $\mathcal{I}_{1,2}$,
\begin{align}
	0 = \Bigg[
	D_q^{(1)}
	- \frac{1}{4} \sum_{i = 1,2} D_{b_i}^2
	-\frac{1}{4}& \ \sum_{\alpha_i = \pm} E_1 \begin{bmatrix}
		1 \\ b_1^{\alpha_1}b_2^{\alpha_2}
	\end{bmatrix}
	\sum_{i = 1,2}\alpha_i D_{b_i}
	- \sum_{i = 1,2} E_1 \begin{bmatrix}
		1 \\ b_i^2
	\end{bmatrix}D_{b_i} \\
	& \ + 2 \bigg(
	E_2 + \frac{1}{2} \sum_{\alpha_i = \pm}E_2 \begin{bmatrix}
		1 \\ b_1^{\alpha_1}b_2^{\alpha_2}
	\end{bmatrix}
	+ \sum_{i = 1,2} E_2 \begin{bmatrix}
		1 \\ b_i^2
	\end{bmatrix}
	\bigg) \Bigg] \mathcal{I}_{1,2} \ .
\end{align}
and
\begin{align}
	\left(D_{b_1}^2 + 4 E_1 \begin{bmatrix}
		1 \\ b_1^2 
	\end{bmatrix}
	- 8 E_2 \begin{bmatrix}
		1 \\ b_1^2
	\end{bmatrix}\right) \mathcal{I}_{1,2}
	= \left(D_{b_2}^2 + 4 E_1 \begin{bmatrix}
		1 \\ b_2^2 
	\end{bmatrix}
	- 8 E_2 \begin{bmatrix}
		1 \\ b_2^2
	\end{bmatrix}\right)\mathcal{I}_{1,2} \ .
\end{align}

At weight-three there are three new equations, 
\begin{align}
	\left[D_{b_1}^3 + 6 E_1 \begin{bmatrix}
				1 \\ b_1^2
			\end{bmatrix} D_{b_1}^2 + \ldots\right] \mathcal{I}_{1,2} = 0\ .
\end{align}
There are two more weight-four equations
\begin{align}\label{MDE-I12-weight-4-1}
	\Bigg(D_{b_1}^{4} + 8E_1 \begin{bmatrix}
		1\\
		b_{1}^{2}
	\end{bmatrix} D_{b_1}^{3}-48E_2 \begin{bmatrix}
		1\\
		b_{1}^{2}
	\end{bmatrix} D_{b_1}^{2}+192E_3 \begin{bmatrix}
		1\\
		b_{1}^{2}
	\end{bmatrix} D_{b_1}-384E_4 \begin{bmatrix}
		1\\
		b_{1}^{2}
	\end{bmatrix} & \ \Bigg) \mathcal{I}_{1,2} \nonumber \\
	= & \ (b_1 \leftrightarrow b_2) \ ,
\end{align}
and a third one which has a tedious form and schematically looks like
\begin{align}
	\left(D^{(2)}_q + \ldots\right) \mathcal{I}_{1,2} = 0\ .
\end{align}
As will be noted shortly, the unflavored index $\mathcal{I}_{1,2}(b_i \to 1)$ satisfies a 3$^\text{rd}$ order modular differential equation, which should also admit a fully flavored refinement. Unfortunately we have not obtained its explicit form.

Let us turn to additional solutions to the modular differential equations. We first point out that the free field character $\mathcal{I}_{(\beta \gamma)^2}$ in front of the Schur index is also a solution to the fully flavored modular differential equations at weight-two and equation (\ref{MDE-I12-weight-4-1}) at weight-four, but not the weight-three and the second one of weight-four. Like in the case of the $SU(2)$ $\mathcal{N} = 4$ theory or $\mathcal{T}_{0,4}$, the free field character $\mathcal{I}_{(\beta \gamma)^2}$ also arises as the residue of the integrand in the contour integral that computes the index $\mathcal{I}_{1,2}$,
\begin{align}
	\mathcal{I}_{1,2} = \oint \left[\prod_{i = 1}^{2} \frac{da_i}{2\pi i a_i}\frac{1}{2}\vartheta_1(2\mathfrak{a}_i)^2\right] \prod_{i = 1}^{2} \prod_{\pm \pm}\frac{\eta(\tau)}{\vartheta_4(\mathfrak{a}_1 \pm \mathfrak{a}_2 \pm \mathfrak{b}_i)}
	\colonequals \oint \prod_{i = 1}^{2} \frac{da_i}{2\pi i a_i} \mathcal{Z}(a, b) \ .
\end{align}
Explicitly,
\begin{align}
	\mathcal{I}_{(\beta \gamma)^2}
	= \mathop{\operatorname{Res}}_{\mathfrak{a}_1 \to \mathfrak{a}_2 + \mathfrak{b}_1 + \frac{\tau}{2}}
	  \mathop{\operatorname{Res}}_{\mathfrak{a}_2 \to - \frac{1}{2}\mathfrak{b}_1 - \frac{1}{2} \mathfrak{b}_2} \mathcal{Z}(a,b) \ .
\end{align}

Drawing an analogy with the results in \ref{section:T04}, \ref{section:TN4}, we conjecture that the free theory of $bc \beta \gamma$ system is actually a module of $\mathbb{V}(\mathcal{T}_{1,2})$. Consequently, we expect that the weight-two and the weight-four equations above arise from actual null states in $\mathbb{V}(\mathcal{T}_{1,2})$, while the weight-three equations and the weight-four equation $(D^{(2)}_q + \ldots)\mathcal{I} = 0$ do not and we shall ignore them from now on.

Next we look for logarithmic solutions. It is easy to check that the unflavored index satisfies an unflavored equation at weight-six\footnote{In \cite{Beem:2017ooy}, a weight-eight equation was listed instead,
\begin{align}
	0 = & \ (D^{(4)}_q - 220 E_4D^{(2)}_q - 2380 E_6 D^{(1)}_q + 6000E_4^2)\mathcal{I}_{1,2}(b \to 1) \ .
\end{align}},
\begin{align}\label{MDE-I12-weight-6-unflavored}
	0 = & \ (D^{(3)}_q - 220 E_4D^{(1)}_q + 700 E_6)\mathcal{I}_{1,2}\ .
\end{align}
The indicial equation based on $\mathcal{I} = q^\alpha ( 1 + \ldots)$ gives
\begin{align}
	0 = & \ (6\alpha - 5)(6\alpha + 1)^2 & \Rightarrow & & \ \alpha = & \ - \frac{1}{6}, - \frac{1}{6}, \frac{5}{6} \ .
\end{align}
Clearly, the $\alpha = 5/6$ solution is the unflavored Schur index. The two linear independent solutions corresponding to $\alpha = - \frac{1}{6}$ are logarithmic ones of the form
\begin{align}
	\mathcal{I}_{\log} = q^{- \frac{1}{6}} \sum_{n} a_nq^{n/2} + q^{- \frac{1}{6}} \log q \sum_{n} a'_n q^{n/2} + q^{\frac{5}{6}} (\log q)^2 \sum_{n} a''_n q^{n/2} \ .
\end{align}

Now we focus on the three flavored modular differential equations of weight-two and weight-four with conjectural associated null states. Under the $S$-transformation (with the critical affine levels $k_i = - 2$ and the $y$-extension as introduced in section \ref{section:beta-gamma}), both the weight-two equations are covariant,
\begin{align}
	S(\text{weight-2}) = \tau^2 (\text{weight-2}) \ .
\end{align}
The weight-four equation (\ref{MDE-I12-weight-4-1}) is almost covariant,
\begin{align}
	S(\text{weight-4}) = \tau^4 (\text{weight-4}) + \tau^3 \frac{12 i}{\pi} (\text{weight-2}')
\end{align}
where the weight-2$'$ equation is defined to be a combination of the two weight-two equations,
\begin{align}
	\left[D_{b_1}^2 - D_{b_2}^2 + 4 \left(E_1 \begin{bmatrix}
						1 \\ b_1^2
					\end{bmatrix} D_{b_1}
					-E_1 \begin{bmatrix}
						1 \\ b_2^2
					\end{bmatrix} D_{b_1}
			\right)
			- 8 \left(E_2 \begin{bmatrix}
						1 \\ b_1^2
					\end{bmatrix} D_{b_1}
					-E_2 \begin{bmatrix}
						1 \\ b_2^2
					\end{bmatrix} D_{b_1}
			\right)\right] \mathcal{I}_{1,2} = 0 \ . \nonumber
\end{align}
Therefore, the (almost) covariance suggests additional logarithmic solutions given by modular transformation of the ($y$-extended) Schur index. More explicitly, under $S$-transformation we have
\begin{align}
	S\mathcal{I}_{1,2}
	= & \ \frac{\eta(\tau)^2}{8\pi^2
		\vartheta_1(2\mathfrak{b}_1)
		\vartheta_1(2\mathfrak{b}_2)
		\vartheta_1(\mathfrak{b}_1 - \mathfrak{b}_2)
		\vartheta_1(\mathfrak{b}_1 + \mathfrak{b}_2)
	}  \nonumber \\
	& \ \times \bigg[
	    - \tau^2  \vartheta''_1(\mathfrak{b}_1-\mathfrak{b}_2)\vartheta_1(\mathfrak{b}_1+\mathfrak{b}_2)
	    - 4 \pi i \tau (\mathfrak{b}_1-\mathfrak{b}_2) \vartheta'_1(\mathfrak{b}_1-\mathfrak{b}_2)\vartheta_1(\mathfrak{b}_1+\mathfrak{b}_2) \nonumber\\
	& \ \qquad + \tau^2 \vartheta''_1(\mathfrak{b}_1+\mathfrak{b}_2)\vartheta_1(\mathfrak{b}_1-\mathfrak{b}_2) 
	  + 4 i \pi \tau (\mathfrak{b}_1+\mathfrak{b}_2) \vartheta'_1(\mathfrak{b}_1+\mathfrak{b}_2)\vartheta_1(\mathfrak{b}_1-\mathfrak{b}_2)  \nonumber\\
	& \ \qquad 
	  -16 \pi ^2  \vartheta_1(\mathfrak{b}_1+\mathfrak{b}_2)\vartheta_1(\mathfrak{b}_1-\mathfrak{b}_2) \bigg]\\
	= & \ - \tau^2 \mathcal{I}_{2,0} -2\mathfrak{b}_1\mathfrak{b}_2\frac{\eta(\tau)^2}{\prod_{i = 1}^2\vartheta_1(2 \mathfrak{b}_i)}
	- \tau\frac{ \eta(\tau)^2}{\prod_{i}\vartheta_1(2 \mathfrak{b}_i)}\sum_{\alpha = \pm} \alpha(\mathfrak{b}_1 +\alpha \mathfrak{b}_2)E_1 \begin{bmatrix}
		1 \\ b_1 b_2^\alpha
	\end{bmatrix} \ . \nonumber
\end{align}
Similarly, after $TS$ transformation,
\begin{align}
	TS\mathcal{I}_{1,2}
	= & \ (-1)^{2/3} \mathcal{I}_{1,2} - (-1)^
	{2/3}S\mathcal{I}_{1,2}\\
	& \  + (-1)^{2/3} \left[2 \tau \mathcal{I}_{2,0} + \frac{ \eta(\tau)^2}{\prod_{i}\vartheta_1(2 \mathfrak{b}_i)}\sum_{\alpha = \pm} \alpha(\mathfrak{b}_1 +\alpha \mathfrak{b}_2)E_1 \begin{bmatrix}
				1 \\ b_1 b_2^\alpha
			\end{bmatrix} \right]\ .
\end{align}

Before moving onto the modular properties of these solutions, we briefly remark on the completeness of the above solutions so far. Although the weight-six fully flavored modular differential equations are currently unavailable, we can look at the partially unflavoring limit $b_i \to b$. In this limit, there is a unique weight-six\footnote{Some equations of lower weights are collected in the appendix \ref{app:fmldes}} flavored modular differential equation that annihilates $\mathcal{I}_{1,2}$, $\mathcal{I}_{(\beta \gamma)^2}$ and the two logarithmic solutions $S\mathcal{I}_{1,2}$ and $TS\mathcal{I}_{1,2}$,
\begin{align}
	0 = & \ \Bigg[ D_q^{(3)} + \frac{1}{24}D_b^4 D_q^{(1)} - \frac{1}{2} D_{q}^{(2)}D_b^2 + \frac{1}{2}E_2 D_q^{(1)}D_b^2 - 14 E_2 D_q^{(2)}
	+ \frac{1}{2} E_1 \begin{bmatrix}
		1 \\ b^2
	\end{bmatrix} D_q^{(1)}D_b^3 \nonumber\\
	& \ + 4 E_1 \begin{bmatrix}
		1 \\ b^2
	\end{bmatrix} (6E_2 - E_2 \begin{bmatrix}
		1 \\ b^2
	\end{bmatrix}) D_q^{(1)}D_b
	+ \left(\frac{1}{6}E_2 - \frac{1}{4}E_2 \begin{bmatrix}
		1 \\ b^2
	\end{bmatrix}\right) D_b^4 \nonumber\\
	& \ + \frac{1}{2}\left(
		3 E_2 E_1 \begin{bmatrix}
			1 \\ b^2
		\end{bmatrix}
		-5 E_2 \begin{bmatrix}
			1 \\ b^2
		\end{bmatrix}E_1 \begin{bmatrix}
			1 \\ b^2
		\end{bmatrix}
		+ 5 E_3 \begin{bmatrix}
			1 \\ b^2
		\end{bmatrix}
	\right)D_b^3 \nonumber\\
	& \ + \frac{1}{3} \left(- 18E_2 E_2 \begin{bmatrix}
		1 \\ b^2
	\end{bmatrix}
	+ 35 E_2 \begin{bmatrix}
		1 \\ b^2
	\end{bmatrix}^2
	+ 38 E_1 \begin{bmatrix}
		1 \\ b^2
	\end{bmatrix}E_3 \begin{bmatrix}
		1 \\ b^2
	\end{bmatrix}
	- 40 E_4 \begin{bmatrix}
		1 \\ b^2
	\end{bmatrix}
	\right)D_b ^2 \nonumber\\
	& \ - \frac{4}{3} \left(
	  15 E_2^2 + 54 E_2 E_2 \begin{bmatrix}
	  	1 \\ b^2
	  \end{bmatrix}
	  + 31 E_2 \begin{bmatrix}
	  	1 \\ b^2
	  \end{bmatrix}^2
	  - 98 E_1 \begin{bmatrix}
	  	1 \\ b^2
	  \end{bmatrix}E_3 \begin{bmatrix}
	  	1 \\ b^2
	  \end{bmatrix}
	  -20 E_4 \begin{bmatrix}
	  	1 \\ b^2
	  \end{bmatrix}
	\right)D_q^{(1)} \nonumber\\
	& \ + \left(
		12 E_2 E_1 \begin{bmatrix}
			1 \\ b^2
		\end{bmatrix}
		- 92 E_2 \begin{bmatrix}
			1 \\ b^2
		\end{bmatrix}E_3 \begin{bmatrix}
			1 \\ b^2
		\end{bmatrix}
		+ 40 E_5 \begin{bmatrix}
			1 \\ b^2
		\end{bmatrix}
	\right)
	D_b \nonumber\\
	& \ + E_2 \left(
		-52 E_1 \begin{bmatrix}
			1 \\ b^2
		\end{bmatrix}E_2 \begin{bmatrix}
			1 \\ b^2
		\end{bmatrix}
		+ 48 E_3 \begin{bmatrix}
			1 \\ b^2
		\end{bmatrix}
	\right)
	D_b
	- \frac{4}{3}\left(
	9E_2^3 - 72 E_2 ^2 E_2 \begin{bmatrix}
		1 \\ b^2
	\end{bmatrix}
	- 87 E_3 \begin{bmatrix}
		1 \\ b^2
	\end{bmatrix}^2 
	\right) \nonumber\\
	& \ - \frac{4}{3}E_2 \left(- 137
	E_2 \begin{bmatrix}
		1 \\ b^2
	\end{bmatrix}
	- 158 E_1 \begin{bmatrix}
		1 \\ b^2
	\end{bmatrix}E_3 \begin{bmatrix}
		1 \\ b^2
	\end{bmatrix}
	+ 18 E_4 \begin{bmatrix}
		1 \\ b^2
	\end{bmatrix}
	\right)\nonumber\\
	& \ + 40\left(5E_2 \begin{bmatrix}
		1 \\ b^2
	\end{bmatrix}E_4 \begin{bmatrix}
		1 \\ b^2
	\end{bmatrix}
	+ E_1 \begin{bmatrix}
		1 \\ b^2
	\end{bmatrix}E_5 \begin{bmatrix}
		1 \\ b^2
	\end{bmatrix}
	+ 7 E_6 \begin{bmatrix}
		1 \\ b^2
	\end{bmatrix}
	\right)\Bigg]\mathcal{I}_{1,2}(b)\ .
\end{align}
The $S$-transformation of this equation produces a set of lower weight flavored modular differential equations whose only non-logarithmic solutions are $\mathcal{I}_{1,2}(b)$ and $\mathcal{I}_{(\beta \gamma)^2}(b)$.

Now let us look at the modular properties of the solutions. One can find a simple basis for the $SL(2, \mathbb{Z})$-orbit of the Schur index,
\begin{align}
	\text{ch}_0 = & \ \mathcal{I}_{1,2},
	& \text{ch}_{\log, 1} = & \ 2\tau \mathcal{I}_{1,2} + \sum_{\alpha = \pm} \alpha (\mathfrak{b}_1 + \alpha \mathfrak{b}_2)\frac{\eta(\tau)^2}{\prod_i\vartheta_1(2\mathfrak{b}_i)} E_1 \begin{bmatrix}
		1 \\ b_1 b_2^\alpha
	\end{bmatrix}\ ,\\
	\text{ch}_{\log, 2} = & \ S\mathcal{I}_{1,2} \ .
\end{align}
In this basis,
\begin{align}
	T \text{ch}_0 = & \ - (-1)^{2/3} \operatorname{ch}_0 \ ,	\\
	T \text{ch}_{\log, 1} = & \ (-1)^{2/3}(-2\text{ch}_0 - \text{ch}_{\log, 1} )\ ,\\
	T \text{ch}_{\log, 2} = & \ (-1)^{2/3}(\text{ch}_0
		+ \text{ch}_{\log, 1} 
		- \text{ch}_{\log, 2}) \ ,
\end{align}
and by construction
\begin{align}
	S \text{ch}_0 = \operatorname{ch}_{\log, 2}, \qquad
	S \text{ch}_{\log, 2} = \operatorname{ch}_0, \qquad
	S \text{ch}_{\log, 1} = \operatorname{ch}_{\log, 1} \ .
\end{align}
In the form of matrix $g\operatorname{ch}_i = \sum_{j} g_{ij}\operatorname{ch}_j$, we have
\begin{align}
	T = e^{\frac{2\pi i}{3}}\begin{pmatrix}
		-1 & 0 & 0 \\
		-2 & - 1 & 0 \\
		1 & 1 & -1
	\end{pmatrix}\ ,\qquad
	S = \begin{pmatrix}
		0 & 0 & 1\\
		0 & 1 & 0 \\
		1 & 0 & 0 
	\end{pmatrix}\ ,
\end{align}
which furnishes a three-dimensional representation of $SL(2, \mathbb{Z})$. The three characters can form an $SL(2, \mathbb{Z})$ invariant partition function
\begin{align}
	Z = 2\operatorname{ch}_0 \overline{\operatorname{ch}_2} + \operatorname{ch}_1 \overline{\operatorname{ch}_1} + 2 \operatorname{ch}_2 \overline{\operatorname{ch}_0} = \sum_{i,j} M_{ij}\operatorname{ch}_i \overline{\operatorname{ch}_j} \ .
\end{align}
Unfortuantely the $S$-matrix does not lead to reasonable fusion coefficients. One can consider new basis $\operatorname{ch}'_i$ such that the pairing matrix $M'_{ij}$ is integral and the fusion coefficients $N_{ij}^k$ are non-negative integers. Such new basis is not unique, and it leads to two possible fusion algebras,
\begin{align}
	[\operatorname{ch}'_0] \times [\operatorname{ch}'_i] = [\operatorname{ch}'_i] \ , \qquad
	[\operatorname{ch}'_1] \times [\operatorname{ch}'_1] = [\operatorname{ch}'_1], \qquad
	[\operatorname{ch}'_2] \times [\operatorname{ch}'_2] = [\operatorname{ch}'_2] \ ,
\end{align}
or,
\begin{align}
	[\operatorname{ch}'_0] \times [\operatorname{ch}'_i] = & \ [\operatorname{ch}'_i],
	\qquad
	[\operatorname{ch}'_1] \times [\operatorname{ch}'_1] = [\operatorname{ch}'_1] , \qquad
	[\operatorname{ch}'_1] \times [\operatorname{ch}'_2] = 2 [\operatorname{ch}'_1], \\
	[\operatorname{ch}'_2] \times [\operatorname{ch}'_2] = & \ 2 [\operatorname{ch}'_1] + [\operatorname{ch}'_2] \ .
\end{align}
it is unclear if such algebras have physical or mathematical meaning.

Besides the Schur index and its modular companions, the residue $\mathcal{I}_{(\beta \gamma)^2}$ transforms in a one-dimensional representation under $SL(2, \mathbb{Z})$,
\begin{align}
	S\mathcal{I}_{(\beta \gamma)^2} = - \mathcal{I}_{(\beta \gamma)^2}, \qquad
	T\mathcal{I}_{(\beta \gamma)^2} = - (-1)^{2/3} \mathcal{I}_{(\beta \gamma)^2} \ ,
\end{align}
satisfying $S^2 = (ST)^3 = \operatorname{id}$.


\subsubsection{The twisted sector}

The defect index $\mathcal{I}_{1,2}^\text{defect}(k = 1)$ is a twisted character, and will satisfy corresponding twisted modular differential equations. Again, these equations can be obtained from all those in the untwisted sector with all the contributions $E_n \big[  \substack{+1 \\ \ldots}  \big]$ from the bifundamentals turned into $E_n \big[  \substack{-1\\ \ldots}  \big]$. For example, at weight-2, there is
\begin{align}
	0 = & \ \bigg[  D_q^{(1)}-\frac{1}{2}D_{b_2}^{2}-\frac{1}{2}\left( E_1\left[ \begin{matrix}
		- 1\\
		b_1b_2\\
	\end{matrix} \right] +E_1\left[ \begin{matrix}
		- 1\\
		\frac{b_1}{b_2}\\
	\end{matrix} \right] \right) D_{b_1}
	-\frac{1}{2}\left( E_1\left[ \begin{matrix}
		- 1\\
		b_1b_2\\
	\end{matrix} \right] -E_1\left[ \begin{matrix}
		- 1\\
		\frac{b_1}{b_2}\\
	\end{matrix} \right] +E_1\left[ \begin{matrix}
		- 1\\
		b_{2}^{2}\\
	\end{matrix} \right] \right) \nonumber  \\
	& \ +2\left( E_2\left[ \begin{matrix}
		- 1\\
		b_1b_2\\
	\end{matrix} \right] +E_2\left[ \begin{matrix}
		- 1\\
		\frac{b_1}{b_2}\\
	\end{matrix} \right] +2E_2\left[ \begin{matrix}
		1\\
		b_{2}^{2}\\
	\end{matrix} \right] +E_2 \right)  \bigg]\mathcal{I}_{1,2} \ ,
\end{align}
and one with $b_1 \leftrightarrow b_2$.

\subsection{Other examples}

In the previous subsections we have discussed a few simplest theories with low $g, n$, where we have studied their flavored modular differential equations and their solutions besides the Schur index. In the following we comment on theories with higher $g, n$. For simplicity, we shall focus on the unflavored index and the unflavored modular differential equations \cite{Beem:2017ooy} they satisfy.

\subsubsection{$\mathcal{T}_{0,5}$}

We start with the theory $\mathcal{T}_{0,5}$. The unflavored Schur index satisfies a weigh-8 modular differential equation,
\begin{align}\label{MDE-I05-weight-8}
	0 = \bigg[D_q^{(4)} - 220E_4 D_q^{(2)}
	- & \  \left(
	3020E_6 + 3840 E_6 \begin{bmatrix}
		-1 \\ 1
	\end{bmatrix}
	\right)D_q^{(1)}\nonumber\\
	& \ - 144\bigg(-35E_8 + 224 E_8 \begin{bmatrix}
			-1 \\ 1
		\end{bmatrix}
		+ 144 E_4 \begin{bmatrix}
			-1 \\ 1
		\end{bmatrix}^2\bigg)\bigg] \mathcal{I}_{0,5} \ .
\end{align}
This equation has four independent solutions. The indicial equation for the anzatz $q^h\sum_{n} a_n q^n$ reads
\begin{align}
	(h - 1)h^3 = 0 \qquad \Rightarrow \qquad h = 1, 0, 0, 0\ .
\end{align}
The integral spacing suggests the presence of two or three logarithmic solutions. Clearly the $h = 1$ solution is given already by the original unflavored Schur index. It turns out that there is an additional non-logarithmic solution given by the unflavored vortex defect index with vorticity $k = 2$,
\begin{align}
	\mathcal{I}_{0,5}^\text{defect}(k = 2) 
	= & \ \frac{-5 \vartheta_4(0) \vartheta^{(2)}_4(0) \left(3 \pi ^2 (12 E_2(q)+5) \vartheta^{(4)}_4(0)-2 \vartheta^{(6)}_4(0)\right)}{1024 \pi ^8 \eta (q)^{12} \vartheta_4(0)^3} \nonumber\\
	 & \ + \frac{\vartheta_4(0)^2 \left(\pi ^2 (12 E_2(q)+5) \vartheta^{(6)}_4(0)-\vartheta_4^{(8)}(0)\right)+5 \vartheta_4(0)\vartheta^{(4)}_4(0)^2}{1024 \pi ^8 \eta (q)^{12} \vartheta_4(0)^3} \nonumber\\
	 & \  + \frac{30 \pi ^2 (12 E_2(q)+5) \vartheta^{(2)}_4(0)^3-30 (\vartheta^{(2)}_4(0))^2 \vartheta^{(4)}_4(0)}{1024 \pi ^8 \eta (q)^{12} \vartheta_4(0)^3} \ .
\end{align}
Note that there are only $\lceil \frac{n + 2g - 2}{2} \rceil = 2 $ defect indices with even vorticity, including the original Schur index $\mathcal{I}_{0,5}$ at $k = 0$.

The remaining two solutions are logarithmic, given by the $S$-transformation of $\mathcal{I}_{0,5}$ and $\mathcal{I}^\text{defect}_{0,5}(k = 2)$, where a basis can be chosen to be
\begin{align}
	\mathcal{I}_{0,5}, \qquad STS \mathcal{I}_{0,5}, \qquad
	TST \mathcal{I}_{0,5}, \qquad
	\mathcal{I}^\text{defect}_{0,5}(k = 2) \ ,
\end{align}
or
\begin{align}
	\mathcal{I}_{0,5}, \qquad STS \mathcal{I}_{0,5}, \qquad
	\mathcal{I}^\text{defect}_{0,5}(k = 2) , \qquad
	TST \mathcal{I}^\text{defect}_{0,5}(k = 2)\ .
\end{align}

There are two independent vortex defect indices with odd vorticity, $\mathcal{I}^\text{defect}_{0,5}(k = 1)$ and $\mathcal{I}^\text{defect}_{0,5}(k = 3)$. Let us first look at the unflavoring limit of the first defect index,
\begin{align}
	\mathcal{I}_{0,5}^\text{defect}(k = 1)
	= & \ - \frac{1}{\eta(\tau)^{12}} (60E_2^2 E_4 - 420 E_2 E_6 + 700 E_8)\nonumber\\
	= & \ q^{\frac{1}{2}}(1 + 48 q + 774 q^2 + 7952 q^3 + 61101 q^4 + 385200 q^5 + \ldots) \ .
\end{align}
It is easy to check that it satisfies an equation in the twisted sector,
\begin{align}
	(D_q^{(4)} - 220 E_4 D_q^{(2)} - 6860 E_6 D_q^{(1)} - 75600 E_8)\mathcal{I}^\text{defect}_{0,5}(k = 1) = 0 \ .
\end{align}
Apparently this equation is just the twisted version of (\ref{MDE-I05-weight-8}), where $E_k \big[  \substack{-1 \\ 1}  \big]$ are replaced by $E_k \big[  \substack{+1 \\ 1}  \big]$ and applying the relation $E_4 ^2 = \frac{7}{3}E_8$.

To study the second defect index $\mathcal{I}^\text{defect}_{0,5}(k = 3)$, one has to turn on flavor fugacities since it does not have a smooth unflavoring limit. The simplest partial flavoring is $b_1 = b, b_{2,3,4,5} = 1$. It turns out that in this limit there are no flavored modular differential equations below weight-eight, and all the weight-eight equations satisfied by $\mathcal{I}^\text{defect}_{0,5}(k = 1)$ will also have $\mathcal{I}^\text{defect}_{0,5}(k = 3)$ as an additional solution. We refrain from showing the detail of these equations due to their complexity.

\subsubsection{$\mathcal{T}_{0,6}$}

For $\mathcal{T}_{0,6}$, the unflavored Schur index $\mathcal{I}_{0,6}$ satisfies a weight-twelve, 6$^\text{th}$ order equation,
\begin{align}
	0 = \Big[D_{q}^{(6)} -545 E_{4} D_{q}^{(4)}-15260 E_{6} D_{q}^{(3)} & \ -164525 E_{4}^{2} D_{q}^{(2)} - 2775500 E_{4} E_{6} D_{q}^{(1)} \nonumber \\
	& \  - 26411000 E_{6}^{2}+1483125 E_{4}^{3} \Big]\mathcal{I}_{0,6} \ .
\end{align}
The indicial equation gives
\begin{align}\label{MDE-I06}
	(5 - 12h)^4 (144h^2 - 120h - 119) = 0 \qquad
	\Rightarrow
	\qquad
	h = \frac{5}{12}, \frac{5}{12}, \frac{5}{12}, \frac{5}{12}, - \frac{7}{12}, \frac{17}{12} \ .
\end{align}
Obviously, the solution with $h = \frac{17}{12}$ correspond to the Schur index. Another non-logarithmic solution comes from the non-trivial defect index: in this case, there are $\lceil \frac{n + 2g - 2}{2} \rceil = 2$ independent defect indices with even $k$, and indeed $\mathcal{I}_{0,6}$ and $\mathcal{I}^\text{defect}_{0,6}(2)$ are the two independent non-logarithmic solutions to the modular differential equation (\ref{MDE-I06}), where the latter corresponds to one of the $h = \frac{5}{12}$. The remaining four solutions are logarithmic obtained from modular transformation of $\mathcal{I}_{0,6}$ and $\mathcal{I}^\text{defect}_{0,6}(2)$. One independent basis can be chosen to be
\begin{align}
	\mathcal{I}_{0,6}, \quad
	S\mathcal{I}_{0,6}, \quad
	TS\mathcal{I}_{0,6}, \quad
	T^2S\mathcal{I}_{0,6}, \quad
	\mathcal{I}_{0,6}^\text{defect}(2), \quad
	S\mathcal{I}_{0,6}^\text{defect}(2) \ ,
\end{align}
spanning the space of solutions.

There are also two linear independent defect indices with odd vorticity, $\mathcal{I}^\text{defect}_{0,6}(k = 1)$ and $\mathcal{I}^\text{defect}_{0,6}(k = 3)$, both with smooth unflavoring limit
\begin{align}
	\mathcal{I}^\text{defect}_{0,6}(k = 1) = & \ q^{\frac{11}{12}}(1 + 19 q+ 64 q^{3/2} + 203 q^2 + 896 q^{5/2} + 2320q^3 + \ldots) \ ,\\
	\mathcal{I}^\text{defect}_{0,6}(k = 3) = & \ q^{\frac{11}{12}}(1 + 64 q^{1/2} + 748 q + 4992 q^{3/2} + 26035 q^2 + 111936^{5/2} + \ldots) \ .
\end{align}
They both satisfy a $6^\text{th}$ order unflavored modular differential equation whose explicit form will not be included here.

\subsubsection{$\mathcal{T}_{g,n = 0}$}

The unflavored index $\mathcal{I}_{2,0}$ of the genus-two theory $\mathcal{T}_{2,0}$ can be written in terms of the standard Eisenstein series,
\begin{align}
	\mathcal{I}_{2,0} = \frac{1}{2} \eta(\tau)^2 \left(E_2 + \frac{1}{12}\right) \ .
\end{align}
It satisfies a 6$^\text{th}$ order equation
\begin{align}
	0 = \Big[D_q^{(6)} - 305 E_4 D_q^{(4)} - 4060E_6 D_q^{(3)}
			+ 20275E_4^2 & \ D_q^{(2)} + 2100E_4 E_6 D_q^{(1)} \nonumber \\
			& \ - 68600(E_6^2 - 49125E_4^3) \Big]\mathcal{I}_{2,0} \ .
\end{align}
Following from seciton \ref{section:defect-indices}, there is an additional vortex defect index
\begin{align}
	\mathcal{I}_{2,0}^\text{defect}(k=2) = \eta(\tau)^2 \ .
\end{align}
Similar to the $g = 0, n = 6$ case, the $SL(2, \mathbb{Z})$ orbit of $\mathcal{I}_{2,0}$ and $\eta(\tau)^2$ form the complete set of solutions of the 6$^\text{th}$ order equation, where an independent basis can be chosen as
\begin{align}
	\mathcal{I}_{2,0}, \quad
	S\mathcal{I}_{2,0}, \quad
	TS\mathcal{I}_{2,0}, \quad
	T^2S\mathcal{I}_{2,0}, \quad
	\mathcal{I}_{2,0}^\text{defect}(2), \quad
	S\mathcal{I}_{2,0}^\text{defect}(2) \ .
\end{align}
Note that since $\eta(\tau)^2$ is a term in the index $\mathcal{I}_{2,0}$ itself, and consequently, the other term $\eta(\tau)^2E_2$ naturally forms another solution.

Similarly, the indices $\mathcal{I}_{3,0}$ and $\mathcal{I}_{4,0}$ of the genus-three and -four theories satisfy a 20$^\text{th}$ and 43$^\text{th}$ order modular differential equation respectively, whose expressions will not be included here. By direct computation, it can be shown that the Schur index itself and the defect indices $\mathcal{I}^\text{defect}_{g,0}(k = \text{even})$ provide a collection of solutions. Note that this equivalently implies that $\eta(\tau)^{2g - 2}, \eta(\tau)^{2g - 2} E_2, \ldots, \eta(\tau)^{2g - 2} E_{2g - 2}$ are $g$ independent solutions to these equations. Their $SL(2, \mathbb{Z})$-orbit will supply additional logarithmic solutions to the equations.

\section*{Acknowledgments}
The authors would like to thank Wolfger Peelaers for sharing important observations and Mathematica notebooks on modular differential equations. Y.P. is supported by the National Natural Science Foundation of China (NSFC) under Grant No. 11905301, the Fundamental Research Funds for the Central Universities, Sun Yat-sen University under Grant No. 2021qntd27.

\appendix

\section{Special Functions}\label{app:specialfuctions}

In this appendix we collect the definitions and a few useful properties of the special functions that appear in the maintext. We often use letters in straight and fraktur font which are related by
\begin{align}
	a = e^{2\pi i \mathfrak{a}}, \qquad
	b = e^{2\pi i \mathfrak{b}}, \qquad \ldots \qquad
	y = e^{2\pi i \mathfrak{y}}, \qquad
	z = e^{2\pi i \mathfrak{z}}\ .
\end{align}

\subsection{Jacobi theta functions}

The Jacobi theta functions are defined as Fourier series
\begin{align}
	\vartheta_1(\mathfrak{z}|\tau) \colonequals & \ -i \sum_{r \in \mathbb{Z} + \frac{1}{2}} (-1)^{r-\frac{1}{2}} e^{2\pi i r \mathfrak{z}} q^{\frac{r^2}{2}} ,\\
	\vartheta_2(\mathfrak{z}|\tau) \colonequals & \sum_{r \in \mathbb{Z} + \frac{1}{2}} e^{2\pi i r \mathfrak{z}} q^{\frac{r^2}{2}} \ ,\\
	\vartheta_3(\mathfrak{z}|\tau) \colonequals & \ \sum_{n \in \mathbb{Z}} e^{2\pi i n \mathfrak{z}} q^{\frac{n^2}{2}},\\
	\vartheta_4(\mathfrak{z}|\tau) \colonequals & \sum_{n \in \mathbb{Z}} (-1)^n e^{2\pi i n \mathfrak{z}} q^{\frac{n^2}{2}} \ .
\end{align}
Through out this paper we denote $q \coloneqq e^{2\pi i \tau}$. For brevity we will frequently omit $|\tau$ in the notation of the Jacobi theta functions. The Jacobi-theta functions can be rewritten as triple product of the $q$-Pochhammer symbol, for example,
\begin{align}\label{theta1-product-formula}
	\vartheta_1(\mathfrak{z}) = - i z^{\frac{1}{2}}q^{\frac{1}{8}}(q;q)(zq;q)(z^{-1};q) \ , \qquad
	\vartheta_4(\mathfrak{z}) = (q;q)(zq^{\frac{1}{2}};q)(z^{-1}q^{\frac{1}{2}};q) \ ,
\end{align}
where $(z;q) \colonequals \prod_{k = 0}^{+\infty}(1 - zq)$.

The functions $\vartheta_i(z)$ almost return to themselves under full-period shifts by $m + n \tau$,
\begin{align}
	\vartheta_{1,2}(\mathfrak{z} + 1) = & - \vartheta_{1,2}(\mathfrak{z}) , & 
	\vartheta_{3,4}(\mathfrak{z} + 1) = & + \vartheta_{3,4}(\mathfrak{z}) , & \\
	\vartheta_{1,4}(\mathfrak{z} + \tau) = & - \lambda \vartheta_{1,4}(\mathfrak{z}), & 
	\vartheta_{2,3}(\mathfrak{z} + \tau) = & + \lambda \vartheta_{2,3}(\mathfrak{z}) , & 
\end{align}
where $\lambda \equiv e^{-2\pi i \mathfrak{z}}e^{- \pi i \tau}$. The above can be combined, for example, into
\begin{align}
	\vartheta_1(\mathfrak{z} + m \tau + n) = (-1)^{m + n} e^{-2\pi i m \mathfrak{z}} q^{ - \frac{1}{2}m^2}\vartheta_1(\mathfrak{z})\ .
\end{align}
Moreover, the four Jacobi theta functions are related by half-period shifts which can be summarized as in the following diagram,
\begin{center}
	\includegraphics[height=100pt]{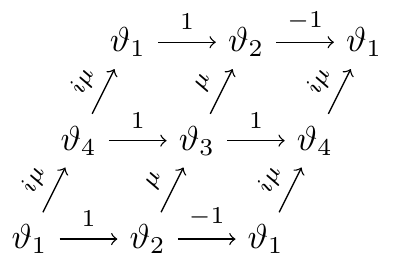}
\end{center}
where $\mu = e^{- \pi i \mathfrak{z}} e^{- \frac{\pi i}{4}}$, and $f \xrightarrow{a} g$ means
\begin{align}
	\text{either}\qquad  f\left(\mathfrak{z} + \frac{1}{2}\right) = a g(\mathfrak{z}) \qquad \text{or} \qquad
	f\left(\mathfrak{z} + \frac{\tau}{2}\right) = a g(\mathfrak{z}) \ ,
\end{align}
depending on whether the arrow is horizontal or (slanted) vertical respectively.

The functions $\vartheta_i(z | \tau)$ transform nicely under the modular $S$ and $T$ transformations, which act, as usual, on the nome and flavor fugacity as $(\frac{\mathfrak{z}}{\tau}, - \frac{1}{\tau})\xleftarrow{~~S~~}(\mathfrak{z}, \tau) \xrightarrow{~~T~~} (\mathfrak{z}, \tau + 1).$ In summary
\begin{center}
	\includegraphics[height=0.2\textheight]{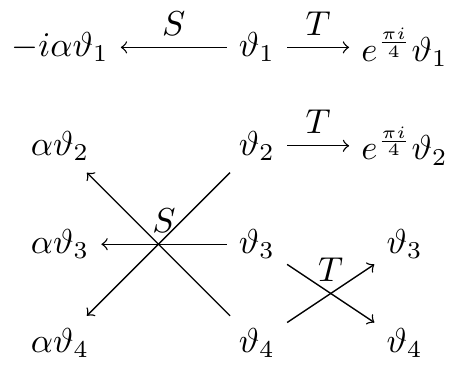}
\end{center}
where $\alpha = \sqrt{-i \tau}e^{\frac{\pi i z^2}{\tau}}$.

\subsection{Eisenstein series}

The twisted Eisenstein series (with characteristics $\big[\substack{\phi\\\theta}\big]$) $E_k\big[\substack{\phi\\\theta}\big]$ are defined as series in $q$,
\begin{align}
	E_{k \ge 1}\left[\begin{matrix}
		\phi \\ \theta
	\end{matrix}\right] \colonequals & \ - \frac{B_k(\lambda)}{k!} \\
	& \ + \frac{1}{(k-1)!}\sum_{r \ge 0}' \frac{(r + \lambda)^{k - 1}\theta^{-1} q^{r + \lambda}}{1 - \theta^{-1}q^{r + \lambda}}
	+ \frac{(-1)^k}{(k-1)!}\sum_{r \ge 1} \frac{(r - \lambda)^{k - 1}\theta q^{r - \lambda}}{1 - \theta q^{r - \lambda}} \ .
\end{align}
Here $0 \le \lambda < 1$ is determined by $\phi \equiv e^{2\pi i \lambda}$, $B_k(x)$ denotes the $k$-th Bernoulli polynomial, and the prime in the sum means that when $\phi = \theta = 1$ the $r = 0$ term should be omitted. We also define
\begin{align}
	E_0\left[\begin{matrix}
		\phi \\ \theta
	\end{matrix}\right] = -1 \ .
\end{align}

The standard (untwisted) Eisenstein series $E_{2n}$ are given by the $\theta, \phi \to 1$ limit of $E_{2n}\big[\substack{\phi \\ \theta}\big]$,
\begin{align}
	E_{2n}(\tau) = E_{2n}\begin{bmatrix}
		+1 \\ +1
	\end{bmatrix} \ .
\end{align}
When $k$ is odd, we have instead
\begin{align}
	E_1 \begin{bmatrix}
		1 \\ e^{2\pi i \mathfrak{z}}
	\end{bmatrix} = \frac{1}{2\pi i} \frac{\vartheta'_1(\mathfrak{z})}{\vartheta_1(\mathfrak{z})} \xrightarrow{\mathfrak{z} \to 0} \frac{1}{2\pi i \mathfrak{z}}, \qquad
	E_{k > 1} \begin{bmatrix}
		+1 \\ +1
	\end{bmatrix} = 0\ .
\end{align}

The Eisenstein series with $\phi = \pm 1$ enjoy a useful symmetry property
\begin{align}\label{Eisenstein-symmetry}
	E_k\left[\begin{matrix}
	  \pm 1 \\ z^{-1}
	\end{matrix}\right] = (-1)^k E_k\left[\begin{matrix}
	  \pm 1 \\ z
	\end{matrix}\right] \ .
\end{align}
They also transform nicely under $z \to qz$ or $z \to q^{\frac{1}{2}}z$, for example,
\begin{align}\label{Eisenstein-shift-property}
	E_n \begin{bmatrix}
		\pm 1\\ z q^{\frac{k}{2}}
	\end{bmatrix}
	= \sum_{\ell = 0}^{n} \left(\frac{k}{2}\right)^\ell \frac{1}{\ell!} E_{n - \ell} \begin{bmatrix}
		(-1)^k (\pm 1)\\
		z
	\end{bmatrix} \ .
\end{align}

The Eisenstein series are closely related to the Jacobi theta function. Let us define an elliptic $P$-function
\begin{align}
	P_2(y) \colonequals - \sum_{n = 1}^{+\infty} \frac{1}{2n}E_{2n}(\tau)y^{2n} \ .
\end{align}
Then we have a simple translation
\begin{align}
	E_k \begin{bmatrix}
		+1 \\ + z
	\end{bmatrix}
	= & \left[e^{- \frac{y}{2\pi i}\mathcal{D}_\mathfrak{z} - P_2(y)}\right]_k \vartheta_1(\mathfrak{z}), &
	E_k \begin{bmatrix}
		-1 \\ + z
	\end{bmatrix}
	= & \left[e^{- \frac{y}{2\pi i}\mathcal{D}_\mathfrak{z} - P_2(y)}\right]_k \vartheta_4(\mathfrak{z}) \\
	E_k \begin{bmatrix}
		+1 \\ -z
	\end{bmatrix}
	= & \left[e^{- \frac{y}{2\pi i}\mathcal{D}_\mathfrak{z} - P_2(y)}\right]_k \vartheta_2(\mathfrak{z}), &
	E_k \begin{bmatrix}
		-1 \\ -z
	\end{bmatrix}
	= & \left[e^{- \frac{y}{2\pi i}\mathcal{D}_\mathfrak{z} - P_2(y)}\right]_k \vartheta_3(\mathfrak{z}) \ ,
\end{align}
where $[\ldots]_k$ means taking the coefficient of $y^k$ when expanding $\ldots$ in $y$-series around $y = 0$, and $\mathcal{D}_\mathfrak{z}^n$ acting on $\vartheta_i$ is defined to be
\begin{align}
	\mathcal{D}_\mathfrak{z}^n \vartheta_i (\mathfrak{z}) \colonequals \frac{\vartheta_i^{(n)}(\mathfrak{z})}{\vartheta_i(\mathfrak{z})} \ .
\end{align}

The Eisenstein series contains simple poles whose residues are easy to work out from the definition. For example,
\begin{align}\label{Eisenstein-residues}
	\mathop{\operatorname{Res}}_{z \to q^{k + \frac{1}{2}}} \frac{1}{z} E_n \begin{bmatrix}
		- 1 \\ z
	\end{bmatrix} = \frac{1}{(n - 1)!} (k + \frac{1}{2})^{n - 1}\ ,\qquad
	\mathop{\operatorname{Res}}_{z \to q^k} \frac{1}{z} E_n \begin{bmatrix}
		+ 1 \\ z
	\end{bmatrix} = \frac{1}{(n - 1)!} k^{n - 1}\ .
\end{align}

The relation between the Eisenstein series and Jacobi theta functions are helpful in working out the modular transformation of the former. In detail, we consider the following transformations,
\begin{align}
	S: \tau \to - \frac{1}{\tau}, \ \mathfrak{z} \to \frac{\mathfrak{z}}{\tau}, \qquad
	\qquad
	T: \tau \to \tau + 1 , \ \mathfrak{z} \to \mathfrak{z} \ .
\end{align}
Under the $S$-transformation,
\begin{align}\label{Eisenstein-S-transformation}
	E_n \begin{bmatrix}
		+1 \\ +z
	\end{bmatrix} \xrightarrow{S} &
	\left(\frac{1}{2\pi i}\right)^n\left[\bigg(\sum_{k \ge 0}\frac{1}{k!}(- \log z)^k y^k\bigg)
	\bigg(\sum_{\ell \ge 0}(\log q)^\ell y^\ell E_\ell \begin{bmatrix}
		+ 1 \\ z
	\end{bmatrix}\bigg)\right]_n\ ,\\
	E_n \begin{bmatrix}
		-1 \\ +z
	\end{bmatrix} \xrightarrow{S} &
	\left(\frac{1}{2\pi i}\right)^n\left[\bigg(\sum_{k \ge 0}\frac{1}{k!}(- \log z)^k y^k\bigg)
	\bigg(\sum_{\ell \ge 0}(\log q)^\ell y^\ell E_\ell \begin{bmatrix}
		+ 1 \\ -z
	\end{bmatrix}\bigg)\right]_n\ ,\\
	E_n \begin{bmatrix}
		1 \\ -z
	\end{bmatrix} \xrightarrow{S} &
	\left(\frac{1}{2\pi i}\right)^n\left[\bigg(\sum_{k \ge 0}\frac{1}{k!}(- \log z)^k y^k\bigg)
	\bigg(\sum_{\ell \ge 0}(\log q)^\ell y^\ell E_\ell \begin{bmatrix}
		-1 \\ +z
	\end{bmatrix}\bigg)\right]_n\ ,\\
	E_n \begin{bmatrix}
		-1 \\ -z
	\end{bmatrix} \xrightarrow{S} &
	\left(\frac{1}{2\pi i}\right)^n\left[\bigg(\sum_{k \ge 0}\frac{1}{k!}(- \log z)^k y^k\bigg)
	\bigg(\sum_{\ell \ge 0}(\log q)^\ell y^\ell E_\ell \begin{bmatrix}
		-1 \\ -z
	\end{bmatrix}\bigg)\right]_n\ ,
\end{align}
where $[ \ldots ]_n$ extracts the coefficient of $y^n$. For readers' convenience, we collect here the $S$-transformation of several lower-weight Eisenstein series,
\begin{align}
	E_1 \begin{bmatrix}
		+ 1 \\ z
	\end{bmatrix} \xrightarrow{S}& \
	\tau E_1 \begin{bmatrix}
		+ 1 \\ z
	\end{bmatrix} + \mathfrak{z} \\
	E_2 \begin{bmatrix}
		+ 1 \\ z
	\end{bmatrix} \xrightarrow{S}& \
	\tau^2 E_2 \begin{bmatrix}
		1 \\ z
	\end{bmatrix}
	- \mathfrak{z}\tau E_1 \begin{bmatrix}
		1 \\ z
	\end{bmatrix}
	- \frac{\mathfrak{z}^2}{2}\\
	E_3 \begin{bmatrix}
		+ 1 \\ z
	\end{bmatrix}
	\xrightarrow{S}& \ 
	\tau^3 E_3 \begin{bmatrix}
		+ 1 \\ z
	\end{bmatrix}
	- \mathfrak{z}\tau^2 E_2 \begin{bmatrix}
		+1 \\ z
	\end{bmatrix}
	+ \frac{1}{2} \mathfrak{z}^2\tau E_1 \begin{bmatrix}
		+ 1 \\ z
	\end{bmatrix}
	+ \frac{\mathfrak{z}^3}{6} \ .
\end{align}
Under the $T$-transformation,
\begin{align}
	E_n \begin{bmatrix}
		+ 1 \\ + z
	\end{bmatrix} \xrightarrow{T}& \ E_n \begin{bmatrix}
		+ 1 \\ + z
	\end{bmatrix}, & 
	E_n \begin{bmatrix}
		- 1 \\ + z
	\end{bmatrix} \xrightarrow{T}& \
	E_n \begin{bmatrix}
		- 1 \\ - z
	\end{bmatrix} \\
	E_n \begin{bmatrix}
		+ 1 \\ - z
	\end{bmatrix} \xrightarrow{T}& \ E_n \begin{bmatrix}
		+ 1 \\ - z
	\end{bmatrix}, & 
	E_n \begin{bmatrix}
		- 1 \\ - z
	\end{bmatrix} \xrightarrow{T}& \ 
	E_n \begin{bmatrix}
		- 1 \\ + z
	\end{bmatrix} \ .
\end{align}
Combined together,
\begin{align}
	E_n \begin{bmatrix}
		-1 \\ z
	\end{bmatrix} \xrightarrow{STS}
	\left(\frac{1}{2\pi i}\right)^n\left[\bigg(\sum_{k \ge 0}\frac{1}{k!}(- \log z)^k y^k\bigg)
	\bigg(\sum_{\ell \ge 0}(\log q - 2\pi i)^\ell y^\ell E_\ell \begin{bmatrix}
		-1 \\ +z
	\end{bmatrix}\bigg)\right]_n\ .
\end{align}

The transformation of the Eisenstein under $SL(2, \mathbb{Z})$ implies that they are generalization of the well known modular forms and Jacobi forms. In the theory of modular/Jacobi forms, there are a few important differential operators that change the modular weight of a form. The Serre derivative $\partial_{(k)}$ is defined to be 
\begin{align}
	\partial_{(k)} \coloneqq q \partial_q + k E_2\ .
\end{align}
It maps a weight-$k$ modular form to a weight-$(k + 1)$ form. For example,
\begin{align}
	\partial_{(2)} E_2 = 5 E_4 + E_2^2 \ ,\qquad
	\partial_{(4)} E_4 = 14 E_6, \qquad
	\partial_{(6)} E_6 = 20 E_8\ .
\end{align}
One can compose the Serre derivative into the modular differential operators $D_q^{(k)}$,
\begin{align}\label{def:MDO}
	D_q^{(k)} \coloneqq \partial_{(2k - 2)} \circ \ldots \circ \partial_{(2)} \circ \partial_{(0)}\ .
\end{align}
Such operator turns a weight-zero form to a weight-$2k$ form. It transform covariantly under the standard $SL(2, \mathbb{Z})$ transformation $\tau \to \tau' \coloneqq \frac{a\tau + b}{c \tau + d}$, 
\begin{align}
	D^{(k)}_{q'} = (c \tau + d)^{2k} D^{(k)}_q  \ .
\end{align}


\section{Vertex operator algebra \label{app:voa}}

In this section we briefly summarize some notions and formula concerning vertex operator algebras (VOAs). For more rigorous account for the subject, see for example \cite{frenkel1989vertex,zhu1996modular}. A VOA $\mathbb{V}$ is characterized by a linear space of states $V$ (i.e., the vacuum module), containing a unique vacuum state $|0\rangle$ and a special state $T$ corresponding to the stress tensor. There is a state-operator correspondence $Y$ that builds a local field $Y(a, z)$ out of any state $a \in V$. We often simply denote the field as $a(z)$ and expand it in a Fourier series\footnote{In math literature the expansion is often taken to be $\sum_{n \in \mathbb{Z}} a_n z^{-n - 1}$.}
\begin{align}
	a(z) \coloneqq Y(a, z) = \sum_{n \in \mathbb{Z} - h_a} a_n z^{- n - h_a} \ ,
  \qquad
  T(z) = \sum_{n \in \mathbb{Z}} L_n z^{-n - 2} \ .
\end{align}
Here the Fourier modes $a_n$ are linear operators that act on $V$, $L_n$ form a Virasoro algebra with central charge $c$, and $h_a$ is the eigenvalue in $L_0 a = h_a a$. The vacuum state $|0\rangle$ is such that $Y(|0\rangle, z) = \operatorname{id}_V$ and $a(0) |0\rangle = a$. For a state $a$ with integer weight $h_a$, one defines its zero mode $o(a) \coloneqq a_0$, whereas $o(a) = 0$ when $h_a$ is non-integral.

To compute torus correlation functions, it is a common practice to consider \cite{zhu1996modular}
\begin{align}
 	a[z] \coloneqq e^{i z h_a}Y(a, e^{i z} - 1) = \sum_{n}a_{[n]}z^{-n - h_a}
\end{align}
where the ``square modes'' $a_{[n]}$ are defined by the expansion. Explicitly,
\begin{align}
  a_{[n]} = \sum_{j \ge n} c(j, n, h_a)a_j
\end{align}
where the coefficients $c$ are given by the coefficients of the expansion
\begin{align}
  (1 + z)^{h - 1}[\log(1 + z)]^n = \sum_{j \ge n} c(j, n, h)z^j\ .
\end{align}
It is worth noting that $o(a_{[-h_a - n]}) = 0$, $\forall n \in \mathbb{N}_{\ge 1}$.

Recursion relations for unflavored torus correlation functions were first studied in \cite{zhu1996modular}, and later generalized to $\mathbb{R}$-graded super-VOAs \cite{Mason:2008zzb} and flavored correlation functions \cite{Krauel:2013lra}. They are the crucial tools for deriving flavored modular differential equations. Consider a $\frac{1}{2}\mathbb{Z}$-graded super-VOA $\mathbb{V}$ containing a $\widehat{\mathfrak{u}}(1)$ current $J$ with zero mode $J_0$, $M$ a module of $\mathbb{V}$ and $a, b \in \mathbb{V}$ are two states of weights $h_a, h_b$. If $J_0 a = 0$, then \footnote{Here all modes are the ``square modes'', which are suitable for torus correlation functions.}\cite{Mason:2008zzb,Beem:2017ooy}
\begin{align}
  \label{recursion1}
  & \ \operatorname{str}_M o(a_{[- h_a]}b)x^{J_0}q^{L_0} \\
  = & \ \operatorname{str}_M o(a_{[-h_a]} |0\rangle)o(b) x^{J_0} q^{L_0} 
    + \sum_{n = 1}^{+\infty} E_{2k}\left[\begin{matrix}
    e^{2\pi i h_a} \\ 1
  \end{matrix}\right]
  \operatorname{str}_M o(a_{[-h_a + 2k]}b)x^{J_0}q^{L_0} \ .
\end{align}
Recall that when $a$ is a conformal descendant, $o(a_{[-h_a]}) = 0$. On the other hand, if the state $a$ is charged with $J_0 a = Qa$ and $Q \ne 0$, then the recursion formula reads \cite{Krauel:2013lra,Beem:202X}
\begin{align}
  \label{recursion2}
  \operatorname{str}_M & \ o(a_{[- h_a]}b)x^{J_0}q^{L_0}
  = \sum_{n = 1}^{+\infty} E_n\left[ \begin{matrix}
    e^{2\pi i h_a} \\ x^Q
  \end{matrix} \right]
  \operatorname{str}_M o(a_{[- h_a+n]}b)x^{J_0}q^{L_0}\ .
\end{align}

Another frequently encounter insertion is $o(L_{[-2]}^k |0\rangle)$. In particular,
\begin{align}
	\operatorname{str} o( (L_{[-2]})^k |0\rangle )q^{L_0 - \frac{c}{24}}
	= \mathcal{P}_{k} \operatorname{str}q^{L_0 - \frac{c}{24}} \ .
\end{align}
Here $\mathcal{P}_k$ denotes a $k$-th order (and weight-$2k$) differential operator on $q$,
\begin{align}
	\mathcal{P}_1 = D_q^{(1)}, \quad
	\mathcal{P}_2 = D_q^{(2)} + \frac{c}{4}E_4 , \quad
	\mathcal{P}_3 = D_q^{(3)} + (8 + \frac{3c}{2})E_4 D_q^{(1)} + 10 c E_6 \ , \quad \ldots \ .
\end{align}


\section{Null states in $\mathfrak{so}(8)_{-2}$}\label{app:su2SQCD}
The Lagrangian of $\mathcal{N}=2$ su(2) super QCD is (we denote both hypermultiplet and its scaler components by $Q$ and $\tilde{Q}$):
\begin{align}
\mathcal{L}=\Im\left(\tau\int d^2\theta d^2\bar{\theta}\operatorname{tr}\left(\Phi^{\dagger}e^{V}\Phi+Q^{\dagger}_i e^{V}Q^i+\tilde{Q}^{\dagger i}e^V \tilde{Q}_i\right)+\tau\int d^2\theta \left(\frac{1}{2}\operatorname{tr}\mathcal{W}^{\alpha}\mathcal{W}_{\alpha}+\sqrt{2}\tilde{Q}^a_{i}\Phi_a^b Q_b^i\right)\right)
\end{align}
Since the fundamental representation of $SU(2)$ is pseudo-real,the hypermultiplet scalars:
\begin{align}
	Q^i_a\qquad \tilde{Q}^a_i\qquad i=1...4\quad a=1,2
\end{align}
which transform under fundamental representation of flavor group $SU(4)$, can be recombined into a single $Q^i_a$, with $i=1,...,8$, transformed under $8_{V}$ of $SO(8)$. From now we collectively use $Q^i_a$, $i=1,..,8$ and $a=1,2$ to denote $Q$ and $\tilde{Q}$. More explicitly, $Q^{i}_a$ refers to $Q$ when $i$ ranges from $1$ to $4$, otherwise to $\tilde{Q}$, when $i$ ranges from $5$ to $8$. The moment map operator of the enhanced flavor group $SO(8)$ is:
\begin{align}
M^{[ij]}=Q^i_a Q^{a j}
\end{align}
It gives $\mathfrak{so}(8)_{-2}$ currents of the corresponding 2d chiral algebra, as is conjectured in \cite{Beem:2013sza}
\begin{align}
J^{[ij]}=\chi\left(M^{[ij]}\right)
\end{align}
where $\chi$ is the map from operators in the same $SU(2)_R$ multiplet with Schur operators to the generators in 2d chiral algebra, as is defined by ....  
There are totally 9 independent null states in 2d chiral algebra $\mathfrak{so}(8)_{-2}$. The first three of them have symmetric indices\cite{Beem:2013sza}:
\begin{align}
	&J^{[1j]}J^{[1j]}+J^{[5j]}J^{[5j]}-\frac{1}{4}J^{[mn]}J^{[mn]}\\
	&J^{[2j]}J^{[2j]}+J^{[6j]}J^{[6j]}-\frac{1}{4}J^{[mn]}J^{[mn]}\\
	&J^{[3j]}J^{[3j]}+J^{[7j]}J^{[7j]}-\frac{1}{4}J^{[mn]}J^{[mn]}
\end{align} 
The other six of null states have totally antisymmetric indices\cite{Argyres:1996eh}:
\begin{align}
&J^{[12]}J^{[56]}-J^{[15]}J^{[26]}+J^{[25]}J^{[16]}\\
&J^{[13]}J^{[57]}-J^{[15]}J^{[37]}+J^{[35]}J^{[17]}\\
&J^{[23]}J^{[68]}-J^{[26]}J^{[38]}+J^{[36]}J^{[28]}\\
&J^{[14]}J^{[58]}-J^{[15]}J^{[48]}+J^{[45]}J^{[18]}\\
&J^{[24]}J^{[68]}-J^{[26]}J^{[48]}+J^{[46]}J^{[28]}\\
&J^{[34]}J^{[78]}-J^{[37]}J^{[48]}+J^{[47]}J^{[38]}
\end{align}
There are four commuting $\mathfrak{su}(2)$ subalgebra in $\mathfrak{so}\left(8\right)_{-2}$. We can choose them to be the Chavalley bases of four simple roots which are not connected in the extended dynkin diagram of $\mathfrak{so}\left(8\right)$. Roughly speaking, generators of $\mathfrak{so}\left(8\right)$ gain charges under the Cartan of four $\mathfrak{su}\left(2\right)$ subalgebra. 
Unfortunately, the generators $J^{[ij]}$ of $\mathfrak{so}\left(8\right)_{-2}$ are not eigenvectors of the four $\mathfrak{su}\left(2\right)$. Therefore we use another definition of $\mathfrak{so}\left(8\right)$:
\begin{align}
M S+S M^{t}=0
\end{align}   
All $8\times 8$ matrices $M$ form a Lie algebra isormorphic to $\mathfrak{so}(8)$. Note that 
\begin{align}
S=\left(\begin{array}{cc}
	 0 & \mathbf{1}_{4\times 4} \\
	 \mathbf{1}_{4\times 4} & 0
\end{array}\right)
\end{align}
The isomorphism between $J^{[ij]}$ and $M$ is:
\begin{align}
J=TMT^{-1}
\end{align}
where
\begin{align}
T=\left(\begin{array}{cc}
	\frac{\mathbf{1}_{4\times 4}}{\sqrt{2}} & \frac{\mathbf{1}_{4\times 4}}{\sqrt{2}}\\
	\frac{i \mathbf{1}_{4\times 4}}{\sqrt{2}} &
	\frac{-i \mathbf{1}_{4\times 4}}{\sqrt{2}}
\end{array}\right)
\end{align}
In the algebra defined by $M$, the four Cartans in commuting $\mathfrak{su}\left(2\right)$ we choose are listed below:
\begin{align}
&h_1=\left(
\begin{array}{cccccccc}
	\frac{1}{2} & 0 & 0 & 0 & 0 & 0 & 0 & 0 \\
	0 & \frac{1}{2} & 0 & 0 & 0 & 0 & 0 & 0 \\
	0 & 0 & 0 & 0 & 0 & 0 & 0 & 0 \\
	0 & 0 & 0 & 0 & 0 & 0 & 0 & 0 \\
	0 & 0 & 0 & 0 & -\frac{1}{2} & 0 & 0 & 0 \\
	0 & 0 & 0 & 0 & 0 & -\frac{1}{2} & 0 & 0 \\
	0 & 0 & 0 & 0 & 0 & 0 & 0 & 0 \\
	0 & 0 & 0 & 0 & 0 & 0 & 0 & 0 \\
\end{array}
\right)\quad h_2=\left(
\begin{array}{cccccccc}
	\frac{1}{2} & 0 & 0 & 0 & 0 & 0 & 0 & 0 \\
	0 & -\frac{1}{2} & 0 & 0 & 0 & 0 & 0 & 0 \\
	0 & 0 & 0 & 0 & 0 & 0 & 0 & 0 \\
	0 & 0 & 0 & 0 & 0 & 0 & 0 & 0 \\
	0 & 0 & 0 & 0 & -\frac{1}{2} & 0 & 0 & 0 \\
	0 & 0 & 0 & 0 & 0 & \frac{1}{2} & 0 & 0 \\
	0 & 0 & 0 & 0 & 0 & 0 & 0 & 0 \\
	0 & 0 & 0 & 0 & 0 & 0 & 0 & 0 \\
\end{array}
\right)\notag\\
&h_3=\left(
\begin{array}{cccccccc}
	0 & 0 & 0 & 0 & 0 & 0 & 0 & 0 \\
	0 & 0 & 0 & 0 & 0 & 0 & 0 & 0 \\
	0 & 0 & \frac{1}{2} & 0 & 0 & 0 & 0 & 0 \\
	0 & 0 & 0 & -\frac{1}{2} & 0 & 0 & 0 & 0 \\
	0 & 0 & 0 & 0 & 0 & 0 & 0 & 0 \\
	0 & 0 & 0 & 0 & 0 & 0 & 0 & 0 \\
	0 & 0 & 0 & 0 & 0 & 0 & -\frac{1}{2} & 0 \\
	0 & 0 & 0 & 0 & 0 & 0 & 0 & \frac{1}{2} \\
\end{array}
\right)\quad
h_4=\left(
\begin{array}{cccccccc}
	0 & 0 & 0 & 0 & 0 & 0 & 0 & 0 \\
	0 & 0 & 0 & 0 & 0 & 0 & 0 & 0 \\
	0 & 0 & \frac{1}{2} & 0 & 0 & 0 & 0 & 0 \\
	0 & 0 & 0 & \frac{1}{2} & 0 & 0 & 0 & 0 \\
	0 & 0 & 0 & 0 & 0 & 0 & 0 & 0 \\
	0 & 0 & 0 & 0 & 0 & 0 & 0 & 0 \\
	0 & 0 & 0 & 0 & 0 & 0 & -\frac{1}{2} & 0 \\
	0 & 0 & 0 & 0 & 0 & 0 & 0 & -\frac{1}{2} \\
\end{array}
\right)\notag\\
\end{align}
The raising operators are chosen to be:
\begin{align}
\left(\begin{array}{cc}
e_{ij} & o\\
0 & -e_{ji}\\
\end{array}\right)
\quad \left(\begin{array}{cc}
0 & e_{ij}-e_{ji} \\
0 & 0\\
\end{array}\right) \quad	0\leq i<j\leq 4
\end{align}
$e_{ij}$ means a four by four matrix only have $1$ at the position $(i,j)$ and $0$ at other positions. The lowering operators are their transpose matrices. We can easily find they are eigenvetors of $h_1$, $h_2$, and $h_3$, $h_4$. To derive the flavored MDE, we use the isomorphism $T$ to transform from the bases $J^{[ij]}$ to the bases $M$.
We give here the flavor modular differential equations respectively to the last three null relations:
\begin{align}
	\bigg[ & - D_{b_1}D_{b_3} + D_{b_1}D_{b_4} - D_{b_2}D_{b_3} + D_{b_2}D_{b_4}\\
& \ - 2 E_1 \begin{bmatrix}
	1 \\ \frac{b_1 b_2 b_3}{b_4}
\end{bmatrix} (D_{b_1} + D_{b_2} + D_{b_3} - D_{b_4}
)
\textcolor{red}{+} 2E_1 \begin{bmatrix}
	1 \\ \frac{b_1 b_2 b_4}{b_3}
\end{bmatrix}(
D_{b_2} + D_{b_1} - D_{b_3} + D_{b_4}
) \\
& \ + 8 \bigg(E_2 \begin{bmatrix}
	1 \\ \frac{b_1 b_2 b_3}{b_4}
\end{bmatrix}
- E_2 \begin{bmatrix}
	1 \\ \frac{b_1 b_2 b_4}{b_3}
\end{bmatrix}
\bigg)\bigg] \mathcal{I}_{0,4} = 0 \ .
\end{align}
And
\begin{align}
	\bigg[ & - D_{b_1}D_{b_3} + D_{b_1}D_{b_4} + D_{b_2}D_{b_3} - D_{b_2}D_{b_4}\\
	& \ - 2 E_1 \begin{bmatrix}
		1 \\ \frac{b_1 b_3}{b_2 b_4}
	\end{bmatrix} (D_{b_1} - D_{b_2} + D_{b_3} - D_{b_4}
	)
	\textcolor{red}{+} 2E_1 \begin{bmatrix}
		1 \\ \frac{b_1 b_4}{b_2 b_3}
	\end{bmatrix}(
	D_{b_1} - D_{b_2} - D_{b_3} + D_{b_4}
	) \\
	& \ + 8 \bigg(E_2 \begin{bmatrix}
		1 \\ \frac{b_1 b_3}{b_2 b_4}
	\end{bmatrix}
	- E_2 \begin{bmatrix}
		1 \\ \frac{b_1 b_4}{b_2 b_3}
	\end{bmatrix}
	\bigg)\bigg] \mathcal{I}_{0,4} = 0 \ .
\end{align}
And
\begin{align}
	\left(D_{b_4}^2 + 4 E_1 \begin{bmatrix}
		1 \\ b_4^2
	\end{bmatrix} D_{b_4}
	- 8 E_2 \begin{bmatrix}
		1 \\ b_4^2
	\end{bmatrix}\right) \mathcal{I}_{0,4}
	= \left(D_{b_3}^2 + 4 E_1 \begin{bmatrix}
		1 \\ b_3^2
	\end{bmatrix} D_{b_3}
	- 8 E_2 \begin{bmatrix}
		1 \\ b_3^2
	\end{bmatrix}\right) \mathcal{I}_{0,4} \ .
\end{align}


\section{Flavored modular differential equations\label{app:fmldes}}

In this appendix we collect a few long equations explicitly that were omited in the maintext.

\subsection{\texorpdfstring{$\mathcal{T}_{1,1}$}{}}

In the class-$\mathcal{S}$ limit $b_i \to b$ of the theory $\mathcal{T}_{1,1}$, the Schur index is given by the formula (\ref{Ign}). It satisfies several flavored modular differential equations of different weights.

At weight-two, there is one equation that corresponds to the total stress tensor $T = T_\text{Sug} + T_{\beta \gamma}$,
\begin{align}
	0 = \left[D^{(1)}_q
			-\frac{D_b^2}{2}
			-\left(2 E_1\begin{bmatrix}
				 1 \\
				 b^2 \\
				\end{bmatrix}
				+ E_1\begin{bmatrix}
				 -1 \\
				 b \\
				\end{bmatrix}\right)D_b
			-2 E_1\begin{bmatrix}
			 -1 \\
			 b \\
			\end{bmatrix} E_1\begin{bmatrix}
			 1 \\
			 b^2 \\
			\end{bmatrix}
			+3 E_2\begin{bmatrix}
			 1 \\
			 b^2 \\
			\end{bmatrix}
			+2 E_2\right]\mathcal{I}_{1,1} \ .
\end{align}

At weight-three, we have
\begin{align}
	0 = \Bigg[& D_b^3
	-4 E_1\begin{bmatrix}
	 1 \\
	 b^2 \\
	\end{bmatrix}D_q^{(1)} 
	+8 E_1\begin{bmatrix}
	 -1 \\
	 b \\
	\end{bmatrix}D_q^{(1)}
	+6  E_1\begin{bmatrix}
	 1 \\
	 b^2 \\
	\end{bmatrix}D_b^2 \nonumber\\
	& \ -16 E_2\begin{bmatrix}
	 -1 \\
	 b \\
	\end{bmatrix}D_b 
	-12 E_2 D_b 
	-12 E_1\begin{bmatrix}
	 -1 \\
	 b \\
	\end{bmatrix} E_1\begin{bmatrix}
	 1 \\
	 b^2 \\
	\end{bmatrix}D_b 
	-32 E_2\begin{bmatrix}
		 1 \\
		 b^2 \\
		\end{bmatrix}D_b \nonumber\\
	& \ -12 E_1\begin{bmatrix}
	 -1 \\
	 b \\
	\end{bmatrix}^2 E_1\begin{bmatrix}
	 1 \\
	 b^2 \\
	\end{bmatrix}
	-24 E_2 E_1\begin{bmatrix}
	 1 \\
	 b^2 \\
	\end{bmatrix}\nonumber\\
	& \ -8 E_1\begin{bmatrix}
	 1 \\
	 b^2 \\
	\end{bmatrix} E_2\begin{bmatrix}
	 -1 \\
	 b \\
	\end{bmatrix}-8 E_1\begin{bmatrix}
	 -1 \\
	 b \\
	\end{bmatrix} E_2\begin{bmatrix}
	 1 \\
	 b^2 \\
	\end{bmatrix} 
	+48 E_3\begin{bmatrix}
	 1 \\
	 b^2 \\
	\end{bmatrix} \Bigg] \mathcal{I}_{1,1} \ ,
\end{align}
and
\begin{align}
	0 = \Bigg[
	D_q^{(1)} D_b
	+2 E_1\begin{bmatrix}
		 -1 \\
		 b \\
		\end{bmatrix}D_q^{(1)}
	-2 E_2\begin{bmatrix}
	 1 \\
	 b^2 \\
	\end{bmatrix}D_b 
	+E_2 D_b
	-2 E_1\begin{bmatrix}
	 -1 \\
	 b \\
	\end{bmatrix} E_2\begin{bmatrix}
	 1 \\
	 b^2 \\
	\end{bmatrix}+6 E_3\begin{bmatrix}
	 1 \\
	 b^2 \\
	\end{bmatrix} \Bigg] \mathcal{I}_{1,1}\ .
\end{align}

At weight-four, there are several fairely complicated equations.
\begin{align}
	0 = \Bigg[  & \ D_b^4
	+
	8E_1\begin{bmatrix}
	 1 \\
	 b^2 \\
	\end{bmatrix}^3  D_b
	+ 48 E_1\begin{bmatrix}
	 -1 \\
	 b \\
	\end{bmatrix} E_1\begin{bmatrix}
	 1 \\
	 b^2 \\
	\end{bmatrix}D_q^{(1)} 
	-72 E_2\begin{bmatrix}
	 1 \\
	 b^2 \\
	\end{bmatrix}D_q^{(1)} 
	\nonumber\\
	& \ +48 D_b \left(E_1\begin{bmatrix}
	 -1 \\
	 b \\
	\end{bmatrix}\right)^2 E_1\begin{bmatrix}
	 1 \\
	 b^2 \\
	\end{bmatrix}
	+264 D_b E_1\begin{bmatrix}
	 -1 \\
	 b \\
	\end{bmatrix} E_2\begin{bmatrix}
	 1 \\
	 b^2 \\
	\end{bmatrix}
	+192 D_b E_1\begin{bmatrix}
	 1 \\
	 b^2 \\
	\end{bmatrix} E_2\begin{bmatrix}
	 1 \\
	 b^2 \\
	\end{bmatrix}\nonumber\\
	& \ +336 D_b E_3\begin{bmatrix}
	 1 \\
	 b^2 \\
	\end{bmatrix}+48 \left(E_1\begin{bmatrix}
	 -1 \\
	 b \\
	\end{bmatrix}\right)^3 E_1\begin{bmatrix}
	 1 \\
	 b^2 \\
	\end{bmatrix}-192 E_2 E_2\begin{bmatrix}
	 1 \\
	 b^2 \\
	\end{bmatrix} \nonumber\\
	 & +336 E_1\begin{bmatrix}
	 -1 \\
	 b \\
	\end{bmatrix} E_1\begin{bmatrix}
	 1 \\
	 b^2 \\
	\end{bmatrix} E_2\begin{bmatrix}
	 1 \\
	 b^2 \\
	\end{bmatrix}+288 E_1\begin{bmatrix}
	 1 \\
	 b^2 \\
	\end{bmatrix} E_3\begin{bmatrix}
	 -1 \\
	 b \\
	\end{bmatrix}-784 E_1\begin{bmatrix}
	 -1 \\
	 b \\
	\end{bmatrix} E_3\begin{bmatrix}
	 1 \\
	 b^2 \\
	\end{bmatrix}\nonumber\\
	& \ -432 E_1\begin{bmatrix}
	 1 \\
	 b^2 \\
	\end{bmatrix} E_3\begin{bmatrix}
	 1 \\
	 b^2 \\
	\end{bmatrix}-2136 E_4\begin{bmatrix}
	 1 \\
	 b^2 \\
	\end{bmatrix}
	+96 D_q^{(1)} E_2\begin{bmatrix}
	 -1 \\
	 b \\
	\end{bmatrix}+288 D_b E_3\begin{bmatrix}
	 -1 \\
	 b \\
	\end{bmatrix}\nonumber\\
	& \ +1472 E_1\begin{bmatrix}
	 -1 \\
	 b \\
	\end{bmatrix} E_3\begin{bmatrix}
	 -1 \\
	 b \\
	\end{bmatrix}-384 E_4\begin{bmatrix}
	 -1 \\
	 b \\
	\end{bmatrix}-48 E_2^2   \Bigg]\mathcal{I}_{1,1} \ ,
\end{align}
\begin{align}
	0 = \Bigg[ & D_b^2 D_q^{(1)} -12 E_1\begin{bmatrix}
	 -1 \\
	 b \\
	\end{bmatrix} E_1\begin{bmatrix}
	 1 \\
	 b^2 \\
	\end{bmatrix}D_q^{(1)}
	-16 E_2\begin{bmatrix}
	 -1 \\
	 b \\
	\end{bmatrix}D_q^{(1)} 
	-10 E_2\begin{bmatrix}
	 1 \\
	 b^2 \\
	\end{bmatrix}D_q^{(1)}
	-8 E_2  E_1\begin{bmatrix}
	 1 \\
	 b^2 \\
	\end{bmatrix}D_b\nonumber
	\\
	& \ -4 E_2 E_1\begin{bmatrix}
	 -1 \\
	 b \\
	\end{bmatrix}D_b 
	+6 E_1\begin{bmatrix}
	 -1 \\
	 b \\
	\end{bmatrix} E_2\begin{bmatrix}
	 1 \\
	 b^2 \\
	\end{bmatrix}D_b 
	+12 E_1\begin{bmatrix}
	 1 \\
	 b^2 \\
	\end{bmatrix} E_2\begin{bmatrix}
	 1 \\
	 b^2 \\
	\end{bmatrix}D_b 
	+18 E_3\begin{bmatrix}
	 1 \\
	 b^2 \\
	\end{bmatrix}D_b 
	\nonumber\\
	& \ +12 E_2 E_2\begin{bmatrix}
	 1 \\
	 b^2 \\
	\end{bmatrix}+24 E_1\begin{bmatrix}
	 -1 \\
	 b \\
	\end{bmatrix} E_1\begin{bmatrix}
	 1 \\
	 b^2 \\
	\end{bmatrix} E_2\begin{bmatrix}
	 1 \\
	 b^2 \\
	\end{bmatrix}+32 E_1\begin{bmatrix}
	 1 \\
	 b^2 \\
	\end{bmatrix} E_3\begin{bmatrix}
	 -1 \\
	 b \\
	\end{bmatrix}\nonumber\\
	& \ -4 E_1\begin{bmatrix}
	 -1 \\
	 b \\
	\end{bmatrix} E_3\begin{bmatrix}
	 1 \\
	 b^2 \\
	\end{bmatrix}-24 E_1\begin{bmatrix}
	 1 \\
	 b^2 \\
	\end{bmatrix} E_3\begin{bmatrix}
	 1 \\
	 b^2 \\
	\end{bmatrix}-74 E_4\begin{bmatrix}
	 1 \\
	 b^2 \\
	\end{bmatrix} \nonumber\\
	& \ 
	+24 E_1\begin{bmatrix}
	 -1 \\
	 b \\
	\end{bmatrix} E_3\begin{bmatrix}
	 -1 \\
	 b \\
	\end{bmatrix}+64 E_4\begin{bmatrix}
	 -1 \\
	 b \\
	\end{bmatrix}
	+4 E_2^2
	-8 E_2 E_1\begin{bmatrix}
	 -1 \\
	 b \\
	\end{bmatrix} E_1\begin{bmatrix}
	 1 \\
	 b^2 \\
	\end{bmatrix}
	\Bigg] \mathcal{I}_{1,1} \ ,
\end{align}
\begin{align}
	0 = \Bigg[ D_q^{(2)}
	 & + 2 E_3\begin{bmatrix}
	 1 \\
	 b^2 \\
	\end{bmatrix}D_b 
	-4 E_2\begin{bmatrix}
	 -1 \\
	 b \\
	\end{bmatrix}D_q^{(1)}
	-4  E_3\begin{bmatrix}
	 -1 \\
	 b \\
	\end{bmatrix}D_b\\
	& \ 
	+\frac{8}{3} E_1\begin{bmatrix}
	 -1 \\
	 b \\
	\end{bmatrix} E_3\begin{bmatrix}
	 -1 \\
	 b \\
	\end{bmatrix}
	+16 E_4\begin{bmatrix}
	 -1 \\
	 b \\
	\end{bmatrix}
	+\frac{2}{3} E_1\begin{bmatrix}
	 -1 \\
	 b \\
	\end{bmatrix} E_3\begin{bmatrix}
	 1 \\
	 b^2 \\
	\end{bmatrix}
	-11 E_4\begin{bmatrix}
	 1 \\
	 b^2 \\
	\end{bmatrix} \Bigg]\mathcal{I}_{1,1} \ . \nonumber
\end{align}

\subsection{\texorpdfstring{$\mathcal{T}_{1,2}$}{}}

Here we consider the $b_i \to b$ limit. There are two equations at weight-three,
\begin{align}
	0 = \Bigg[ D_b^3-32 E_1\begin{bmatrix}
	 1 \\
	 b^2 \\
	\end{bmatrix}D_q^{(1)} 
	& +16 E_1\begin{bmatrix}
	 1 \\
	 b^2 \\
	\end{bmatrix}D_b^2 
	 +48  E_1\begin{bmatrix}
	 1 \\
	 b^2 \\
	\end{bmatrix}^2D_b
	-96 E_2\begin{bmatrix}
	 1 \\
	 b^2 \\
	\end{bmatrix}D_b \nonumber\\ 
	& \ -32 E_2 E_1\begin{bmatrix}
	 1 \\
	 b^2 \\
	\end{bmatrix}-288 E_1\begin{bmatrix}
	 1 \\
	 b^2 \\
	\end{bmatrix} E_2\begin{bmatrix}
	 1 \\
	 b^2 \\
	\end{bmatrix}+96 E_3\begin{bmatrix}
	 1 \\
	 b^2 \\
	\end{bmatrix} \Bigg] \mathcal{I}_{1,2} \ ,
\end{align}
and
\begin{align}
	0 = \Bigg[  D_b D_q 
	-6 D_b E_2\begin{bmatrix}
	 1 \\
	 b^2 \\
	\end{bmatrix}
	& +4 E_2 D_b \nonumber\\
	& \ +12 E_2 E_1\begin{bmatrix}
	 1 \\
	 b^2 \\
	\end{bmatrix}-12 E_1\begin{bmatrix}
	 1 \\
	 b^2 \\
	\end{bmatrix} E_2\begin{bmatrix}
	 1 \\
	 b^2 \\
	\end{bmatrix}+12 E_3\begin{bmatrix}
	 1 \\
	 b^2 \\
	\end{bmatrix} \Bigg]\mathcal{I}_{1,2} \ .
\end{align}

At weight-four, there are two equations,
\begin{align}
  0 = \Bigg[	& D_q^{(2)} -\frac{1}{8} D_b^2 D_q^{(1)}
	-\frac{3}{2} E_1\begin{bmatrix}
	 1 \\
	 b^2 \\
	\end{bmatrix}D_b D_q^{(1)}
	+4 E_2\begin{bmatrix}
	 1 \\
	 b^2 \\
	\end{bmatrix}D_q^{(1)}
	+2 E_2 D_q^{(1)}\nonumber
	\\
	& \ 
	+\frac{1}{4}  E_2\begin{bmatrix}
	 1 \\
	 b^2 \\
	\end{bmatrix}D_b^2
	+3 E_1\begin{bmatrix}
	 1 \\
	 b^2 \\
	\end{bmatrix} E_2\begin{bmatrix}
	 1 \\
	 b^2 \\
	\end{bmatrix}D_b 
	\\
	& \ -8 E_2 E_2\begin{bmatrix}
	 1 \\
	 b^2 \\
	\end{bmatrix}-24 E_1\begin{bmatrix}
	 1 \\
	 b^2 \\
	\end{bmatrix} E_3\begin{bmatrix}
	 1 \\
	 b^2 \\
	\end{bmatrix}-24 E_4\begin{bmatrix}
	 1 \\
	 b^2 \\
	\end{bmatrix}-E_2^2-31 E_4  \Bigg] \mathcal{I}_{1,2} \ , \nonumber
\end{align}
and
\begin{align}
	0 = \Bigg[  D_q^{(2)} & -D_b D_q^{(1)} E_1\begin{bmatrix}
	 1 \\
	 b^2 \\
	\end{bmatrix}+4 E_2\begin{bmatrix}
	 1 \\
	 b^2 \\
	\end{bmatrix}D_q^{(1)}
	+2 E_2 D_q^{(1)}
	-\frac{1}{2} E_2\begin{bmatrix}
	 1 \\
	 b^2 \\
	\end{bmatrix}D_b^2 
	+\frac{1}{2} E_2 D_b^2\\
	& \ +2 E_2 E_1\begin{bmatrix}
	 1 \\
	 b^2 \\
	\end{bmatrix}D_b 
	+6 E_3\begin{bmatrix}
	 1 \\
	 b^2 \\
	\end{bmatrix}D_b 
	-8 E_2 E_2\begin{bmatrix}
	 1 \\
	 b^2 \\
	\end{bmatrix}-24 E_4\begin{bmatrix}
	 1 \\
	 b^2 \\
	\end{bmatrix}-4 E_2^2-16 E_4  \Bigg]\mathcal{I}_{1,2} \ . \nonumber
\end{align}

\clearpage

{
\bibliographystyle{utphys}
\bibliography{ref}
}

\end{document}